\documentclass[twocolumn,
amsmath,amssymb,
 aps,
 prd,
superscriptaddress,
altaffilletter,
10pt
]{revtex4-2}

\usepackage{graphicx}\usepackage{dcolumn}\usepackage{bm}\usepackage{siunitx}
\usepackage{amsmath}
\usepackage{ifthen}
\usepackage[colorlinks = true, linkcolor = blue, breaklinks=true]{hyperref}
\usepackage{xcolor}
\usepackage{isotope}
\usepackage{multirow}
\usepackage{datetime}

\usepackage{lineno}

\usepackage[super]{nth}
\usepackage{float}
\usepackage{booktabs}
\usepackage{xspace}
\usepackage[symbol]{footmisc}
\usepackage{siunitx}
\usepackage[nolist]{acronym} \usepackage{lineno}
\usepackage{orcidlink}
\usepackage{cuted}
\usepackage{ulem}

\newcommand{\amu}{\ensuremath{a^{}_{\mu}}\xspace}
\newcommand{\amuSM}{\ensuremath{a^{SM}_{\mu}}\xspace}

\newcommand{\amuHLO}{\ensuremath{a^{HLO}_{\mu}}\xspace}
\newcommand{\gm}{\ensuremath{g\!-\!2}\xspace}
\newcommand{\gmtwo}{\gm}
\newcommand{\oa}{\ensuremath{\omega^{}_a}\xspace}
\newcommand{\oam}{\ensuremath{\omega^{m}_a}\xspace}
\newcommand{\ocycl}{\ensuremath{\omega^{}_c}\xspace}

\newcommand{\op}{\ensuremath{\omega^{}_p}\xspace}

\newcommand{\opprime}{\ensuremath{\omega'^{}_p}\xspace}
\newcommand{\opprimeatTexp}{\ensuremath{\omega'^{}_p(\Tr)}\xspace}
\newcommand{\opprimetilde}{\ensuremath{\tilde{\omega}'^{}_p}\xspace}
\newcommand{\opprimetildeofT}{\ensuremath{\tilde{\omega}'^{}_p(T)}\xspace}
\newcommand{\opprimetildeatTexp}{\ensuremath{\tilde{\omega}'^{}_p(\Tr)}\xspace}

\newcommand{\FNALexperiment}{\ac{FNAL} Muon \gm Experiment\xspace}
\newcommand{\RunOne}{Run-1\xspace}
\newcommand{\RunTwo}{Run-2\xspace}
\newcommand{\RunThree}{Run-3\xspace}
\newcommand{\RunTwoThree}{Run-2/3\xspace}
\newcommand{\RunFour}{Run-4\xspace}
\newcommand{\RunFournoRF}{noRF\xspace}
\newcommand{\RunFive}{Run-5\xspace}
\newcommand{\RunFiveX}{xRF\xspace}
\newcommand{\RunFiveXY}{xyRF5\xspace}
\newcommand{\RunSix}{Run-6\xspace}
\newcommand{\RunSixXY}{xyRF6\xspace}

\newcommand{\RunFourFiveSix}{Run-4/5/6\xspace}

\newcommand{\pp}{\mathrm{cp}}
\newcommand{\muBohr}{\ensuremath{\mu^{}_\text{B}}\xspace}

\newcommand{\opprimeJPatTexp}{\ensuremath{\omega'^{}_{p,\mathrm{JP}}(\Tr)}\xspace}
\newcommand{\opprimeUSatTexp}{\ensuremath{\omega'^{}_{p,\mathrm{US}}(\Tr)}\xspace}
\newcommand{\opmeasUS}{\omega_{p,\mathrm{US}}^{\pp}\xspace}

\newcommand{\opmeasJP}{\omega_{p,\mathrm{JP}}^{\pp}\xspace}

\newcommand{\Tr}{\ensuremath{T^{}_{r}}\xspace}
\newcommand{\TJP}{\ensuremath{T^{}_\mathrm{JP}}\xspace}
\newcommand{\TUS}{\ensuremath{T^{}_\mathrm{US}}\xspace}

\newcommand{\deltasetup}{\Delta\delta^\mathrm{setup}}
\newcommand{\deltaT}{\delta^{T}}

\newcommand{\deltatJP}{\delta^{t}_\mathrm{JP}}
\newcommand{\deltatUS}{\delta^{t}_\mathrm{US}}
\newcommand{\deltab}{\delta^{b}}

\DeclareMathOperator{\sinc}{sinc}

\newcommand{\Rmu}{\ensuremath{{\mathcal R}^{}_\mu}\xspace}
\newcommand{\Rmuprime}{\ensuremath{{\mathcal R}'^{}_\mu}\xspace}

\newcommand{\authornote}[1]{{\let\thempfn\relax \footnotetext[0]{$\diamond${ }#1}}}

\newcommand{\wagroupZep}{G1\xspace}
\newcommand{\wagroupTyler}{G2\xspace}
\newcommand{\wagroupMurong}{G3\xspace}
\newcommand{\wagroupOnKim}{G4\xspace}
\newcommand{\wagroupSJTU}{L2\xspace}
\newcommand{\wagroupUKY}{E1\xspace}
\newcommand{\wagroupBU}{L1\xspace}
\newcommand{\wagroupITA}{L3\xspace}
\newcommand{\wagroupEuropa}{L3\xspace}
\newcommand{\Roa}{\ensuremath{R\left( \oa \right)}\xspace}

\newcommand{\alphabeta}{\ensuremath{\alpha/\beta}\xspace}
\newcommand{\Aphi}{\ensuremath{A/\phi}\xspace}
\newcommand{\BNLFitModel}{BNL\xspace}

\newcommand{\amuFNALprecisionppb}{\SI{127}{ppb}\xspace}

\newcommand{\amuworld}{\ensuremath{0.\,001\,165\,920\,715(145)}\xspace}

\newcommand{\amuWP}{\ensuremath{0.\,001\,165\,920\,33(62)}\xspace}

\newcommand{\SMref}{Aoyama:2012wk,Volkov:2019phy,Volkov:2024yzc,Aoyama:2024aly,Parker:2018vye,Morel:2020dww,Fan:2022eto,Czarnecki:2002nt,Gnendiger:2013pva,Ludtke:2024ase,Hoferichter:2025yih,RBC:2018dos,Giusti:2019xct,Borsanyi:2020mff,Lehner:2020crt,Wang:2022lkq,Aubin:2022hgm,Ce:2022kxy,ExtendedTwistedMass:2022jpw,RBC:2023pvn,Kuberski:2024bcj,Boccaletti:2024guq,Spiegel:2024dec,RBC:2024fic,Djukanovic:2024cmq,ExtendedTwistedMass:2024nyi,MILC:2024ryz,Bazavov:2024eou,Keshavarzi:2019abf,DiLuzio:2024sps,Kurz:2014wya,Colangelo:2015ama,Masjuan:2017tvw,Colangelo:2017fiz,Hoferichter:2018kwz,Eichmann:2019tjk,Bijnens:2019ghy,Leutgeb:2019gbz,Cappiello:2019hwh,Masjuan:2020jsf,Bijnens:2020xnl,Bijnens:2021jqo,Danilkin:2021icn,Stamen:2022uqh,Leutgeb:2022lqw,Hoferichter:2023tgp,Hoferichter:2024fsj,Estrada:2024cfy,Deineka:2024mzt,Eichmann:2024glq,Bijnens:2024jgh,Hoferichter:2024bae,Holz:2024diw,Cappiello:2025fyf,Colangelo:2014qya,Blum:2019ugy,Chao:2021tvp,Chao:2022xzg,Blum:2023vlm,Fodor:2024jyn}

\begin{document}

\title{
Final Report on the Measurement of the Positive Muon Anomalous Magnetic Moment at Fermilab to 127\,ppb
}
\renewcommand{\thefootnote}{\fnsymbol{footnote}}
\affiliation{Argonne National Laboratory, Lemont, Illinois, USA}
\affiliation{Boston University, Boston, Massachusetts, USA}
\affiliation{Brookhaven National Laboratory, Upton, New York, USA}
\affiliation{Budker Institute of Nuclear Physics, Novosibirsk, Russia}
\affiliation{Center for Axion and Precision Physics (CAPP) / Institute for Basic Science (IBS)}
\affiliation{Cornell University, Ithaca, New York, USA}
\affiliation{Fermi National Accelerator Laboratory, Batavia, Illinois, USA}
\affiliation{INFN, Laboratori Nazionali di Frascati, Frascati, Italy}
\affiliation{INFN, Sezione di Napoli, Naples, Italy}
\affiliation{INFN, Sezione di Pisa, Pisa, Italy}
\affiliation{INFN, Sezione di Roma Tor Vergata, Rome, Italy}
\affiliation{INFN, Sezione di Trieste, Trieste, Italy}
\affiliation{Department of Physics and Astronomy, James Madison University, Harrisonburg, Virginia, USA}
\affiliation{Institute of Physics and Cluster of Excellence PRISMA++, Johannes Gutenberg University Mainz, Mainz, Germany}
\affiliation{Korea Advanced Institute of Science and Technology (KAIST)}
\affiliation{Michigan State University, East Lansing, Michigan, USA}
\affiliation{North Central College, Naperville, Illinois, USA}
\affiliation{Northern Illinois University, DeKalb, Illinois, USA}
\affiliation{Regis University, Denver, Colorado, USA}
\affiliation{School of Physics and Astronomy, Shanghai Jiao Tong University, Shanghai, China}
\affiliation{Tsung-Dao Lee Institute, Shanghai Jiao Tong University, Shanghai, China}
\affiliation{Department of Physics and Astronomy, Trinity University, San Antonio, Texas, USA}
\affiliation{Institut f\"ur Kern- und Teilchenphysik, Technische Universit\"at Dresden, Dresden, Germany}
\affiliation{Universit\`a del Molise, Campobasso, Italy}
\affiliation{Universit\`a di Udine, Udine, Italy}
\affiliation{Department of Physics and Astronomy, University College London, London, United Kingdom}
\affiliation{University of Illinois at Urbana-Champaign, Urbana, Illinois, USA}
\affiliation{University of Kentucky, Lexington, Kentucky, USA}
\affiliation{University of Liverpool, Liverpool, United Kingdom}
\affiliation{Department of Physics and Astronomy, University of Manchester, Manchester, United Kingdom}
\affiliation{Department of Physics, University of Massachusetts, Amherst, Massachusetts, USA}
\affiliation{University of Michigan, Ann Arbor, Michigan, USA}
\affiliation{University of Mississippi, University, Mississippi, USA}
\affiliation{University of Virginia, Charlottesville, Virginia, USA}
\affiliation{University of Washington, Seattle, Washington, USA}
\affiliation{City University of New York at York College, Jamaica, New York, USA}
\author{D.~P.~Aguillard\orcidlink{0000-0002-5556-1169}} \affiliation{University of Michigan, Ann Arbor, Michigan, USA}
\author{T.~Albahri\orcidlink{0000-0002-1191-5463}} \affiliation{University of Liverpool, Liverpool, United Kingdom}
\author{D.~Allspach\orcidlink{0009-0008-2417-5338}} \affiliation{Fermi National Accelerator Laboratory, Batavia, Illinois, USA}
\author{J.~Annala} \affiliation{Fermi National Accelerator Laboratory, Batavia, Illinois, USA}
\author{K.~Badgley\orcidlink{0000-0002-3942-5108}} \affiliation{Fermi National Accelerator Laboratory, Batavia, Illinois, USA}
\author{S.~Bae{\ss}ler\orcidlink{0000-0001-7732-9873}} \affiliation{University of Virginia, Charlottesville, Virginia, USA}
\author{L.~Bailey\orcidlink{0009-0004-7290-8963}} \affiliation{Department of Physics and Astronomy, University College London, London, United Kingdom}
\author{E.~Barlas-Yucel\orcidlink{0000-0001-9562-2125}} \altaffiliation[Now at ]{Fermi National Accelerator Laboratory, Batavia, Illinois, USA.} \affiliation{University of Illinois at Urbana-Champaign, Urbana, Illinois, USA}
\author{T.~Barrett\orcidlink{0000-0002-0807-5846}} \affiliation{Cornell University, Ithaca, New York, USA}
\author{E.~Barzi\orcidlink{0000-0001-5829-2147}} \affiliation{Fermi National Accelerator Laboratory, Batavia, Illinois, USA}
% ...
\footnotetext[2]{Deceased.}
\renewcommand{\thefootnote}{\arabic{footnote}}
\collaboration{The Muon \gmtwo Collaboration} \noaffiliation
\vskip 0.25cm

\begin{abstract}
This report details the final measurement of the muon magnetic anomaly, $a_{\mu}=(g_{\mu}-2)/2$, by the Muon \gm experiment at Fermi National Accelerator Laboratory (FNAL), using positive muons collected from 2021 to 2023.
The value of $a_{\mu}$ is determined from the ratio of the anomalous spin precession frequency to the shielded proton precession frequency in the muon storage ring magnetic field, combined with external constants known at the 22 ppb level.
The new dataset, containing over $2.5$ times the statistics of our previous results, yields
$a_{\mu}=116\,592\,0710(162)\times 10^{-12}$ (\SI{139}{ppb}), or $a_{\mu}=116\,592\,0705(148)\times 10^{-12}$ (\SI{127}{ppb}) when combined with our previous results.
The new experimental world average, dominated by the measurements at FNAL, is
$a_{\mu}^{\text{Exp}}=116\,592\,0715(145)\times 10^{-12}$ (\SI{124}{ppb}).
\end{abstract}

\maketitle

\tableofcontents

\section{Introduction}
Measurements of magnetic moments of charged leptons have been crucial in the development of the \ac{SM} of particle physics. The g-factor, $g^{}_l$, of a charged lepton $l=(e, \mu, \tau)$, plays a central role in these measurements as it relates the lepton's spin $\vec{S}$ to its magnetic moment
\begin{equation*}
    \vec{\mu}^{}_l = g^{}_l\frac{q^{}_l}{2m^{}_l} \vec{S},
\end{equation*}
where $q^{}_l$ is the charge and $m^{}_l$ the mass of the lepton. For a structureless, spin-$\frac{1}{2}$ particle like the electron, the Dirac theory predicted $g^{}_e \equiv 2$~\cite{Dirac:1928}, after Uhlenbeck and Goudsmit had already postulated $g^{}_e = 2$ to explain the anomalous Zeeman effect \cite{Uhlenbeck:1925pqz}. Two decades later, experimental results~\cite{PhysRev.72.1256.2, Kusch1948} revealed $g^{}_e \neq 2$ and motivated Schwinger's updated theoretical value of $g^{}_e = 2 + \frac{\alpha}{\pi}$ due to the leading order radiative correction~\cite{Schwinger1948}. This formed the basis for the introduction of the magnetic anomaly, $a^{}_{l}= (g^{}_l\!-\!2)/2$, and the understanding that $a^{}_l$ is generated by the interaction of the lepton with virtual particles. The magnetic anomaly, also known as the anomalous magnetic moment, is thus a sensitive probe to test the \ac{SM} and has played a vital role in the development of modern quantum field theories. 
The anomalous magnetic moment is sensitive to the full particle content of nature and the interactions among these particles, and thus the completeness of the \ac{SM} can be tested by comparing its prediction with the experimental measurement.

The observed experimental deviation from $g^{}_e = 2$ and Schwinger's explanation motivated decades of research to improve the precision of both the experimental measurement and the \ac{SM} value of the magnetic anomaly of charged leptons. Measurements of the electron anomaly have now reached a precision of \SI{112}{ppt}~\cite{PhysRevLett.130.071801}, by far the most precise among any elementary particle. However, the sensitivity to \ac{BSM} physics through loop-induced contributions from a possible new particle with mass $M$ is proportional to $m^2_l/M^2$ in most models, enhancing the sensitivity of the muon to \ac{BSM} physics by $m^2_\mu / m^2_e \approx 43,000$ compared to the electron.  Since the electron anomaly is less sensitive to \ac{BSM} physics and heavily dominated by QED contributions, equating the experimental measurement and theoretical prediction provides a determination of the fine structure constant, $\alpha$, which enters as a parameter in the prediction. The result can be compared with independent determinations from atom-interferometry measurements in Cs~\cite{Parker:2018vye} and Rb~\cite{Morel:2020dww}.

Early experiments with the muon at Columbia University Nevis Laboratory~\cite{PhysRev.105.1415,PhysRev.109.973,PhysRev.118.271} and the University of Liverpool~\cite{Cassels_1957} showed $g^{}_\mu \approx 2$, confirming a similar behavior for the muon as for the electron. Precision improvements continued at \ac{CERN} with direct measurements of \amu that were pioneered at the CERN-I experiment~\cite{Charpak1965}. 
The next generation CERN-II experiment used a storage ring for the first time to measure $a^{}_\mu$ with \SI{270}{ppm} precision~\cite{BAILEY1968287}. The final CERN-III experiment~\cite{Bailey1979} used pion injection and pioneered the magic-momentum technique, for which electric focusing does not affect the spin precession, concluding the experiments at \ac{CERN} with a precision of \SI{7}{ppm}.
The \ac{BNL} E821 experiment further improved the magic-momentum storage ring technique by switching to direct muon injection and several years of data taking improved the precision by more than an order of magnitude to \SI{0.54}{ppm}~\cite{PhysRevD.73.072003}. The E821 results revealed a tension relative to the \ac{SM} value, which, with improved theoretical calculations, grew to around $3.5$ standard deviations~\cite{keshavarzi:2018mgv, Davier:2017zfy}.

This observed discrepancy and the sensitivity of the magnetic anomaly to particles and their interactions were the motivation for both the latest iterations of storage ring experiments, the Muon \gm Experiment~\cite{grange2018muong2technicaldesign} at \ac{FNAL} and the Muon \gm/EDM experiment~\cite{10.1093/ptep/ptz030} at \ac{J-PARC}, and the formation of the Muon \gm \ac{TI} \footnote{\url{https://muon-gm2-theory.illinois.edu/}}, an international collaboration working to improve the recommended \ac{SM} value for \amu. Details on the progress and current challenges for the determination of the \ac{SM} value in the past few years are described in Section~\ref{sec:theorycomparison}.

The Muon \gm Experiment at \ac{FNAL} was designed and commissioned with the goal of improving the experimental precision by a factor of four compared to the \ac{BNL} E821 experiment, originally intending to provide a conclusive statement on the hints of \ac{BSM} physics. It is the latest experiment using the storage ring technique at the magic momentum. Many improvements in the muon production and transport beam lines, the pulsed muon injection systems, and the muon beam, decay positron, and magnetic field detectors, compared to the \ac{BNL} E821 experiment, have been crucial to reaching and surpassing the design goals. The experiment started taking quality physics data in 2018 (\RunOne) and the first result \cite{Run1PRL, Run1PRDomegaa, Run1PRAField, Run1PRAB} with a precision of \SI{0.46}{ppm} based on about \SI{5}{\%} of the full statistics, was a strong confirmation of the \ac{BNL} E821 result. Combined with subsequent data taken in 2019 and 2020 (\RunTwoThree), the second result yielded a more than two-fold improvement of the precision to \SI{0.2}{ppm} \cite{PhysRevLett.131.161802, PhysRevD.110.032009} and confirmed the experimental value of \amu. The most recent result \cite{PRL_Run-4/5/6} added the data taken in the final three years through \RunFourFiveSix in 2021/2022/2023 and comprises about \SI{70}{\%} of the full statistics. The combined dataset for \RunOne to \RunSix achieved \SI{98}{ppb} statistical uncertainty through the acquired muon statistics and reduced the total systematic uncertainty to \SI{78}{ppb}. This reduction in statistical and systematic uncertainty led to the latest milestone in the measurement of \amu for a single experiment with a precision of \amuFNALprecisionppb, surpassing our design goal by ten percent.

The final result from the Muon \gm experiment provides the most precise measurement of the muon magnetic anomaly. The achieved experimental precision and the stability of the experimental result over the decades serve as a fundamental benchmark for any future extension to the \ac{SM}. This report provides details of this latest measurement of \amu and the associated data analysis, focusing on improvements in the hardware and analysis compared to previous results~\cite{Run1PRL, PhysRevLett.131.161802}. Section~\ref{sec:experiment} describes the experimental technique and the analysis scheme, while Sec.~\ref{sec:datasets} summarizes the characteristics of the main datasets \RunOne to \RunSix. Sections~\ref{sec:oam}--\ref{sec:op} detail the analyses for the anomalous spin precession frequency, \oam, the beam dynamics corrections, and the magnetic field, \op, respectively, with a focus on the analysis changes and improvements in \RunFourFiveSix. Section~\ref{sec:amu} provides the details for the determination of \amu for the latest datasets \RunFourFiveSix, the combination of all datasets, and the new world average. Section~\ref{sec:consistencyOfAmu} summarizes several consistency checks that were performed to check the \amu results versus various changes in the experimental settings. Section~\ref{sec:theorycomparison} provides an overview of the current status of the \amu evaluation in the \ac{SM} while Sec.~\ref{sec:conclusion} summarizes this latest result from the Muon \gm experiment at \ac{FNAL} and gives a brief outlook for the relevant future experimental results from the collaboration and other, closely related experiments.
 
\section{The Muon \gm experiment}\label{sec:experiment}
\label{sec:experiment_principle}

This section first outlines the principle of measuring the muon magnetic anomaly and then describes the experiment in detail.
Polarized muons are stored in a very uniform vertical magnetic field, $\vec{B}$, where they circulate at the cyclotron frequency, \ocycl, and their spins precess at the spin precession frequency, $\omega_{s}$. In the ideal case of a muon on a circular orbit perpendicular to a uniform, vertical magnetic field, in the absence of an electric field $\vec{E}$ and assuming no muon electric dipole moment $\vec{d}$, the anomalous precession frequency $\vec\omega_a=\vec\omega_s-\vec \omega_c$ takes the simple form
\begin{equation}
    \vec{\omega}^{}_a = - a_{\mu}\frac{q^{}_\mu}{m^{}_\mu}\vec{B},
    \label{eq:amuideal}
\end{equation} 
where $q^{}_\mu$ and $m^{}_\mu$ are the charge and mass of the muon, respectively. The magnetic field strength $B = |\vec{B}|$ is expressed in terms of the Larmor frequency of shielded protons in a spherical water sample at a reference temperature, $\Tr=\SI{25}{\celsius}$
\begin{equation}
\opprime(\Tr) = \frac{2\mu'_\text{p}(\Tr)}{\hbar} B,
    \label{eq:omega_pB}
\end{equation}
where $\mu'_\text{p}(\Tr)$ is the shielded proton magnetic moment at \Tr. In the experiment, the relevant magnetic field to which the muons are exposed is denoted by $\tilde B$ (or equivalently by \opprimetildeatTexp), to indicate that the magnetic field is weighted by the muon distribution in the storage region. The ratio of these two frequencies \oa and \opprimetildeatTexp defines
\begin{equation}
    \Rmuprime(\Tr) \equiv \frac{\oa}{\opprimetildeatTexp}.
    \label{eq:Rmu}
\end{equation}

Combining Eqs.~\eqref{eq:amuideal}--\eqref{eq:Rmu} together with the definition of the Bohr magneton, $\muBohr=e\hbar/2m^{}_e$, the muon's anomalous magnetic moment is derived via

\begin{linenomath*}
\begin{equation}
\begin{aligned}
a_{\mu} = \Rmuprime(\Tr) \frac{\mu'_\text{p}(\Tr)}{\muBohr} \frac{m_{\mu}}{m_\text{e}}.
\label{eq:amueqnewcodata}
\end{aligned}
\end{equation}
\end{linenomath*}

$\Rmuprime(\Tr)$ is measured in the experiment, whereas the ratios $\mu'_\text{p}(\Tr)/\muBohr$ and $m_{\mu}/m_e$ are precisely known from other experiments~\footnote{$\mu'_p/\mu^{}_B=\SI{1.5209931551(62)e-3}{}$;\newline $m_{\mu}/m_e=206.768\,2827(46)$, both from~\cite{mohr2024codatarecommendedvaluesfundamental}.}.

Equations~\eqref{eq:amuideal}--\eqref{eq:amueqnewcodata} refer to the idealized storage of muons with $\vec{\beta} \cdot \vec{B} = 0$, where $\vec{\beta}$ is the velocity vector of the muon divided by the speed of light. 
In practice, muons enter the ring with finite transverse momenta, which requires weak vertical focusing to maintain storage.  Furthermore, vertical betatron oscillations lead to the muon motion not being purely perpendicular to the magnetic field. 
These effects, if unaccounted for, result in a measured frequency $\oam$ that deviates from the ideal frequency \oa. 
Additional terms are required to extend the simple relation in Eq.~\eqref{eq:amuideal} to
\begin{eqnarray}
\vec{\omega}^{}_{a} & \approx &
- \frac{q^{}_\mu}{m^{}_\mu}\left[ \amu\vec{B}\nonumber - a_{\mu}\left(\frac{\gamma}{\gamma+1}\right) (\vec{\beta} \cdot \vec{B})\vec{\beta} \right.\\
&& - \left. \left(\amu - \frac{1}{\gamma^2-1}\right)\frac{\vec{\beta}\times \vec{E}}{c} \right],
   \label{eq:amurealistic}
\end{eqnarray}
so that to first order $\omega^{m}_{a}$ is $|\vec{\omega^{}_a}|$ (some additional corrections are needed), the frequency measured in the experiment, $\gamma$ is the Lorentz factor of the muon, and, in our experimental setup, $\vec{E}$ is a weakly vertical-focusing and horizontally de-focusing quadrupole-dominated electric field. The approximate sign in Eq.~\eqref{eq:amurealistic} is due to the fact that there are additional effects that lead to further deviations of \oam discussed below. To first order, both the second and third terms in Eq.~\eqref{eq:amurealistic} vanish for an ideal muon precisely at the magic momentum and with a zero pitch angle. However, because the actual stored muons possess a finite momentum spread and undergo vertical betatron oscillations ($\vec{\beta}\cdot\vec{B} \neq 0$), the actual contributions of these terms are accounted for as corrections. An important feature of Eq.~\eqref{eq:amuideal} is that $\omega_a$ is independent of the muon momentum. Note that the presence of an electric dipole moment of the muon would further modify Eq.~\eqref{eq:amurealistic}.

\paragraph*{Muon Production and Injection}
While Eq.~\eqref{eq:amueqnewcodata} forms the basis of the measurement of \amu outlined above, the experimental setup requires many additional components for the actual measurement described below. The experiment uses polarized muons near the magic momentum of \SI{3.094}{GeV/c} ($\gamma=29.3$) for which the third term on the right-hand side of Eq.~\eqref{eq:amurealistic} vanishes.

\begin{figure*}[tb]
\centering
\includegraphics[width=0.88\textwidth]{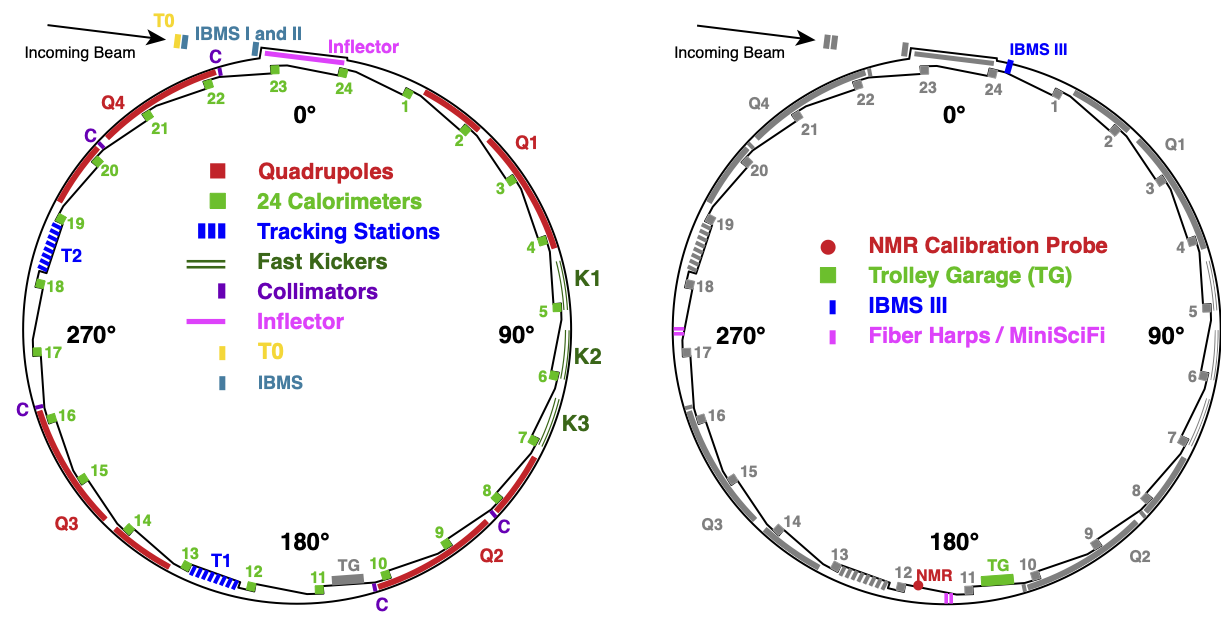}
\caption{Schematics of the Muon \gm storage ring with the various systems described in the text. The left panel represents the systems that are used during the regular data taking with stored muon beam while the right panel indicates additional systems that are used for special data taking modes (e.g. the field mapping, field calibration, systematic studies etc.). }\label{fig:ring_schematics}
\end{figure*}

The muons are produced by the following sequence at \ac{FNAL}.  Intense bunches of $\sim 10^{12}$ \SI{8}{\giga\electronvolt} protons are formed in the Recycler Ring with slightly different characteristic temporal shapes.  These bunches impinge on the former Tevatron anti-proton target made from Inconel~\cite{Church:1993ga}. 
Positive pions (and protons) at 3.1\,GeV/$c$ are captured by a 279\,m-long beamline, along which $\simeq80\%$ of $\pi^+$ decay to muons. The particles next enter the  505\,m circumference Delivery Ring (DR), where they circulate four times, allowing the remaining pions to decay away, and the (slower) protons to be swept out by a kicker.  The relatively pure, highly polarized, muon beam 
enters the \ac{SR}~\cite{SR-NIM} through a superconducting inflector magnet~\cite{YAMAMOTO200223} mounted in a hole in the outer yoke of the storage magnet that cancels the main magnetic field to provide a quasi field-free injection channel, delivering muons tangentially to the circumference of the \ac{SR}. Figure~\ref{fig:ring_schematics} shows bird's-eye views of the muon injection and \ac{SR} with the experimental systems, which will be described below. The left panel in Fig.~\ref{fig:ring_schematics} shows the detectors used for the normal data taking with stored muons while the right panel shows additional systems for calibrations or systematic studies. 

The timing information and intensity profiles of the incoming muon bunches are measured immediately upstream of the \ac{SR} by the T0 detector, a 1-mm-thick plastic scintillator with photo-multiplier readout at both ends. The horizontal and vertical profiles are measured by the \ac{IBMS}~\cite{osti_2333018}, consisting of three stations of scintillating fiber detectors. IBMS-1 is located downstream of the T0 detector, and IBMS-2 is placed immediately upstream of the inflector entrance.  The IBMS-3 station is located about \SI{1}{\meter} downstream of the inflector exit, inside the vacuum system. It is normally retracted but can be inserted into the \ac{SR} to provide horizontal beam profiles over multiple turns to aid with beam tuning. 

The muon beam exits the inflector with a \SI{77}{\milli\meter} radial offset from the ideal orbit. A fast kicker magnet~\cite{schreckenberger_fast_2021}, placed \SI{90}{\degree} in azimuth relative to the inflector magnet, then deflects the muon ensemble onto the storage orbit. \Acp{ESQ}~\cite{SEMERTZIDIS2003458}, located in $4$ symmetric regions around the azimuth and covering \SI{43}{\percent} of the storage ring circumference, provide the vertical storage for the muons. Initially, muons are injected in bunches with durations of less than \SI{120}{\nano\second}, creating a modulation of hits from muon decay positrons in individual detectors with the cyclotron period ($T_{c} \approx \SI{149}{\nano\second}$). The resulting intensity signal is referred to as the \ac{FR}. Due to the momentum spread of the stored muons ($\delta=\Delta p/p_0 \approx \SI{0.15}{\percent}$), this bunched structure gradually disperses over the time of muon storage of about \SI{700}{\micro\second}, roughly ten times their lab-frame lifetime of \SI{64.4}{\micro\second}. 
This storage window is called a fill.
Set by the Fermilab accelerator cycle, the experiment receives two trains of 8 fills over the course of $\sim$\SI{1.4}{\second}. For a small fraction of the data collected, the experiment received only one train of 8 fills per \SI{1.4}{\second}. Fills within each train are separated by \SI{10}{\milli\second}.

\paragraph*{Storage Ring}
Efficient muon injection and storage is hampered because of the mismatch between the inflector’s $\num{18} \times \num{56}\,\si{\square\milli\meter}$ small aperture compared to the incoming beam’s phase space, and the storage ring’s acceptance with its \SI{90}{\milli\meter} diameter storage aperture.
This mismatch leads to significant beam oscillations in the storage ring. The stored muons exhibit betatron oscillations in both the horizontal and vertical directions. The oscillation frequencies are determined by the effective field index~\footnote{The field index n is defined by $n = R^{}_0/v B^{}_0 \partial E^{}_y/\partial y$, where $R^{}_0$ is the central orbit radius, $v$ is the muon velocity, $B^{}_0$ is the magnetic field, and $\partial E^{}_y / \partial y$ is the gradient in the effective vertical electric field.} $n \approx 0.108$, which results from the uniform magnetic dipole field combined with the vertical focusing and horizontal defocusing electric fields from the \acp{ESQ}, and betatron tune shifts~\cite{Valetov:2026chromaticity}. The oscillation amplitudes are driven by both the optical mismatch at injection and the subsequent kicker deflection. These beam oscillations need to be accounted for in the analysis of \oa. 
The effective acceptance of the detectors is a function of both decay position in the storage ring $(x,y,\varphi)$, the energy of the positron, and its decay angle. Consequently, the betatron motions become imprinted on the measured precession frequency, as discussed below. 

The \ac{ESQ} system underwent a significant hardware upgrade with the development and installation of a \ac{RF} system~\cite{Kim_2020}. 
A modulated high voltage of $\sim$\SI{1}{kV} is applied at the \ac{CBO} frequency in the first \SI{6}{\micro\second} after injection on top of the $\sim$\SI{18}{kV} nominal \ac{ESQ} static voltage.   
The effect reduces the \ac{CBO} amplitude of the muon ensemble and reduces lost muons. The system applies a horizontal dipole \ac{RF} electric field at the correct phase relative to the betatron oscillations, exploiting the momentum-dependent radial positions of muons to resonantly damp the coherent motion. The \ac{RF} system was commissioned at the end of \RunFour and operated throughout the rest of \RunFourFiveSix with various combinations of the applied horizontal and vertical \ac{RF} fields.

\paragraph*{Anomalous spin precession frequency \oam}
The experiment measures the frequency $\oam$ by recording the decay positrons from the muon decay. The positrons' hit times and energies are measured in 24 PbF$_2$ electromagnetic calorimeter stations \cite{Kaspar_2017,KHAW2019162558,ANASTASI201786} placed radially inward from the storage region and evenly spaced around in azimuth. The Cherenkov light produced in the crystals is detected by 
silicon photomultipliers (SiPMs). In the muon rest frame, parity violation causes positrons to be emitted preferentially parallel to the muon spin. Upon applying the Lorentz boost to the laboratory frame, positrons emitted along the muon's momentum vector attain the highest energies. Therefore, the detection rate of high-energy positrons is maximized when the muon spin and momentum are parallel. Since this effect depends strongly on the measured positron energy, the detectors require highly accurate energy calibration.

A laser-based monitoring system as described in~\cite{Anastasi_2019} is used to flash each calorimeter crystal in a variety of time sequences with calibrated short light pulses. These data are used to establish gain stability corrections over short-, intermediate-, and long-time scales. 

The extraction of \oam comes from a fit of the time spectrum of decay positrons. While the final fit is more complicated, its main features are described by the five-parameter expression
\begin{equation}\label{eq:omega_fiveparameterfit}
    N(t) = N^{}_0 \cdot e^{-t/\gamma\tau^{}_\mu} \Big(1 + A(E^{}_{\textrm{th}}) \cos (\oam t - \phi^{}_0) \Big),
\end{equation}
where $N^{}_0$ is proportional to the number of detected decay positrons, $\gamma\tau^{}_\mu$ the lifetime of the boosted muon, and $\phi^{}_0$ the average phase of the muon spins relative to the beam direction at our chosen $t=0$. The energy-threshold dependent asymmetry parameter, $A(E^{}_{\textrm{th}})$, governs the amplitude of oscillation about the average exponential for muon decay. 

\paragraph*{Magnetic Field}
The frequency \opprimetildeatTexp in the denominator of Eq.~\eqref{eq:Rmu} is determined from the precise measurement of the magnetic field, \opprimeatTexp, weighted with the stored muon distribution, $M$, denoted by $\langle \opprimeatTexp\times M\rangle \approx \opprimetildeatTexp$. 
Here, angular brackets denote an average over the entire muon storage region. The approximate sign is due to the need for small corrections arising from transient magnetic fields. The magnetic field measurement is executed in a chain of several steps. It uses an in-vacuum trolley system~\cite{trolley} with 17 \ac{NMR} probes to map the field in the storage region every 3--7 days. 
Additionally, about 400 fixed \ac{NMR} probes \cite{swanson2025fixedprobestoragering}, installed above and below the storage region, track field drifts continuously. 
Finally, a high-purity, calibrated, water-based \ac{NMR} probe \cite{FlayPP} provides absolute calibration in terms of the Larmor frequency of protons in a spherical water sample. 
These three steps combined provide \opprimeatTexp.

A set of three additional tracking systems inside the storage ring provides important information on the muon ensemble and beam dynamics. The first system consists of two straw tracker stations~\cite{King_2022} located in-vacuum in front of two calorimeters and separated by \SI{90}{\degree}, which track muon decay positrons headed for the calorimeters. Extrapolation of these trajectories to the decay vertices provides the muon beam distribution $M$ as a function of time. While the straw trackers measure positrons from muon decays and thus do not interfere with the muon beam, the other two tracking systems are used only during dedicated run periods, as they directly measure the stored muons and thus degrade storage. The second system utilizes two scintillating fiber-detector stations, each comprising horizontal and vertical planes of scintillating fibers, which can be inserted into the storage region for semi-destructive measurements of the stored muon distribution.

New for the \RunFourFiveSix running periods was the development of the \ac{MiniSciFi} detector system.  It was designed to measure the stored muon beam with minimal disturbance using three thin (\SI{0.25}{\milli\meter}) scintillating fibers read out by SiPMs.
The detector was inserted into the storage ring during special systematic runs and retracted during normal production runs. Two configurations addressed complementary goals. First, a horizontal fiber configuration sampled the circulating beam intensity, which is used for \ac{FR} momentum reconstruction, thereby enabling an independent cross-check of beam dynamics corrections. Second, a vertical fiber configuration scanned and reconstructed the horizontal beam profile throughout the fill, providing an independent cross-check of the tracker and calorimeter beam distribution measurements.

The measured \oam requires several corrections to account for the $\vec{\beta}\cdot\vec{B}$ and $\vec{\beta}\times\vec{E}$ terms in Eq.~\eqref{eq:amurealistic}, and for other beam dynamics effects that cause slow time-dependent changes in the muon ensemble over the fill. 
The measured quantity $\langle \opprimeatTexp\times M\rangle$ must be corrected for two fast field transients, which are not accessible by the nominal \ac{NMR} measurements. Accounting for these effects, the master formula, Eq.~\eqref{eq:Rmu}, the true frequencies \oa and \opprimetildeatTexp are expanded to

\begin{eqnarray} \label{eq:R}
   \Rmuprime(\Tr) = \frac{\oam\left(1+C^{}_{e}+C^{}_{p}+C^{}_{pa}+C^{}_{dd}+C^{}_{ml}\right)}{\langle \opprimeatTexp\times M\rangle(1+B^{}_K+B^{}_Q )}.
\end{eqnarray}

The corrections applied to \oam in the numerator account for five effects driven by the dynamics of the beam, some acting directly and others through time variations in the average phase of the stored muon ensemble: i)  $C^{}_e$ corrects for the incomplete cancellation of the $\vec{\beta}\times\vec{E}$ term since not all muons are on the magic momentum due to the momentum spread of the stored muons, ii) $C^{}_p$ accounts for the reduction of the measured frequency due to the actual vertical betatron motion of the stored muons, iii) $C^{}_{pa}$ corrects for the phase-acceptance effect arising from the coupling of injected muon phases with detector acceptance, iv) $C^{}_{dd}$ accounts for the momentum dependence of the boosted muon lifetime combined with spin phase and muon momentum correlations, and v) $C^{}_{ml}$ reflects time-dependent muon losses combined with changes in the average phase of lost muons that can bias the measured frequency.

The \ac{NMR} probes employed to determine the quantity $\langle \opprimeatTexp\times M\rangle$ in the denominator of Eq.~\eqref{eq:R} are not sensitive to fast changes in the magnetic field. Therefore, two corrections accounting for magnetic field transients induced by the fast switching kickers, $B^{}_K$, and the \ac{ESQ}, $B^{}_Q$, are needed to arrive at the relevant magnetic field experienced by the stored muons.

While the analysis and determination of corrections and systematic uncertainties rely heavily on the data itself, three simulation packages, \texttt{gm2ringsim} based on \texttt{Geant4}~\cite{AGOSTINELLI2003250, ALLISON2016186, 1610988}, \texttt{COSY INFINITY}~\cite{MAKINO2006346, Berz:2026cosyinfinity, Makino:2026vicpo11}, and \texttt{BMAD}~\cite{SAGAN2006356}, were used to validate analysis tools and extrapolate measured beam distributions beyond the storage ring. Each package offers unique strengths:  \texttt{gm2ringsim} models material interactions with high fidelity, \texttt{COSY INFINITY} verifies long-term beam effects using symplectic tracking, and \texttt{BMAD} models the injection line and storage ring using field maps and multipole expansions. A detailed description and comparison of the packages is given in Ref.~\cite{Run1PRAB}.

\section{Datasets}\label{sec:datasets}
The \FNALexperiment collected data over six years, from 2018 to 2023, with annual accelerator shutdowns during the summer months, thereby dividing the data collection into distinct running periods labeled \RunOne through \RunSix. The full dataset was analyzed and published in three major blocks of increasing statistics with respect to the full dataset: \RunOne (\SI{5}{\percent}),  \RunTwoThree (\SI{26}{\percent}), and \RunFourFiveSix (\SI{69}{\percent}). Details of the \RunFourFiveSix data taking and analysis are a main focus of this publication.

A key difference between the three major datasets was the continual improvement of experimental conditions and reduction of systematic uncertainties. In \RunOne, the experiment operated with suboptimal \ac{ESQ} conditions due to damaged high-voltage resistors, which were replaced before \RunTwo. The beam storage conditions were improved in \RunThree by optimizing the kicker strength. The \ac{ESQ} \ac{RF} system was commissioned at the end of \RunFour and operated throughout \RunFive and \RunSix, reducing both the \ac{CBO} amplitude and muon losses by approximately a factor of five.

The \RunFourFiveSix dataset is divided into smaller, lettered subsets (e.g., 4AB, 4CD, 4E, \dots), and further grouped into four distinct categories corresponding to different beam dynamics conditions arising from various \ac{ESQ} operation modes. The first part, labeled \RunFournoRF, represents data taken without the \ac{RF} system. The second part, \RunFiveX, includes data with horizontal ($x$) \ac{RF} fields applied to reduce horizontal \ac{CBO}. The remaining data are split into \RunFiveXY and \RunSixXY, corresponding to \RunFive and \RunSix, respectively, where both horizontal and vertical ($xy$) \ac{RF} fields were applied. Despite identical \ac{RF} configurations, these last two datasets are analyzed separately due to observed differences in beam dynamics. Table~\ref{tab:conditions} summarizes each dataset's key parameters and conditions. 

\begin{table*}
    \centering
    \caption{Key parameters for the \RunOne, \RunTwoThree, and the four major groupings of the \RunFourFiveSix dataset periods. The RF mode indicates which RF components were used in the \ac{ESQ} system: no RF (--), horizontal only (x), or both horizontal and vertical (xy). The effective field index $n$ is also listed. The number of analyzed decay positrons is shown, as well as the number of field periods (defined as time intervals between consecutive field maps or between a field map and magnet ramp-down).
    }
    \begin{tabular}{lccccccccc}
        \toprule
         Dataset & \ac{ESQ} & \ac{RF} & Kicker & Field & Positrons & Magnetic field \\
         & (kV) & Mode & (kV) & index &  ($10^9$) & periods\\
         \midrule
         \RunOne       & 18.3 & -- & 125 to 137  &  0.107 to 0.120 & 15.4 & 16 \\\RunTwoThree  & 18.2 & -- & 142 and 161 & 0.107 to 0.108  & 70.9 & 69 \\\midrule
         noRF          & 18.2 & -- & 161         & 0.108           & 86.0 & 71 \\xRF           & 18.2 & x  & 161         & 0.108           & 49.3 & 40 \\xyRF5         & 18.2 & xy & 161         & 0.108           & 47.8 & 37 \\xyRF6         & 18.2 & xy & 161         & 0.108           & 39.1 & 46 \\\bottomrule
    \end{tabular}
    \label{tab:conditions}
\end{table*}

\section{Anomalous precession frequency measurement: \oam}\label{sec:oam}
Determination of the muon precession frequency involves the reconstruction of the data collected from the $24$ calorimeters spaced uniformly around the inside of the storage ring. These data were hardware-blinded by shifting the GPS-stabilized master clock that drives the 800 MHz digitizers~\cite{Sweigart:2016jty} away from the nominal 40 MHz into the range 39.997 to 39.999 MHz, a 50 ppm range.  A second blinded clock, GPS stabilized completely independently, mixed with the master clock provided blinded monitoring of the clock stability.

Fits to the reconstructed time evolution of the oscillating rate of that data provide the value of \oam, the measured precession frequency. A combination of factors involving intensity variations, dynamical beam motion, and calorimeter acceptance lead to significant complexity within that evolution. The analysis effort therefore involved five independent teams that utilized four different reconstruction methods and ten variations of fitting methodology, which provided a significant level of cross-checking and robustness. 
Each team developed one or more complete analysis pipelines by combining a chosen reconstruction method with a particular fitting strategy, resulting in the eight analysis groups whose configurations are summarized in Table~\ref{tab:analysis_choices_wa}.

The five teams each developed their analyses with an independent, unknown, software-based blinding offset on their value of \oam.  Only after each analysis group had completed their analysis, followed by independent review of the analysis details and systematic uncertainty evaluation, did the groups shift to a common, unknown software blinding offset for final comparisons and cross-checks.

A critical component of these analyses involves the compensation of effects that vary systematically as a function of time after muon injection in a fill through corrections or modeling.
As an example, consider pileup events in which two lower-energy positrons strike a calorimeter nearby in time and reconstruct as a higher-energy positron.  
Low-energy positrons have a larger curvature compared to high-energy positrons. As a result, on average, they come from muon decays that occur closer to the calorimeter than do high-energy positrons. Consequently, the pileup events introduce a different precession phase into the ensemble of reconstructed positrons than true high-energy positrons contribute.
Because the pileup rate varies as the square of the beam intensity, it decays with half the muon lifetime, and therefore the ensemble phase average drifts with time.  Left uncorrected, such a drift directly biases the measured \oam.  

As a second example, a systematic variation of gain over the course of a fill would result in an effective shift of the energy threshold of the accepted positrons. Left uncorrected, that variation  also leads to a bias in \oam. The decay positions, and hence the precession phase, of the ensemble of accepted positrons would shift as a function of time into fill.

This section provides a brief summary of the reconstruction methods and associated data corrections, which are described in more detail in the \RunOne~\cite{Run1PRDomegaa} and the \RunTwoThree~\cite{PhysRevD.110.032009} publications.  It will then detail the different approaches to the modeling of the time series data for fitting for \oam, and the results of those fits.

\begin{table*}[tb]
\centering
\caption{Summary of the analysis choices made by the 8 different analysis groups in the \RunFourFiveSix analysis. The (statistically optimal) A-Method analyses in bold are included in the final average value of \oa. Other analysis methods (T/Q) are performed as cross-checks for those analyses. For the analyses included in the final \oam average, the number of degrees of freedom corresponds to the fit used in averaging. The group labels refer to the style of reconstruction used. The following text explains different approaches.}
\begin{tabular}{cccccccccccc}
\toprule
\multirow{2}{*}{{Group}} &
  \multicolumn{1}{p{0.09\textwidth}}{\centering {Recon- struction}} &
  \multirow{2}{*}{{Analysis}} &
  \multicolumn{1}{p{0.08\textwidth}}{\centering {Fast Rotation}} &
  \multirow{2}{*}{{Pileup}} &
  \multicolumn{1}{p{0.09\textwidth}}{\centering {Fit model}} &
  \multicolumn{1}{p{0.08\textwidth}}{\centering {Cosine terms}} &
  \multicolumn{1}{p{0.09\textwidth}}{\centering {Envelope modeling}} &
  \multicolumn{4}{c}{{Degrees of freedom}} \\ 
 &
   &
  \multicolumn{1}{c}{} &
  \multicolumn{1}{c}{} &
  \multicolumn{1}{c}{} &
  \multicolumn{1}{c}{} &
  \multicolumn{1}{c}{} &
  \multicolumn{1}{c}{} &
  NoRF &
  xRF &
  xyRF5 &
  xyRF6 \\ \midrule
\textit{\wagroupZep}    & Global   & T/\textbf{A}       & Rand.    & Empirical      & Nonlinear        & \alphabeta      & Splines          & 4119 & 4119 & 4119 & 4119 \\
                        &          &                    &          &                &   Moment \\
\textit{\wagroupTyler}  & Global   & T/\textbf{A}       & Rand.    & Empirical      & Linear           & \alphabeta      & Parameterize & 4100 & 4101 & 4107 & 4100 \\
\textit{\wagroupMurong} & Global   & RT/\textbf{RA}     & Rand.    & Empirical      & \BNLFitModel     & \Aphi           & Parameterize & 4121 & 4123 & 4122 & 4121 \\
\textit{\wagroupOnKim}  & Global   & ST/\textbf{SA}     & Rand.    & Empirical      & \BNLFitModel     & \alphabeta      & Parameterize & 4006 & 4006 & 4006 & 4006 \\
\textit{\wagroupBU}     & Local I  & T/A/RT/\textbf{RA} & Kernel   & Empirical      & \BNLFitModel     & \Aphi           & Parameterize & 4140 & 4142 & 4141 & 4140 \\
\textit{\wagroupSJTU}   & Local I  & T/\textbf{A}       & Rand.    & Empirical      & Sideband         & \alphabeta      & GPR              & 4108 & 4108 & 4108 & 4108 \\
\textit{\wagroupITA}    & Local II & T/A/RT/\textbf{RA} & Rand.    & Semi-Emp.      & \BNLFitModel     & \Aphi           & Parameterize & 4140 & 4140 & 4140 & 4140 \\
\textit{\wagroupUKY}    & Energy   & Q/RQ               & Kernel   & --             & \BNLFitModel     & \Aphi           & Parameterize & 4093 & 4093 & 4093 & 4093 \\                               & Flow     & \\ \bottomrule
\end{tabular}
\label{tab:analysis_choices_wa}
\end{table*}

\subsection{Positron and Energy-flow Reconstruction}
\paragraph*{Reconstruction Methods}
Three of the four calorimeter reconstruction methods take a positron-centric electromagnetic shower reconstruction approach, while the fourth uses the time flow of the total energy in the calorimeters.  The former methods reconstruct the arrival time and energy of individual positrons from the electromagnetic shower waveforms in each calorimeter. These methods must be corrected for pileup.  The energy flow method leverages the growing parity-asymmetry with energy, complemented by calorimeter acceptance that favors higher-energy positrons, resulting in the total energy observed in the calorimeters retaining an imprint of the muon precession. This strategy does not need to distinguish between single and multiple positrons in a calorimeter, so it does not require explicit handling of pileup.  All methods begin with the approximately \SI{40}{\nano\second} time windows extracted from the calorimeter waveforms by the DAQ system when activity exceeds \SI{50}{\MeV}, generally referred to as ``islands''.

All three positron-centric reconstruction methods utilize template fits to the waveform data to extract the deposited energy and time in a given crystal.  The templates are constructed from the data for each individual crystal~\cite{osti_1515050}.  Two of these methods begin reconstruction with template fits to the waveform islands in individual crystals for each fill, with an approximately 60~\unit{\MeV} minimum energy requirement, but differ in their crystal-clustering algorithms that build the complete shower.  

The first, dubbed Local I, clusters the peaks found in different crystals by the template fits based on their time separation.   For the comparisons, the time-differences are normalized by the time separation resolution~\cite{Binney:2022hqc}, which varies as $\sigma_{t}(E_{\rm eff}) = 1.25\sqrt{2(0.032^{2}) + 1.93^{2}/E_{\rm eff}[\unit{\MeV}]}$~\unit{ns}. The effective energy $E_{{\rm eff}}=(E_{1}+E_{2})/\sqrt{(E_{1}^{2}+E_{2}^{2})/2}$. This dependence gives a time resolution of about \SI{90}{\ps} above \SI{1}{\GeV}.  The clustering criteria balanced minimal distortion of the energy spectrum against maximizing pileup reduction. The final criteria result in clustering of all crystals within $6\sigma_{t}(E_{{\rm eff}})$; additionally, a crystal with a $\Delta t/\sigma_{t}(E_{{\rm eff}})$ in the $6-8$ range relative to one or more clusters gets combined within the cluster with which it has the smallest $\Delta t/\sigma_{t}(E_{{\rm eff}})$.

The second clustering algorithm~\cite{osti_2246737}, Local II, sorts the hits according to their energy in descending order. Starting from the most energetic hit, the seed, the algorithm clusters the hits if they meet a resolution-based time separation criterion similar to the Local I method. Every hit above 100 MeV already belonging to a cluster participates in adjoining new hits into the same cluster. The process repeats with a new seed crystal until all hits above 100 MeV have been selected. Remaining hits belonging to any seed crystal are merged with the seed's cluster if closer than 10 ns of time separation. After a first clustering pass, the new algorithm refits the waveform in each crystal using a pair of templates to test for the presence of a second shower. This fit pins the time of the earlier template to the cluster seed. The method accepts a second waveform down to a time difference of half a sampling period, or \SI{0.625}{\ns}, and an energy of at least \SI{25}{MeV}, with some additional criteria on the energies of the two pulses (see~\cite{osti_1515050}). Any newly identified shower contributions are then time- and energy-corrected and input into the new clustering algorithm for a second pass.

\begin{table}[tb]
\centering
\begin{tabular}{ccccc} \hline\hline
Method   & template   & crystal   & clustering \\
         & fits       & threshold &            \\ 
\hline
Local I  & individual & \SI{60}{MeV}    & temporal   \\
Local II & both & \SI{60}{MeV}    & temporal       \\
Global   & in cluster & none      & spatial    \\
\hline\hline
    \end{tabular}
    \caption{The key conceptual differences between the three clustering algorithms. Template fits refers to whether templates are fit independently in each crystal, and then clustered, or are first clustered and fit simultaneously with a common time. The crystal threshold refers to the minimum energy for a crystal to be used in clustering.  In the case of no threshold, crystal energies are allowed to fluctuate negatively.  The clustering methods refer to whether the clustering algorithm utilizes spatial information or temporal information.}
    \label{tab:reconSummary}
\end{table}

The third, Global, reconstruction method~\cite{osti_1581403} begins with the highest-energy crystal in a calorimeter time window and simultaneously fits templates to all crystals in a $3\times3$ array around that seed crystal.  The fit constrains the nine templates to a common time, while the amplitudes in each float freely.  To accommodate pedestal fluctuations, the amplitudes can take negative values. After subtracting the fit templates from the waveform, the algorithm repeats until no significant energies remain.  Overlapping clusters within the \SI{1.25}{\ns} sampling period receive energy from the common crystals partitioned between them when the seed separation exceeds one crystal width. Otherwise, those clusters are merged.

These three methods represent a range of complementary optimization choices, and consistency in the precession frequencies among those methods provides further confidence in the result. Such choices include handling the lowest-energy crystals and balancing bremsstrahlung photon inclusion (for optimal energy reconstruction) against pileup cluster separation. Table~\ref{tab:reconSummary} summarizes the key differences between these three methods.

The final reconstruction method involves reconstruction of the energy flow in the event. Online data collection summed four muon fills of data in real time into 37.5~\unit{\ns} bins in each crystal.  To accommodate the small drop and recovery in the pedestal due to the flash of particles at injection, which would otherwise dominate the signal, the final offline reconstruction performs a running pedestal subtraction~\cite{Run1PRDomegaa} while resumming the energy into \SI{75}{\ns} bins.  To contribute to that resummation, the pedestal-subtracted energy in each \SI{37.5}{\ns} bin must exceed \SI{500}{\MeV}. This criterion eliminates contributions from deuteron contamination~\footnote{Deuterons produced at the primary target can become stored in the Delivery Ring and, with low, but non-negligible probability, leak into the SR and strike calorimeters on the first turn after the inflector.} within the stored muon beam.  For the data in this analysis, the online energy flow time window covered \SI{600}{\us} of the muon fill, almost its full extent, compared to the truncation at 300~\unit{\us} for the data in previous publications~\cite{Run1PRDomegaa,PhysRevD.110.032009}.

\begin{figure}[tb]
\begin{center}
\includegraphics[width=0.48\textwidth]{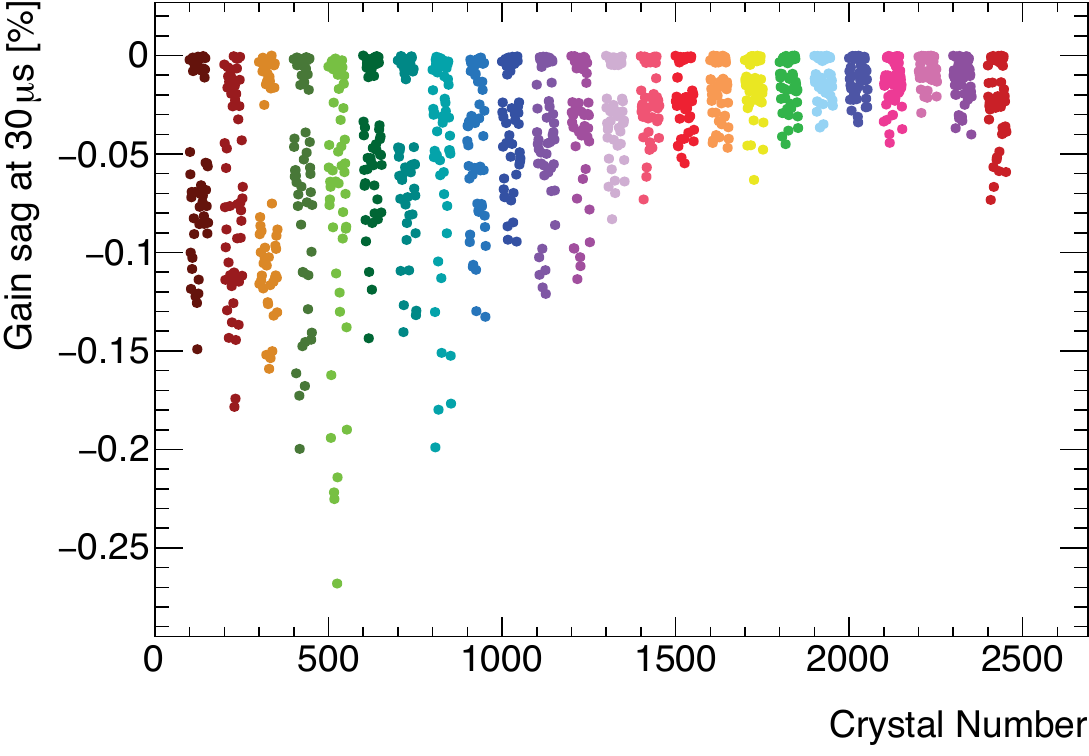}
\caption{The in-fill gain sag measured with the laser system for each crystal in the 24 calorimeter stations near the start time of the fits to determine the muon precession frequency. The crystal numbers are grouped as (Calorimeter Station)$\times 100 +$(Local Crystal Number), with the former running from 1 to 24, and the latter from 0 to 53.  The colors help to differentiate the 24 different stations. Recovery times are typically in the \SI{4}{}--\SI{8}{\us} range.  The activity within a crystal at injection time drives the variation in the observed sags: within a calorimeter, the largest gain sags occur in crystals closest to the stored beam and in the calorimeter stations just downstream of the beam injection point. Negligible gain sags are seen in the crystals farthest from the beam and in the stations farthest from injection.  The consistency of the precession frequencies obtained for the individual calorimeters indicates that the gain corrections are working properly.}
\label{fig:ifgAmp}
\end{center}
\end{figure}

\paragraph*{Gain monitoring}
As noted above, a systematic variation in gain over the course of a muon fill will introduce a bias in \oam. Monitoring and calibration of the calorimeter gains utilizes a pulsed laser system that simultaneously delivers light via a fiber network  to each of the $1296$ crystals~\cite{lasersystem,Run1PRDomegaa}, as well as data samples enriched with minimum-ionizing muons that pass through the calorimeters. The laser system allowed controlled double pulsing of the calorimeter stations, which helped correct two time-dependent gain effects: an initial sag in the gain while the power supply capacitors recovered from the flash of incident particles accompanying beam injection, and a sag in the gain of an individual crystal as individual \ac{SiPM} pixels recover after an electromagnetic shower.  The exponential recovery times for the crystals following the flash were determined using the double-pulsing configuration and fell into two groups, roughly at \SI{4}{\us} and \SI{8}{\us}.  The amplitudes were determined by pulsing the lasers in 1 of every 11 fills initially, systematically sweeping the pulses over the fill.  The rate was reduced to 1 in every 22 fills for the end of Run 4 and beyond.

Figure~\ref{fig:ifgAmp} shows the average amplitude sag at \SI{30}{\us}, near the start time of the fits for the muon spin precession frequency. The calorimeters just downstream of beam injection, where most of the flash occurs, show the largest effect.  For the \ac{SiPM} recovery, studies varying the double-pulse separation and the laser intensity provided the necessary corrections. Laser flashes of all crystals, taken each fill after the beam dump, provided long-term tracking and correction of the gains. 
Finally, the minimum-ionizing peak from muons scattered out of the beam provided the energy scale for each crystal.  Unbiased measurement of the precession frequency requires gain stability, not high resolution, and this muon calibration scheme sufficed.

\begin{figure}[!htbp]
\begin{center}
\includegraphics[width=0.4\textwidth]{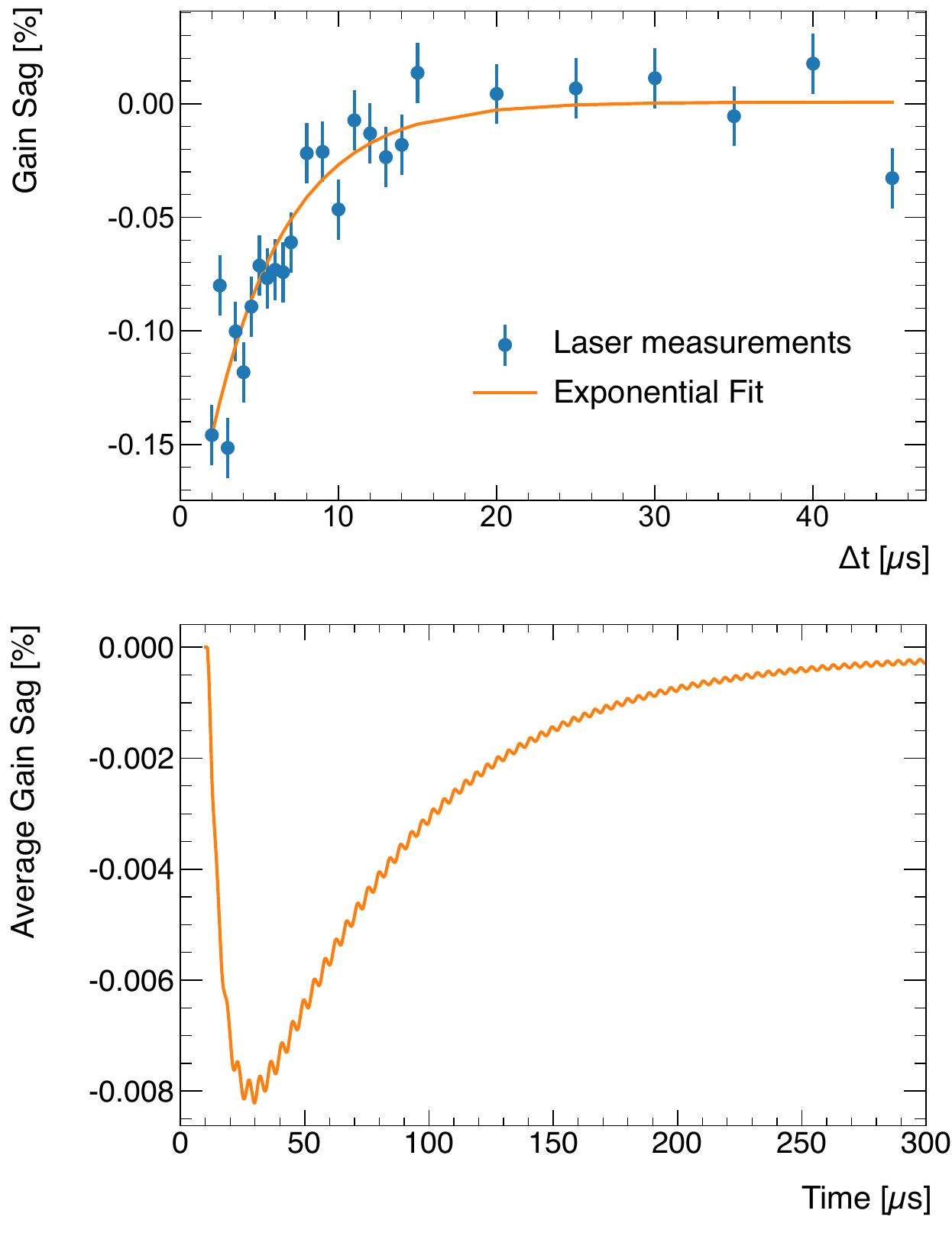} 
\caption{Top: A measure of the gain sag in a representative crystal as a function of the time between two laser pulses from standalone measurements executed after data-taking. The lifetimes obtained from these studies agreed well with those from laser studies, interspersed with data running for the in-fill gain correction. Bottom: The average gain sag seen within a crystal during a fill, determined from simulation of EM showers within a muon fill.  The simulation takes as input the sag parameters for individual crystals obtained from the laser-based studies. The small oscillation lags the anomalous spin precession oscillation, which results in a larger sensitivity to this effect than the in-phase modulation (assumed in previous publications) would.}
\label{fig:itdp}
\end{center}
\end{figure}

The analyses for the \RunOne~\cite{Run1PRDomegaa} and \RunTwoThree~\cite{PhysRevD.110.032009} results noted evidence for a residual long-lifetime effect that could be explained by a gain variation oscillating at the anomalous muon precession frequency. The variation would have an amplitude of order $\times 10^{-4}$ that damped at a rate of the beam muon lifetime (\SI{64.4}{\us}). The amplitude differed between the local and global reconstructions.  Because other mechanisms could also result in this behavior, but with opposite signs in the shift of \oam, data corrections were not applied by all analyses. It contributed only to the systematic uncertainty in those results. The difference in effect size between reconstructions has since been traced to a pileup-related feature in the Local reconstruction methods, which has been corrected for this analysis.

The effect common to both global and local reconstructions is indeed gain-related and results from the same power supply capacitance recovery as the in-fill gain sag.  In this case, the electromagnetic showers from the positrons themselves induce sag, which affects the gain of the following showers.  Because the sag contribution from successive showers can accumulate,  as Fig.~\ref{fig:itdp} illustrates, the history of previous showers contributes to the gain sag seen by a given shower.  Crucially, this effect naturally induces a phase lag relative to the muon precession phase, and that phase lag drives an intensity-dependent bias in the extraction of the precession frequency. That bias ranges from 21 to 46~ppb across the analyses included in the final \oam average, and over the Runs 4 through 6 datasets.  Table~\ref{tab:ITDPcorrections} presents the corrections now applied to previous results, which correspond to higher intensities on average, to account for this effect.  While initially identified using the \RunFourFiveSix data, the effect was verified through an intensive standalone study with the laser system for one of the calorimeter stations.  Those studies involved masking different activity patterns onto the crystal array, which allowed untangling of the sag contributions directly within a crystal versus correlated contributions among crystals sharing a power supply.

\begin{table}[b]
\caption{The corrections applied to the data subsets in the \RunOne and \RunTwoThree analyses to account for the intensity-dependent gain sag that led to the residual slow term. The systematic uncertainties quoted here are the increase to the previously published value, added linearly. }
\begin{center}
\begin{tabular}{llc}
\toprule
Publication & Dataset & Correction [ppb] \\
\hline
\multirow{4}{*}{\RunOne}
 & 1a & $56.2 \pm 29.9$ \\
 & 1b & $60.3 \pm 32.0$ \\
 & 1c & $54.1 \pm 28.8$ \\
 & 1d & $40.3 \pm 26.1$ \\
\hline
\multirow{3}{*}{\RunTwoThree}
 & 2  & $55.3 \pm 29.0$ \\
 & 3a & $40.6 \pm 20.0$ \\
 & 3b & $48.7 \pm 24.8$ \\
\bottomrule
\end{tabular}
\end{center}
\label{tab:ITDPcorrections}
\end{table}

\begin{figure*}[tb]
\begin{center}
\includegraphics[width=0.97\textwidth]{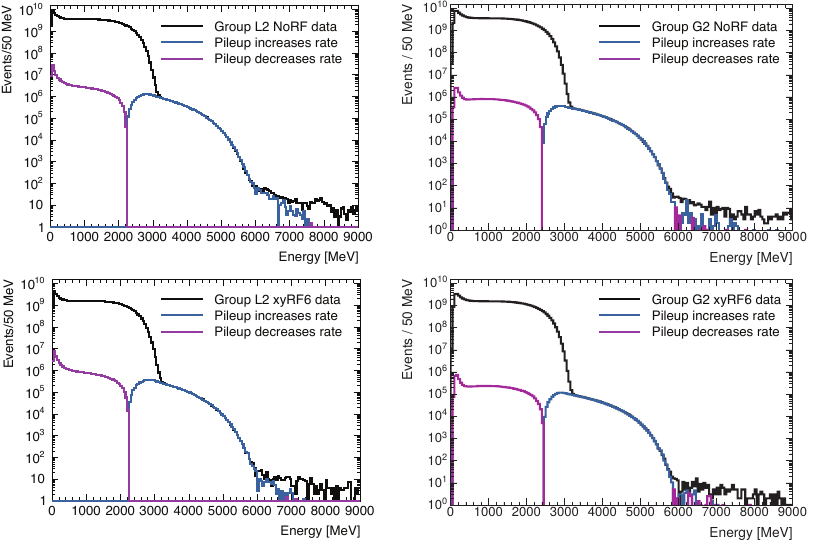} 
\caption{The energy spectra for positron candidates (black) and the predicted combined double and triple pileup spectra that increase the observed rate (blue) or decrease the observed rate (violet) for the \wagroupSJTU analysis (Local I reconstruction-based ) (left) and the \wagroupTyler analysis (Global reconstruction-based) (right).  The data correspond to \RunFournoRF dataset (top) and the \RunSixXY dataset (bottom).  Note that the excess in data above 6~\unit{\GeV} is dominated by cosmic-ray interactions, whose low rate has a negligible effect on the results.  The pileup normalization is absolute.  The \RunSixXY results show a lower fractional rate than the \RunFournoRF because of the reduced intensity.  Otherwise, the application of the RF itself has little effect on pileup.}
\label{fig:pileup}
\end{center}
\end{figure*}

\subsection{$\omega_a$ determination}
\subsubsection{Multi-positron Pileup correction}
Preparation of the positron-based data for the fits to determine \oam first corrects for the residual pileup events that the reconstruction methods could not resolve into separate positron candidates.  The data-driven correction involves overlaying each positron candidate with activity one cyclotron period later.  Because the cyclotron period is significantly shorter than the anomalous precession and CBO frequencies, that second window provides a representative random sampling of activity in the calorimeter.  The first ``empirical'' algorithm involves overlay of the raw digitization samples from the two time windows. The overlaid events proceed then through the standard reconstruction~\cite{osti_1581403,Foster:2023hpz} methods.  The second ``semi-empirical'' algorithm instead overlays reconstructed positron candidates~\cite{PhysRevD.110.032009}.  To assess triple pileup, a third window located two cyclotron periods later gets overlaid as well.  Appropriate weighted combinations of these samples compensate for the presence of pileup itself in the original and later time windows and provide an absolute normalization.  Figure~\ref{fig:pileup} shows the energy spectra in the \RunFournoRF and \RunSixXY datasets, which correspond to the highest and lowest intensity datasets in this analysis.  The figure also shows the combined double- and triple-pileup spectra predicted for both Local I and Global analyses. These comparisons show agreement at the 2\% or better level in the nonphysical region above about 3.2~\unit{\GeV}.

As noted earlier, the energy flow reconstruction requires no correction for pileup.

\subsubsection{Beam-related dynamical effects}
\label{sec:bdoa}
\begin{figure*}[tb]
\begin{center}
\includegraphics[width=0.97\textwidth]{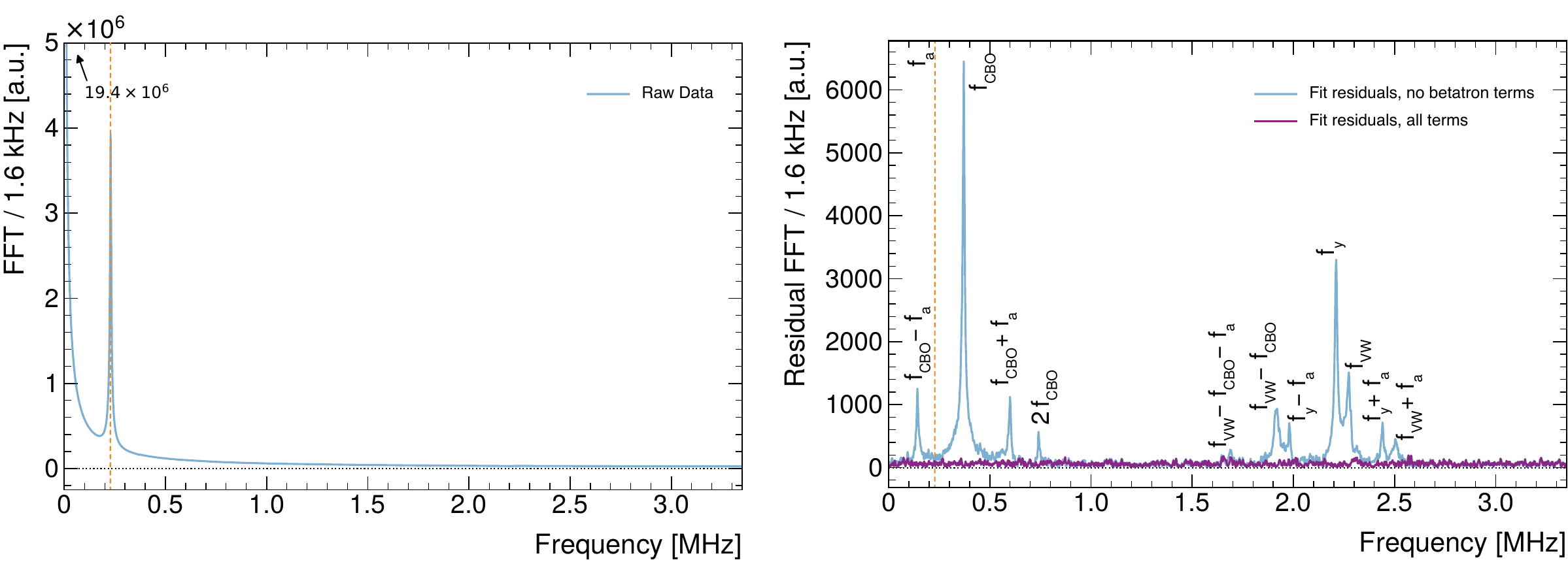} \\
\includegraphics[width=0.97\textwidth]{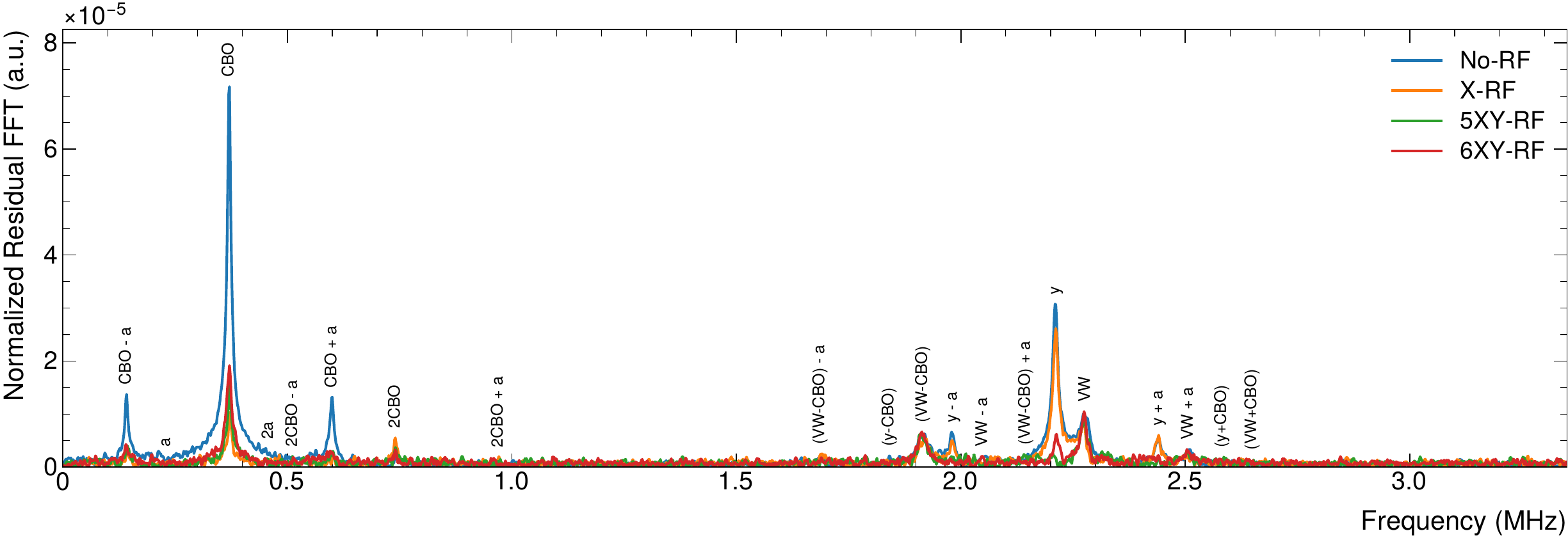}
\caption{Top left: \acf{FFT} of the combined data for Runs 4, 5, and 6.  The exponential behavior due to the muon lifetime and the peak at the anomalous muon spin precession frequency, \oam, broadened by the finite sampling time, dominates the data.  Top right: \ac{FFT} of the residual differences between the summed datasets and fit using the simple model of Eq.~\eqref{eq:omega_fiveparameterfit} (blue) and between the summed datasets and the sum of the fully-modeled fits to the four individual datasets (purple, see text).  Beam-dynamical motion dominates the peaks in the former. These effects are well-described by the full fit model, resulting in an essentially featureless final \ac{FFT} at the full statistics of the \RunFourFiveSix data. Table~\ref{tab:freqs} defines the listed frequencies.  The orange lines indicate the position of the anomalous spin precession frequency $f_a$. Bottom: The \ac{FFT} of the residual differences between the individual datasets and a fit using the simple model of Eq.~\eqref{eq:omega_fiveparameterfit}, illustrating the effects of application of the \ac{RF} to the \ac{ESQ}.  The amplitudes of betatron frequencies related to horizontal motion drop noticeably from the No-RF data (blue) to the later datasets with the horizontal (X) RF applied.  The amplitudes of the vertical frequencies then also drop from the No-RF and X-RF datasets (orange) to the XY-RF datasets (green and red).}
\label{fig:ffts}
\end{center}
\end{figure*}

{\it Cyclotron motion} Equation~\eqref{eq:omega_fiveparameterfit} describes the dominant behavior of the number of positrons detected as a function of time stored in a muon fill. The \acf{FFT} of the pile-up corrected data for the combined \RunFourFiveSix datasets in the left panel of Fig.~\ref{fig:ffts} illustrates this.  However, the dynamical nature of the stored muon beam, coupled with the acceptance of the individual calorimeters, introduces a variety of effects that fits to the data must incorporate~\cite{Run1PRDomegaa,Run1PRAB}.  An \ac{FFT} of the residuals to a fit to that combined data using Eq.~\eqref{eq:omega_fiveparameterfit} as the fit model, also shown in Fig.~\ref{fig:ffts}, reveals many frequencies related to the muon beam dynamics. While these frequencies correlate only weakly with \oa in the fits, the precision of the available statistics means they cannot be neglected. Excluding the dominant frequencies would introduce a bias of up to hundreds of ppb.  The largest contribution comes from the horizontal \acl{CBO} term. Table~\ref{tab:freqs} tabulates the sources of the frequencies apparent in Fig.~\ref{fig:ffts}.

The cyclotron frequency \ocycl of the stored muon beam would, without suppression, introduce many sideband frequencies, as well as its fundamental contribution, that would have to be modeled by the \oam fit.  The periodic sampling by each calorimeter also aliases some beam-dynamical frequencies below the Nyquist frequency.
Combining the data into bins with a width equal to the average cyclotron period $T_{c}=149.2$~\unit{\ns} suppresses that frequency.  Binning at the cyclotron period $T_c$ is almost equivalent to applying the frequency domain sinc suppression function 
\begin{equation}
E(\omega) = \sinc\left(\frac{\omega T_{c}}{2}\right). \label{eq:suppress}
\end{equation}
This function vanishes at $\ocycl = 2\pi/T_{c}$ and its harmonics, and would eliminate the cyclotron frequency and its harmonics.  However, variation in the rate across a time bin due to the muon lifetime causes a small residual amplitude to persist at those frequencies.  In addition, the spread in beam energy results in a spread of cyclotron frequencies, and the sinc function does not completely suppress frequencies near $\ocycl$.  As a result, a residual imprint of the cyclotron motion remains even in the binned data.

\begin{figure}[tb]
\begin{center}
\includegraphics[width=0.48\textwidth]{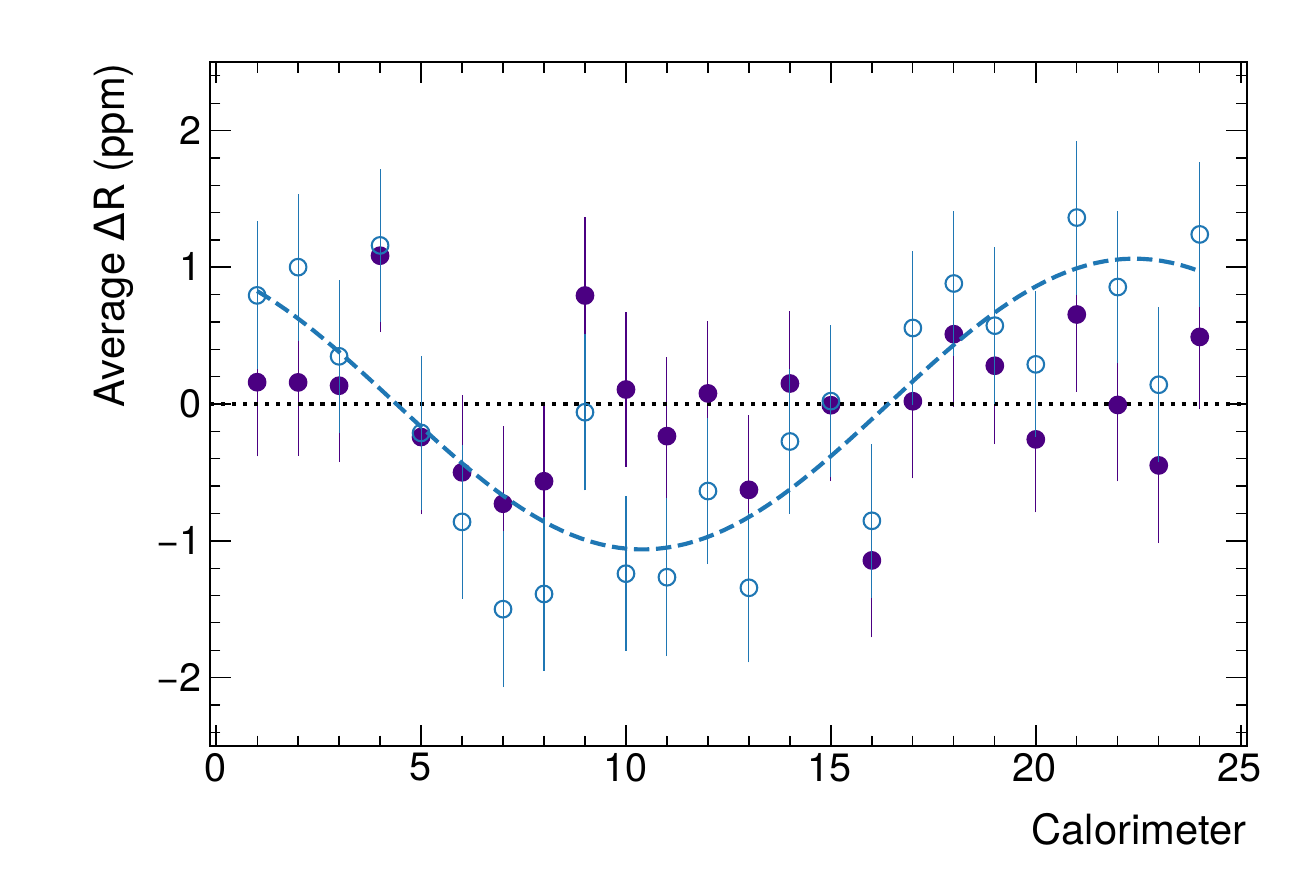}
\caption{Relative values of \oam extracted from data in each calorimeter without time randomization (blue points) and with (purple filled points).  The $\chi^2$ for a constant fit to the data without and with the time randomization is 64.5 and 19.5, respectively, for 23 degrees of freedom.  The fit (dashed blue line) shown for sinusoidal behavior without the time randomization, which is expected from the cyclotron-related aliasing, has a $\chi^2$ of 20.3 for 22 degrees of freedom.}
\label{fig:Rwave}
\end{center}
\end{figure}

The effect of this residual stamp on \oam appears when comparing the \oam obtained from fitting the calorimeters individually. The cyclotron frequency region gets aliased down to the $\omega=0$ region. This aliasing results in a small shift in the \oa value measured at individual calorimeter stations, which varies with the cyclotron phase, and hence as a function of calorimeter position around the ring.  Rather than attempting to model the residual or relying upon cancellation in the calorimeter sum, the effect is further suppressed using time randomization, as follows.   Randomly shifting each measured time over a range of $\pm T_{c}/2$ introduces an additional factor of the sinc function $E(\omega)$ in Eq.~\eqref{eq:suppress} for a total suppression of $\sinc^{2}$.  The final suppression factor of the amplitude at the cyclotron frequency due, for example, to the finite lifetime effect, becomes $(T_{c}/2\pi\gamma{\tau_{\mu}})^2 \sim 10^{-7}$. Figure~\ref{fig:Rwave} shows the variation of \oam extracted without and with the randomization step for the \RunFour data, with the expected wave-like behavior evident before randomization.

The energy-flow analysis and one of the positron-fit methods discussed below must use alternative binning due to hardware and algorithmic constraints, respectively.  All other analyses apply the 149.2~\unit{\ns} binning and randomization.

{\it Synchrotron motion} Although binning mitigates cyclotron motion, accurate \oam extraction requires extending the basic precession signal of Eq.~\eqref{eq:omega_fiveparameterfit} to include the dynamical behavior of the stored muon beam. A first-principles analysis~\cite{tyler_thesis} of the beam and muon decay dynamics coupled with the detector acceptance confirms the empirical models used in previous publications.  It also provides the basis for a new, complementary formulation of the model and offers a qualitative understanding of the relative strengths of the oscillatory frequencies of the beam shown in Fig.~\ref{fig:ffts}.  The modeling confirms three main categories of oscillatory signatures in the data.  The first is the main physics signal oscillation at the frequency \oa.  The second arises from the beam betatron motion, which causes the positron illumination at a calorimeter to oscillate horizontally and vertically.  Because the positron acceptance varies across the calorimeter and because the positron energy correlates with impact position, the changing illumination imprints the betatron frequencies and their harmonics, $\omega_i$, onto the data.  Finally, the anomalous frequency \oa creates sidebands on the betatron-related frequencies.  Because the betatron motion decoheres over the course of a muon fill, the oscillatory amplitudes change versus time. These effects, integrated over energy, determine the total positron rate as a function of time, described~\cite{tyler_thesis} as the real part of
\begin{widetext}
\begin{align} \label{eq:fullModel}
\rho(t)
& = N_0 \left[1-\Lambda(t)\right] e^{-t/\gamma\tau_{\mu}} \notag \\
  &\times\left\{1 + A_a\cos(\omega_a t - \phi_0)+\sum_{n=1}^{24}\sum_{{\mathbf j}\ne 0}f_n e^{-i(2nj_c\pi/24)}\left( A_{\mathbf{j}}^{(n)}(t) e^{i\omega_{\mathbf{j}} t}
   + \left[\frac{A_a^{(n)}}{2} A_{\mathbf{j}}^{(n)}(t) + \Delta A_{\mathbf{j}\pm a}^{(n)}(t) \right]e^{i\omega_{\mathbf{j}\pm a}t}\right)\right\}
\end{align}
\end{widetext}
where $\mathbf{j} = (j_x, j_y, j_c)$ represents the indices in the Fourier series over the horizontal and vertical betatron oscillation frequencies $\omega_x$ and $\omega_y$, respectively, and over the cyclotron frequency $\omega_c$, $\omega_{\mathbf{j}} = j_x\omega_x + j_y\omega_y + j_c\omega_c$.  The frequency $\omega_{\mathbf{j}\pm a} = \omega_{\mathbf{j}}\pm \omega_a$.  The outer summation indexed by $n$ combines data from the 24 calorimeters.  The inner summation incorporates the harmonics of the beam frequencies as well as their beating with $\omega_a$.  The parameter $N_0$ represents the total number of muons observed and $f_n$ the fraction of events observed in calorimeter $n$, while $\Lambda(t)$ describes the loss of beam muons through processes other than decay. The asymmetries $A_a$ and $A_a^{(n)}$ represent the average asymmetry overall and in calorimeter $n$, respectively. They depend on the acceptance as a function of positron energy.  The amplitude envelopes $A_{\mathbf{j}}^{(n)}(t)$ for oscillation at the frequency combinations $\omega_{\mathbf{j}}$ and $\omega_{\mathbf{j}\pm a}$ incorporate effects from the variation of detector acceptance across the beam storage volume and the decoherence of the muon motion within the beam. The additional sideband terms $\Delta A_{\mathbf{j}\pm a}^{(n)}(t)$ accommodate asymmetries between the $\pm\omega_a$ sidebands of a given beam frequency, which correspond to a small time dependence of the beam ensemble averaged phase $\phi_a$ resulting from small variations of the makeup of the ensemble over a muon fill.

The phase factors $e^{-i(2nj_c\pi/24)}$ represent the relative cyclotron phase at each calorimeter.  With these factors made explicit, the cancellation of the asymmetry amplitudes in the sum $\sum_{n=1}^{24} f_n A_{\mathbf{j}}^{(n)}(t)e^{-i(2nj_c\pi/24)}$ becomes transparent. If the acceptances and rates observed by each calorimeter were identical, the sum over the exponential phase terms would cancel exactly, and the beam modulations of the observed rate would vanish. However, variations in the material shielding the calorimeters lead to small differences in acceptance as a function of positron energy.  The cancellation therefore becomes approximate, resulting in a nonzero residual contribution in the calorimeter sum that is still much smaller than the size of the modulation observed in an individual calorimeter. As discussed in detail below, the different analyses employed a variety of methods to measure or model the envelopes and phase advance.

Note that for clarity, effects of the muon momentum spread in the beam and time-of-flight variations for the positrons from muon decay have been neglected here, though these play a role in several of the beam dynamics corrections (see Sections~\ref{sec:bd} and \ref{sec:timeVaryingEnsemble}).

\begin{table}[b]
\caption{Expected beam dynamics frequencies that play a role in modeling the spin precession data for an effective field index $n=0.108$ in a continuous quadrupole approximation (see~\cite{Run1PRDomegaa}).  The measured frequencies coincide with these at the 1\% level. \label{tab:freqs}}
\begin{center}
\begin{tabular}{ccccc}\toprule
Physical frequency  & Name   & Expression                 & Frequency & Period \\
                    &             &                           & (MHz)     & (\unit{\us}) \\ \midrule
Anomalous precession& $f_{\rm a}$  & $\frac{e}{2\pi m} \amu B$&0.229      &4.37          \\
Cyclotron           & $f_{\rm c}$  & $\frac{v}{2\pi R_0}$     &6.71       &0.149         \\
Horizontal betatron & $f_{\rm x}$  & $\sqrt{1-n}\,f_{\rm c}$  &6.34       &0.158         \\
Vertical betatron   & $f_{\rm y}$  & $\sqrt{n}\,f_{\rm c}$    &2.20       &0.453         \\
Horizontal \ac{CBO}      & $f_{\rm CBO}$& $f_{\rm c} - f_{\rm x}$  &0.37       &2.68          \\
Vertical waist      & $f_{\rm VW}$ & $f_{\rm c} - 2f_{\rm y}$ &2.31       &0.433         \\
\bottomrule
\end{tabular}
\end{center}
\label{default}
\end{table}

The overall scale of the amplitude  $A_{\mathbf{j}}^{(n)}(t)$ for a harmonic of one of the fundamental frequencies couples strongly to the shape of the acceptance as a function of the transverse beam position. Table~\ref{tab:harmonics} shows how the expected amplitudes for the first three harmonics of $\omega_x$ depend on the power of the radial coordinate in a Taylor expansion of the acceptance map as a function of muon decay position. The results show that odd harmonics couple predominantly to odd powers in the acceptance map, and even harmonics to even powers.  Vertical frequency harmonics behave analogously. The frequency contributions in data agree qualitatively with these expectations. For example, at leading order, the acceptance behaves linearly in the radial direction and quadratically in the vertical.  The fundamental horizontal betatron frequency, $f_x$, and the second harmonic of the vertical betatron oscillation, $2f_y$, aliased down to $f_{\rm CBO}$ and $F_{VW}$ respectively, exhibit the largest amplitudes in the data, as predicted.

\begin{table}
  \centering
  \caption{The amplitude predictions from the analytical analysis in Ref.~\cite{tyler_thesis} for $A_x = A^{(n)}_{\mathbf{j}}$ with $\mathbf{j} = (j_x,0,0)$ in Eq.~\ref{eq:fullModel} of the horizontal ($x$) oscillations in the accepted positron data that results from the horizontal betatron oscillations of the beam. These amplitudes determine relative amplitudes of the peaks related to that motion in the FFT of the residuals shown in Fig.~\ref{fig:ffts}.  The coordinate $x_{e}$ represents the offset of the beam oscillation relative to the center of the storage volume (normalized to the storage volume size).  Because $x_{e}$ is generally close to zero, it suppresses contributions relative to $A_{x}$.  As a result, the purely $A_x$ terms dominate, and odd harmonics couple predominantly to odd powers in the acceptance map, and even harmonics to even powers.}
  \begin{tabular}{clll}
    \toprule
      Series & \multicolumn{3}{c}{Harmonic of frequency $\omega_x$} \\
      Term   & 1                                                                                   & 2                                            & 3  \\    \midrule
      1      & $\quad\quad\frac{A_x}{2}$                                                           & $\quad\quad\,\, 0$                           & $\,\,\,\, 0$  \\
      2      & $2x_e\left(\frac{A_x}{2}\right)$                                                    & $\quad\,\,\,\,\left(\frac{A_x}{2}\right)^2$  & $\,\,\,\, 0$ \\
      3      & \,\,\,\,\,\,$3\left(\frac{A_x}{2}\right)^3 + 3x_e^2\left(\frac{A_x}{2}\right)\quad$ & $3x_e\left(\frac{A_x}{2}\right)^2\quad$      & $\left(\frac{A_x}{2}\right)^3$ \\
    \bottomrule
  \end{tabular}
  \label{tab:harmonics}
\end{table}

\subsubsection{Fitting for \oam}
\begin{figure*}[tb]
    \centering
    \includegraphics[width=\linewidth]{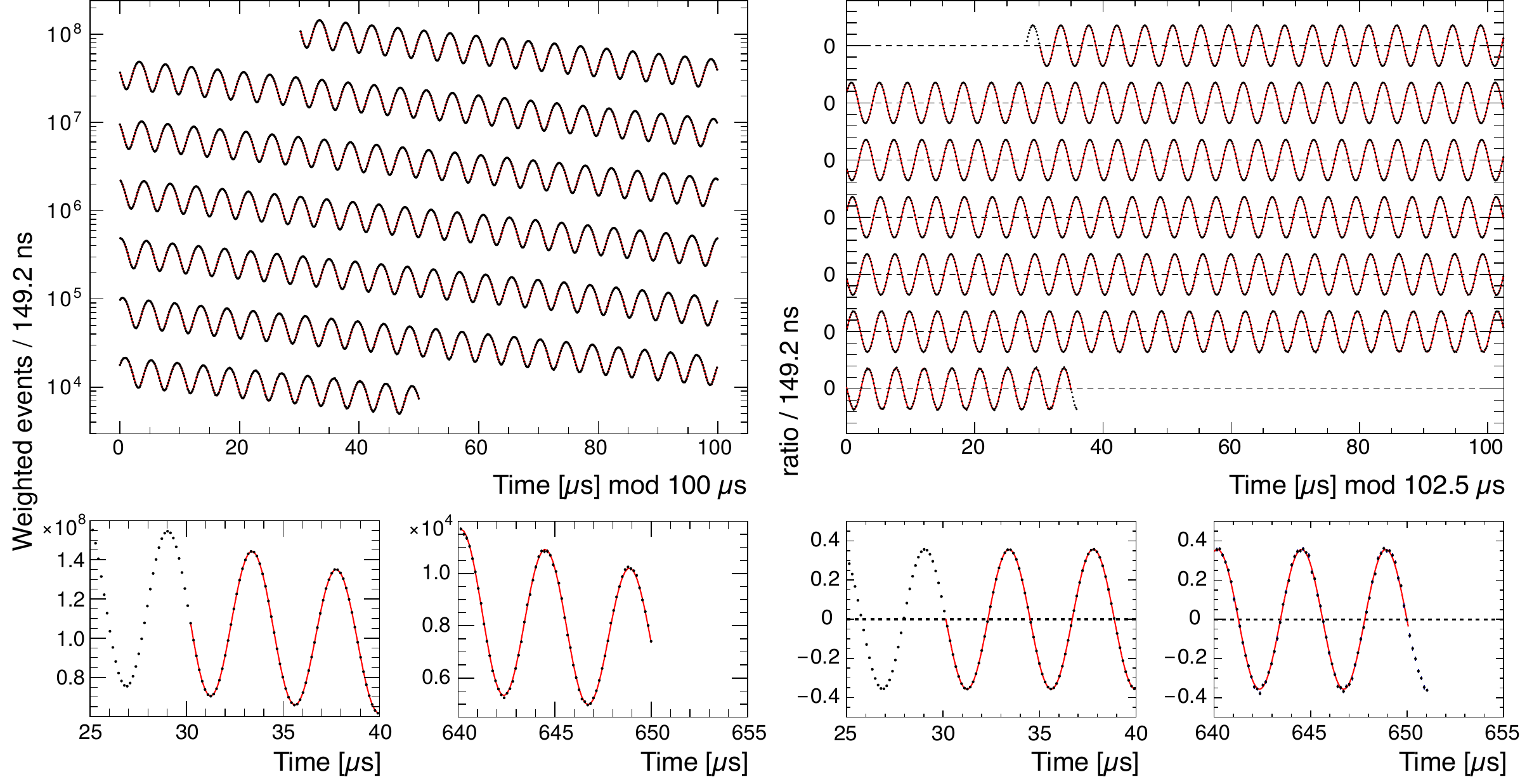}
    \caption{Left: Asymmetry-weighted time series summed over all data from Runs 4, 5, and 6 (black points) with the sum of the fit functions from the four datasets overlaid (red curve).  Right: Asymmetry-weighted ratio method time series from the Run 4 data (black points) with its best fit function overlaid (red curve). The top plots show the full time series, wrapped for viewability, while the plots below zoom in on the beginning and end regions of the fitted data.}
    \label{fig:wiggle}
\end{figure*}
Extraction of \oam involves fits to binned time series of data, summed over all calorimeters, performed with four distinct variations.  This section provides the details for the analysis approaches summarized in Table~\ref{tab:analysis_choices_wa}.

\begin{figure}[tb]
\centering
\includegraphics[width=0.48\textwidth]{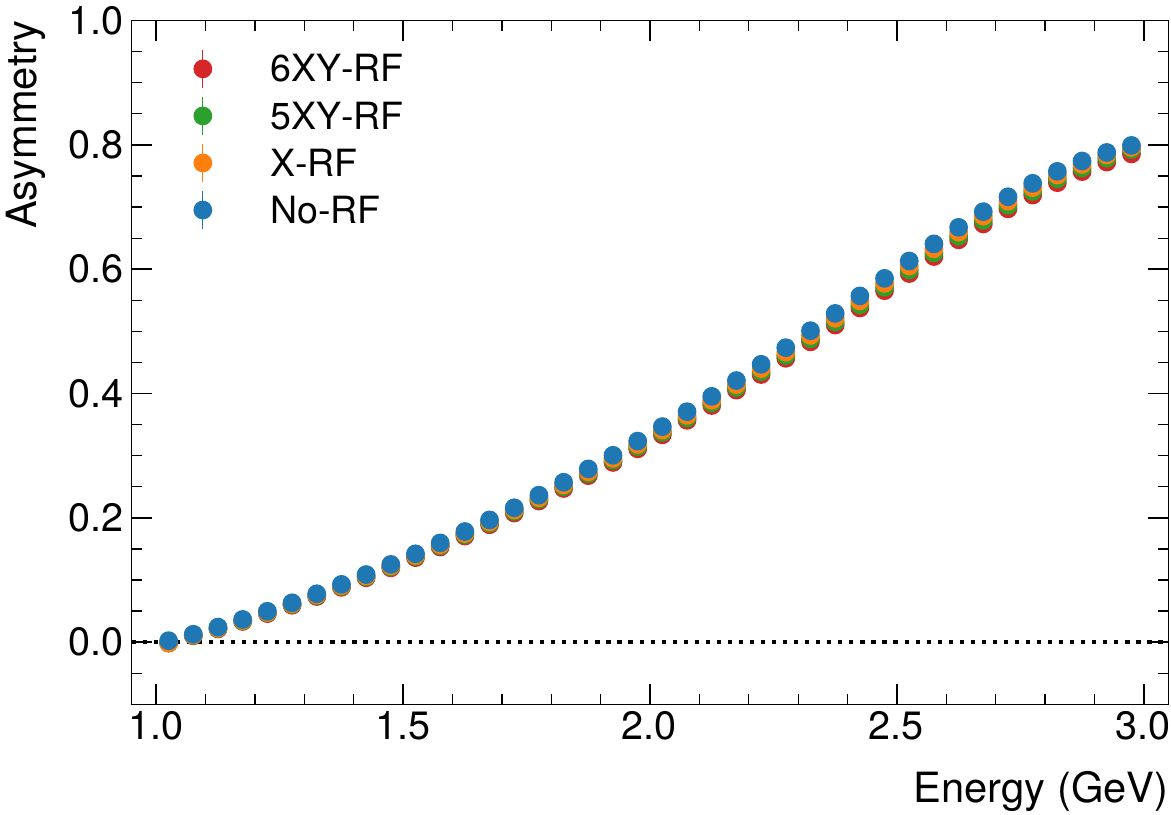}
\caption{Asymmetry values measured from the data used in the asymmetry-weighted fits in the global reconstruction analyses.  Weights for the local reconstruction analyses are very similar.}
\label{fig:asymm}
\end{figure}

Two variations bin the data with a width of the cyclotron period of 149.2~\unit{\ns}, as discussed above.  The first of these simply bins the time series for the positron candidates, as shown in Fig.~\ref{fig:wiggle}, using either a unit weight (the threshold or T method) or with a weight of the oscillation amplitude $A$ based on the positron energy, dubbed the asymmetry (A) method.  The asymmetry-weighting maximizes the \oam statistical sensitivity, and utilizes the asymmetry measured versus positron energy from the data.  Figure~\ref{fig:asymm} shows the measured asymmetries for the global reconstruction analyses.  All positron-based methods provide asymmetry-weighted fits for the final determination of \oam, with unit-weighted fits providing a consistency check.

The second, ratio (R) method, randomly subdivides the data into four groups to separate the precession oscillation from the exponential decay.  By shifting times early or late by one-half an \oa period in two of the categories and combining (see Ref.~\cite{Run1PRDomegaa,osti_1581404,Foster:2023hpz} for details), the oscillatory behavior can be isolated from the exponential (Fig.~\ref{fig:wiggle}b).  While the first method has slightly better statistical power, the ratio method has reduced sensitivity to slow effects such as gain recovery.

A new stroboscopic (S) method subdivides the positron data into 18 subsets based on the horizontal \ac{CBO} phase of each event.  The horizontal \ac{CBO} period of $T_{\ac{CBO}}=2.7$~\unit{\us}, results in a time binning of 150~\unit{\ns}.  This binning is close to the nominal muon cyclotron period to suppress cyclotron frequencies in the data, which are further suppressed by the same time randomization described earlier.  The final \oam determination includes input from the stroboscopic method.

The energy flow (Q) method also utilizes both a summed-energy approach and a ratio method (RQ).  Given the (blinded) \SI{1.25}{ns} sampling period of the digitizers, the Q method could not bin directly in the cyclotron period $T_{C}$.  Instead, it used 150~\unit{\ns} bins, the multiple of the digitization frequency closest to $T_{C}$. Because the Q method performs a running energy sum, there is no ``candidate time'' to randomize away the residual cyclotron frequency in the data. Instead, each time bin is randomly combined with its neighbors using a kernel weighting of 0.25:0.5:0.25.  Other corrections in this method are documented elsewhere~\cite{ChakrabortyR,PeckJ}. This method serves as a cross-check with very different systematic uncertainties.

Fits to the binned histograms determine \oam, and mainly use one of two complementary forms in the rate model to capture the imprint of the beam oscillations discussed in Section~\ref{sec:bdoa}. The first incorporates beam dynamical effects in the nonlinear form
\begin{multline}
N(t) =  N_{0}e^{-t/\gamma\tau_{\mu}}\eta_{N}(t)[1-\kappa\Lambda(t)] \\
     \times \left\{1 + A_{a}\eta_{A}(t)\cos[\oam t - \phi_{0} + \eta_{\phi}(t)]\right\}.
     \label{eq:waform1}
\end{multline}
In this expression, $\eta_N(t)$ captures the oscillation in the overall positron acceptance.  Because the change in acceptance changes the ensemble average positron energy, the terms $\eta_A(t)$ and $\eta_{\phi}(t)$ accommodate the resulting oscillations in the amplitude of the oscillation (the asymmetry) and in the phase of the oscillation, respectively.  All three of these $\eta$ terms also describe the decay in the amplitudes of the oscillations as the muon beam betatron motion decoheres. 
This form was used in the analyses published earlier~\cite{Run1PRDomegaa,PhysRevD.110.032009}.  The explicit implementations by the different analysis groups are described below.  The frequencies of importance for the fits are
\begin{align} \label{eqn:freqs}
    \{\omega_i\} &= \{\omega_{\rm CBO},2\omega_{\rm CBO},\omega_{\rm VW}, \omega_y, \omega_{{\rm CBO}\pm{\rm VW}}, \\ 
                 & \quad\quad\omega_{{\rm CBO}\pm y}, \omega_{a\pm\rm{VW}}, \omega_{y\pm a} \}, \nonumber
\end{align}
where the individual frequencies are defined in Table~\ref{tab:freqs} and the frequency combinations $\omega_{a\pm b} = \omega_a \pm \omega_b$.

The second form utilizes the linearized description, inspired by Eq.~\eqref{eq:fullModel}, with
\begin{multline}
N(t) =  N_{0}e^{-t/\gamma\tau_{\mu}}[1-\kappa\Lambda(t)][1+A\cos(\oam t - \phi_0)] \\
     \times \sum_{\omega_i} \left[\xi_{\omega_i}(t) + \xi_{\omega_i + \omega_a}(t) + \xi_{\omega_i - \omega_a}(t) \right],
     \label{eq:waform2}
\end{multline}
where the sum is over the frequencies $\omega_i$ given above.
In both expressions, \oam and $A_{a}$ represent the muon anomalous precession frequency and the average weak-decay asymmetry, respectively. The phase $\phi_{0}$ is the phase difference between the muon momentum and spin directions extrapolated back to the time of injection.  The phase and asymmetry average over the ensemble of muons.  $N_{0}$ provides the overall normalization and $\gamma\tau_{\mu}$ represents the average boosted muon lifetime (about 64.4~\unit{\us}).  Note that the linear and nonlinear representations of the beam-related corrections have small differences at second order in the small perturbations.  As a result, the $N_0$ normalizations are expected to be similar, but not necessarily identical, between the two methods.  Studies of fits in which the models were built on the muon momentum dependence of lifetime, asymmetry, and beam motion, then integrated over the measured beam momentum distribution, yielded essentially identical fit results.  These studies confirmed that the use of the average quantities is sufficient. The $\Lambda(t)$ term accounts for loss of beam muons, mainly through scattering off collimators, and is measured using correlated, minimum-ionizing signals in three or more of the calorimeter stations~\cite{Run1PRDomegaa}.  The $\kappa$ scale factor, which the fit determines, accounts for the overall loss rate and detection efficiency. 
At the precision level of the experiment, the nonlinear and linear forms should yield equivalent results.  For example, letting the $\omega_{i}\pm\oam$ sideband amplitudes in the second summation of Eq.~\eqref{eq:waform2} evolve independently recovers the phase advance behavior in $\eta_{\phi}(t)$ in Eq.~\eqref{eq:waform1} from the frequency evolution of $\omega_{i}$.  As Fig.~\ref{fig:ffts} shows, neglecting these terms from the fit model leaves strong frequency peaks in the \ac{FFT} of the fit residuals.  Neglecting these terms would shift the measured \oam frequency by about \SI{800}{ppb} in the \RunFournoRF dataset, while only about \SI{80}{ppb} in the \RunFiveXY and \RunSixXY datasets in which the amplitudes of the coherent beam motions were significantly damped.

The analysis groups took a variety of approaches to modeling the time modulations introduced by beam motion.  Table~\ref{tab:analysis_choices_wa} summarizes the combinations.   

\paragraph{Nonlinear models} For the nonlinear parameterization of Eq.~\eqref{eq:waform1}, the rate modulation term takes the following forms.

\paragraph{BNL + parameterizations} The first approach, initially developed empirically for the \ac{BNL} muon $g-2$ measurements~\cite{PhysRevD.73.072003}, takes the product form
\begin{equation}
    \eta_{N}(t) = \prod_{\{\omega_i\}}N_{\omega_i}(t),
\end{equation}
where the frequencies are defined in Table~\ref{tab:freqs}.  The parameterizations of the individual terms are variants of
\begin{widetext}
\begin{align*}
{\rm CBO}:\quad &N_{\omega_{\rm CBO}}(t) = 1 + A_{0_{\rm CBO}}A_{\rm CBO}(t)[\cos(\omega_{\rm CBO}^0t + Ae^{-t/\tau_A} -\phi_{\rm CBO})],\\
{\rm 2CBO}:\quad &N_{\omega_{\rm 2CBO}}(t) = 1 + A_{0_{\rm 2CBO}}A^2_{\rm CBO}(t)[\cos(2(\omega_{\rm CBO}^0t + Ae^{-t/\tau_A}) -\phi_{\rm 2CBO})],\\
{\rm VW}: \quad &N_{\rm VW}(t) = 1 + A_{\rm VW}[\cos(\omega_{\rm VW}t-\phi_{\rm VW}) + \cos(\omega_{\rm VW}t + A_{\rm VW}^{\rm PA} e^{-t/\tau^{\rm PA}_{\rm VW}})\\
{\rm CBO}\pm{\rm VW}, {\rm CBO}\pm y: \quad & N_{\omega_{a\pm b}}(t) = 1 + A_{a\pm b}(t)[\cos(\omega_{a+b}t - \varphi_{a+b}) + r\cos(\omega_{a-b}t - \varphi_{a-b})] \\
{\rm others}:\quad &N_{\omega_i}(t) = 1 + A_i(t)\cos(\omega_i^0t-\phi_i)
\end{align*}
\end{widetext}
The envelopes $A_i(t)$ and $A_{a\pm b}(t)$ are parametrized  as exponentials $A_i^0 e^{-t/\tau_i}$ except for the dominant horizontal \ac{CBO} term, which includes an additive constant.  The phase advance $\eta_{\phi}(t)$ and asymmetry modulation terms have similar forms. The envelope parameters float in the fits for \oam. Note that the ratio method fits, which further randomize the times to eliminate the vertical frequencies, use a simplified version of this parameterization with the vertical frequency terms set to unity.  The Q method fits utilize a similar parameterization.

\paragraph{Moments + splines} A second form ($\alpha/\beta$) for the nonlinear model derives from an analysis of moments of the beam motion~\cite{osti_1581403,Run1PRDomegaa}.  It takes the form
\begin{align*}
\eta_{N}(t) &= [1+n_{\omega_{\rm CBO}}(t) + n_{2\omega_{\rm CBO}}(t)][1+n_{\omega_y}(t)+n_{\omega_{\rm VW}}(t)] \\
&+ n_{\omega_{{\rm VW}-{\rm CBO}}}(t) + n_{\omega_{{\rm VW}+{\rm CBO}}} + n_{\omega_{y-{\rm CBO}}} + n_{\omega_{y+{\rm CBO}}}.
\end{align*}
In this model, the individual terms take the form
\begin{equation}
    n_{\omega_i}(t) = \alpha_i(t)\cos(\omega_i t) + \beta_i(t)\sin(\omega_i t).
    \label{eqn:alpha_beta}
\end{equation}
Note that allowing an independent time dependence between $\alpha_i(t)$ and $\beta_i(t)$ accommodates the phase advance behavior captured by exponential in the $N_{\omega_{\rm CBO}}(t)$ and $N_{\omega_{\rm VW}}(t)$ terms.  While trigonometrically equivalent to the $A\cos(\omega t + \phi)$ (\Aphi) form, the \alphabeta form lends itself well to a data-driven determination of the envelopes because at larger times when the amplitude goes to zero, the $\phi$ term becomes ill-defined.  By fixing a reference frequency $\omega_i$, fits to windows of order \SI{10}{}-\SI{15}{\us} in size determine  the $\alpha_i(t)$ and  $\beta_i(t)$ envelopes.  The envelopes determined from these ``sliding window''fits can then be smoothed and fed into the \oam fit, or used as a cross-check for parameterizations of the envelopes within the \oam fit.  With this nonlinear model, cubic splines were used to smooth and interpolate the envelopes for the \oam fit.

In the nonlinear models given above, the product of the oscillatory terms generates sidebands, such as the two sidebands on either side of the dominant \ac{CBO} peak in the \ac{FFT} of the residuals plot in Fig.~\ref{fig:ffts}. These products naturally create sidebands of equal amplitudes, which necessitate the introduction of the additional asymmetry and phase modulation terms, $\eta_A(t)$ and $\eta_\varphi(t)$, to capture the full dynamics.  

\paragraph{Linear models} In the linearized form of the fit function [Eq.~\eqref{eq:waform2}], the sidebands are included explicitly in the sum, each with an independent amplitude. As a result, separate asymmetry and phase modulation terms are no longer required.
The analyses utilizing this model again take   
\begin{equation}
    \xi_{\omega_i}(t) = \alpha_i(t)\cos(\omega_i t) + \beta_i(t)\sin(\omega_i t).
\end{equation}
There were two model variants using this form.

\paragraph{Gaussian Process Regression (GPR)} In a similar fashion to the moments+spline method, the data-driven window fits measured the envelope parameters $\alpha_i(t)$ and $ \beta_i(t)$, but now interpolated them with a Gaussian Process Regression method for use in the fit for \oam.  

\paragraph{Parameterization} Each envelope in this approach is parameterized as a product of an exponential and a polynomial, with the power of the polynomial for each envelope, at most cubic, determined by requiring that each higher order term improve the fit $\chi^2$ by at least a unit, given the additional degree of freedom. 

\begin{figure*}[tb]
    \centering
    \includegraphics[width=0.9\linewidth]{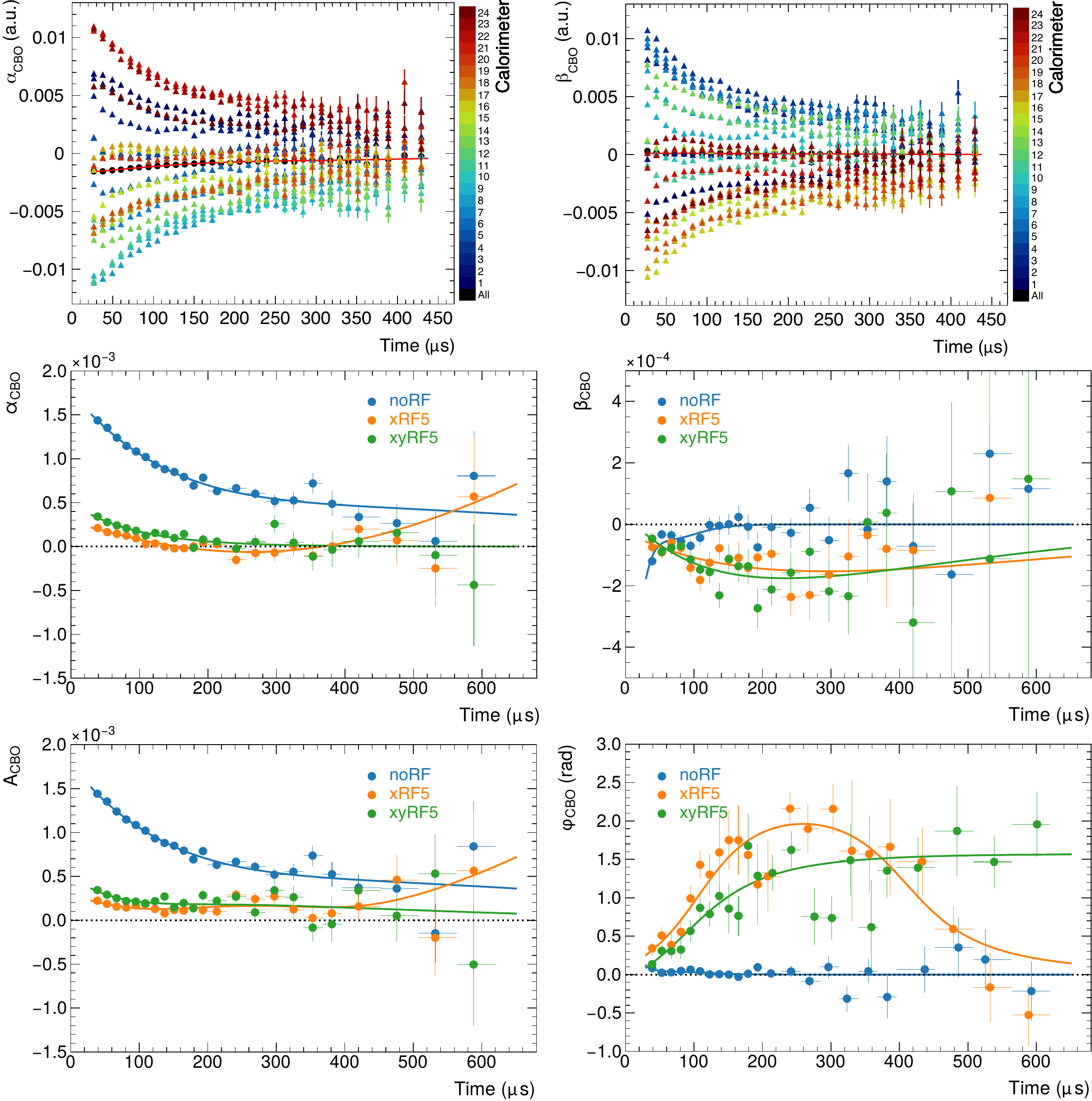}
    \caption{Top: The $\alpha_{\rm CBO}(t)$ (left) and $\beta_{\rm CBO}(t)$ (right) envelopes extracted from fits to the noRF data in successive time windows (sliding windows) for each of the 24 calorimeters (triangles) and for the sum of the calorimeters (black circles) from analysis \wagroupSJTU.  The spread of the envelopes across the calorimeters arises from the aliasing of the horizontal betatron oscillations at frequency $\omega_x$ by the cyclotron frequency to the \ac{CBO} frequency $\omega_{\rm CBO}$.  In the sum of the calorimeter data, these envelopes largely cancel, leaving only a small residual. The lines indicate the Gaussian Process Regression interpolation used in the fits for that analysis.
    Middle: The $\alpha_{\rm CBO}(t)$ (left) and $\beta_{\rm CBO}(t)$ envelopes from sliding window fits for the noRF, xRF5, and xyRF5 data by analysis \wagroupTyler.  For points at late times, results from several windows have been averaged for clarity of display.  Note in particular the amplitude drop in the envelopes with application of the horizontal RF damping.  The lines indicate the shapes parameterized within the \oam fits for that analysis.  
    Bottom: The middle envelopes and their parameterizations converted to the amplitude (left) and phase advance (right) envelopes.}
    \label{fig:slidingwindows}
\end{figure*}

As examples of the envelope behavior, Fig.~\ref{fig:slidingwindows} compares the data-driven (from \wagroupSJTU) and parameterized (from \wagroupTyler) envelopes for individual calorimeters and for the azimuthal sum from two analyses.
The per-calorimeter envelopes show the effects of the relative betatron phases at each station,
which cause the envelopes to significantly cancel in the calorimeter sum.   
The figure also demonstrates a reduction in envelope size due to RF damping of the beam's coherent betatron oscillations.

The model for the ratio method fits uses Eq.~\eqref{eq:waform1} to compute a ratio using the same algorithm applied to the data to produce the distribution shown in Fig.~\ref{fig:wiggle}.

The fits utilize data in the time window \SI{30.1384}{\us} to \SI{650.0644}{\us}. The start time corresponds to a node in the anomalous spin precession, which studies have shown minimizes the overall systematic effects arising from uncertainties in the calorimeter gains and beam dynamics corrections.  The ESQs have also settled into their storage potential by that time, resulting in stable beam motion and thus minimizing beam-motion-induced biases. This window corresponds to $4155$ bins, and the number of fit parameters ranges from $7$ (for the stroboscopic method) to $50$ (for the linearized asymmetry-weighted fits).  For clarity while fitting and comparing results, all analyses express $R(\oa)$ as
\begin{equation}
\oam = (2\pi \cdot0.2291\,{\rm{MHz}})\left[1+\left(R+\Delta R_{\rm blind}\right)\times 10^{-6}\right],
\label{eq:Rdef}
\end{equation}
which is also where the software blinding shift $\Delta R_{\rm blind}$ is applied.  That shift has been removed in the results shown below.

\subsubsection{Fit results} \label{sec:wa-fit-results}

\begin{table*}[tb]
    \centering
    \caption{The five major parameters for the NoRF \oa dataset for each group. The analyses in bold are included in the final average of \oam.  Note that the uncertainties on $N_0$ are of order $O(10^3)$. The asymmetry parameter depends strongly on positron momentum, leading to significant variation in the average asymmetry fit parameter across different weighting methods and momentum ranges. The $R(\oa)$ values have had hardware and software blindings removed.}
    \label{tab:wa-fit-results-major-parameters-norf}
\begin{tabular}{llllllll}
\toprule
{Group}                 & {Analysis}  & {$R(\oa)$ (ppm)} & {$N_0$}              & {$A_0$}       & {$\phi_{0}$ (rad)} & {$\gamma \tau_{\mu}$ (\SI{}{\micro\second})} & reduced $\chi^2$\\ \midrule
\textit{\wagroupZep}    & T           & -98.833(200)     & $1.54\times 10^{8}$  & 0.368141(7)   & 4.118853(32)       & 64.400(8)                                    &  0.988   \\
                        & \textbf{A}  & -98.729(180)     & $7.00\times 10^{7}$  & 0.361410(7)   & 4.119000(29)       & 64.405(8)                                    &  0.979   \\ \midrule
\textit{\wagroupTyler}  & T           & -98.833(201)     & $1.54\times 10^{8}$  & 0.368146(7)   & 4.118850(33)       & 64.375(19)                                   &  0.988   \\
                        & \textbf{A}  & -98.729(180)     & $7.00\times 10^{7}$  & 0.361320(8)   & 4.119370(29)       & 64.416(12)                                   &  0.991   \\ \midrule
\textit{\wagroupMurong} & RT          & -98.792(205)     & --                   & 0.362000(7)   & 4.119300(33)       & --                                           &  1.004   \\
                        & \textbf{RA} & -98.722(182)     & --                   & 0.355000(6)   & 4.119700(30)       & --                                           &  1.005   \\ \midrule
\textit{\wagroupOnKim}  & ST          & -98.818(204)     & $1.54\times 10^{8}$  & 0.362040(7)   & 4.119200(33)       & 64.403(3)                                    &  1.005   \\
                        & \textbf{SA} & -98.726(183)     & $6.92\times 10^{7}$  & 0.355000(6)   & 4.120000(30)       & 64.406(3)                                    &  1.010   \\ \midrule
\textit{\wagroupBU}     & T           & -98.823(203)     & $1.52\times 10^{8}$  & 0.366536(7)   & 4.119040(33)       & 64.402(1)                                    &  1.004   \\
                        & A           & -98.753(182)     & $6.93\times 10^{7}$  & 0.356835(7)   & 4.119550(30)       & 64.403(1)                                    &  1.013   \\
                        & RT          & -98.832(203)     & --                   & 0.380670(7)   & 4.118950(33)       & --                                           &  1.003   \\
                        & \textbf{RA} & -98.765(183)     & --                   & 0.356850(6)   & 4.119500(30)       & --                                           &  1.012   \\ \midrule
\textit{\wagroupSJTU}   & T           & -98.869(199)     & $1.52\times 10^{8}$  & 0.372598(7)   & 4.118675(32)       & 64.403(1)                                    &  0.997   \\
                        & \textbf{A}  & -98.819(179)     & $7.05\times 10^{7}$  & 0.362871(6)   & 4.119187(29)       & 64.403(1)                                    &  1.014   \\ \midrule
\textit{\wagroupITA}    & T           & -98.789(200)     & $1.52\times 10^{8}$  & 0.371800(10)  & 4.118580(32)       & 64.403(1)                                    &  1.003   \\
                        & A           & -98.747(180)     & $6.92\times 10^{7}$  & 0.356530(10)  & 4.119000(30)       & 64.403(1)                                    &  1.021   \\
                        & RT          & -98.787(207)     & --                   & 0.365690(10)  & 4.119000(30)       & --                                           &  0.993   \\
                        & \textbf{RA} & -98.746(187)     & --                   & 0.356620(10)  & 4.119500(30)       & --                                           &  1.000   \\ \midrule
\textit{\wagroupUKY}    & Q           & -98.911(208)     & $5.12\times 10^{11}$ & 0.244866(5)   &-2.545462(34)       & 64.450(187)                                  &  1.041   \\
                        & RQ          & -98.916(208)     & --                   & 0.244866(5)   &-2.545462(34)       & --                                           &  1.007   \\ \bottomrule
\end{tabular}
\end{table*}

\begin{table*}[tb]
\caption{The major fit results for the three RF-on datasets for the methods included in the final \oa averaging procedure (A, RA, and SA-Methods). The values for the NoRF dataset can be seen in Table \ref{tab:wa-fit-results-major-parameters-norf}. The values of $R(\oa)$ listed here have no beam dynamics corrections or combination with magnetic field values, and therefore some small variation across datasets is expected.  They have had both hardware and software blindings removed. The $\gamma\tau$ uncertainties depend strongly on what other slowly-varying parameters, such as the muon loss scaling factor, are floating in any given fit. }
\centering
\begin{tabular}{llllllll}\toprule
{Dataset} & {Group/Method}    & {$R(\oa)$ (ppm)} & {$N_0$}  & {$A_0$}  & {$\phi_{0}$ [rad]} & {$\gamma \tau_{\mu}$ (\SI{}{\micro\second})} & reduced $\chi^2$ \\ \midrule
xRF       & \wagroupZep/A     & -98.087(239)     & 3.99E+07 & 0.359194(8)  & 4.118620(39)           & 64.401(9)                            &  0.977  \\
          & \wagroupTyler/A   & -98.108(239)     & 3.99E+07 & 0.359072(8)  & 4.118610(39)           & 64.424(27)                           &  1.002  \\
          & \wagroupMurong/RA & -98.139(242)     & --       & 0.353000(8)  & 4.118900(40)           & --                                   &  1.002  \\
          & \wagroupOnKim/SA  & -98.091(244)     & 3.94E+07 & 0.353000(8)  & 4.119000(40)           & 64.413(18)                           &  1.002  \\
          & \wagroupBU/RA     & -98.222(243)     & --       & 0.354430(8)  & 4.118600(40)           & --                                   &  1.008  \\
          & \wagroupSJTU/A    & -98.165(238)     & 4.02E+07 & 0.360473(8)  & 4.118289(39)           & 64.389(2)                            &  1.010  \\
          & \wagroupITA/RA    & -98.202(243)     & --       & 0.354180(7)  & 4.118630(41)           & --                                   &  1.002  \\
xyRF5     & \wagroupZep/A     & -98.693(244)     & 3.84E+07 & 0.358743(8)  & 4.115853(40)           & 64.399(8)                            &  1.001  \\
          & \wagroupTyler/A   & -98.692(245)     & 3.83E+07 & 0.358564(10) & 4.115870(40)           & 64.396(25)                           &  1.000  \\
          & \wagroupMurong/RA & -98.680(246)     & --       & 0.352700(8)  & 4.116000(40)           & --                                   &  0.992  \\
          & \wagroupOnKim/SA  & -98.712(249)     & 3.79E+07 & 0.353000(8)  & 4.116000(41)           & 64.397(9)                            &  0.992  \\
          & \wagroupBU/RA     & -98.734(249)     & --       & 0.353910(8)  & 4.115900(41)           & --                                   &  0.998  \\
          & \wagroupSJTU/A    & -98.722(243)     & 3.86E+07 & 0.359930(8)  & 4.115569(40)           & 64.393(1)                            &  1.005  \\
          & \wagroupITA/RA    & -98.713(248)     & --       & 0.353550(10) & 4.114470(60)           & --                                   &  0.995  \\
xyRF6     & \wagroupZep/A     & -98.643(275)     & 3.04E+07 & 0.357396(9)  & 4.118226(45)           & 64.400(9)                            &  0.994  \\ 
          & \wagroupTyler/A   & -98.645(275)     & 3.04E+07 & 0.357346(9)  & 4.118200(45)           & 64.344(34)                           &  0.997  \\ 
          & \wagroupMurong/RA & -98.600(279)     & --       & 0.351300(9)  & 4.119000(46)           & --                                   &  0.999  \\ 
          & \wagroupOnKim/SA  & -98.609(280)     & 3.01E+07 & 0.351000(9)  & 4.119000(46)           & 64.369(37)                           &  1.002  \\ 
          & \wagroupBU/RA     & -98.694(280)     & --       & 0.352930(9)  & 4.118200(46)           & --                                   &  1.003  \\ 
          & \wagroupSJTU/A    & -98.762(273)     & 3.06E+07 & 0.358758(9)  & 4.117896(45)           & 64.387(3)                            &  0.993  \\ 
          & \wagroupITA/RA    & -98.664(280)     & --       & 0.352562(10) & 4.118310(50)           & --                                   &  0.998  \\ \bottomrule
\end{tabular}          

\label{tab:wa-fit-results-a-method-only}
\end{table*}

\begin{figure}[tb]
    \centering
    \includegraphics[width=\linewidth]{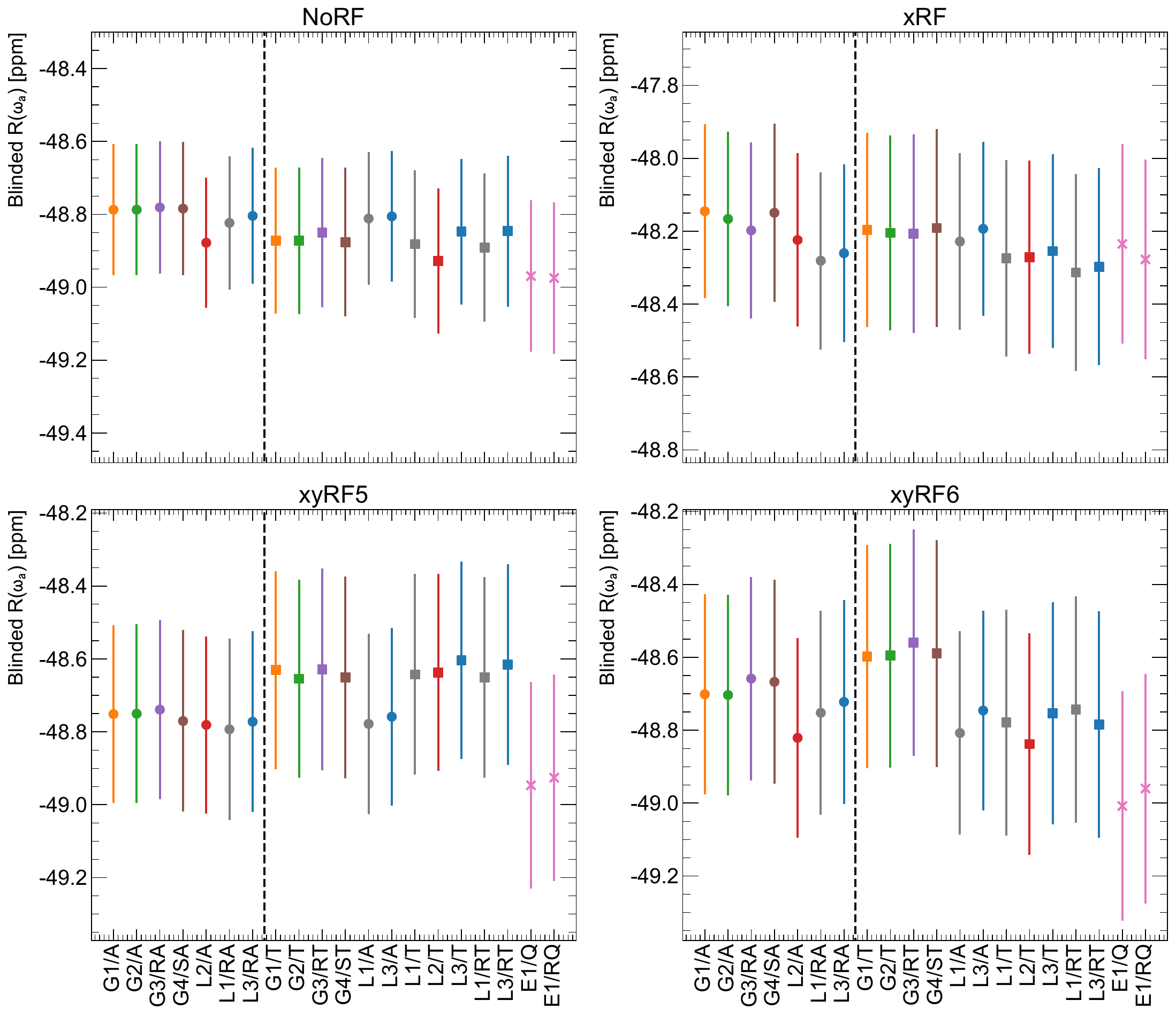} 
    \caption{Value of the commonly blinded \Roa for all datasets. The marker indicates the weighting used when histogramming the data, while the color corresponds to the analysis group. The values to the left of the dashed line in each panel are included in the final average value of \oam. Other values are included as a cross-check.}
    \label{fig:wa-all-datasets-all-methods}
\end{figure}

Table~\ref{tab:wa-fit-results-major-parameters-norf} presents the results from the full fits for the basic five parameters shown in Eq.~\eqref{eq:omega_fiveparameterfit} for the NoRF datasets for each of the analysis methods listed in Table~\ref{tab:analysis_choices_wa}. 
Examples of a full representative fit result are presented in Appendix~\ref{app:representative_fit}.

Table~\ref{tab:wa-fit-results-a-method-only} presents the results from the asymmetry-weighted analyses for the other three major datasets.
The results show excellent consistency between the different analysis groups and between the different analysis methods, as Fig.~\ref{fig:wa-all-datasets-all-methods}, Table~\ref{tab:wa-fit-results-major-parameters-norf}, and Table~\ref{tab:wa-fit-results-a-method-only} illustrate. Table~\ref{tab:wa-fit-results-major-parameters-norf} also demonstrates that fits to the asymmetry-weighted data provide the best statistical power on \oam.  Note that the positron-based data sets used by the different analysis groups are highly correlated, with variation in the statistics at most at the few-percent level, so one does not expect fluctuations between analyses at the level of the statistical uncertainties from the fits. As noted below, the evaluation of the consistency included consideration of the statistical correlations.
The reduced $\chi^2$ values in Table~\ref{tab:wa-fit-results-major-parameters-norf} and Table~\ref{tab:wa-fit-results-a-method-only} all agree with unity within statistical expectations, indicating the fit models describe the data well. The FFTs of the fit residuals presented in Fig.~\ref{fig:ffts} show no additional structure.

\paragraph{Statistical correlation and bootstrap samples}

\begin{figure}[tb]
    \centering
    \includegraphics[width=0.95\linewidth]{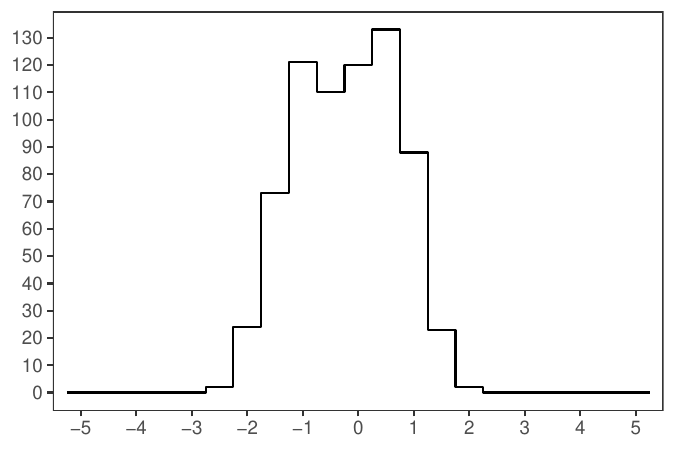}
    \caption{The pairwise difference of the \Roa values obtained from each analysis, normalized to the allowed statistical difference for that pair.  The distribution has a mean of -0.21 and standard deviation of 0.90.}
    \label{fig:normalizedDifferences}
\end{figure}

To understand the correlations between the different analysis methods, $200$ bootstrapped variations of each of the major datasets were created \cite{10.1214/aos/1176344552}. These were created by resampling, with replacement, data subsets accumulated approximately every 10 minutes. 
The statistically allowed differences between analysis groups and methods are determined by calculating the standard deviation of the difference, $\Delta R_{\omega_a}$, across the bootstrap samples for each pair. For the unit-weighted and asymmetry-weighted results from the same dataset, the quadrature difference in uncertainties also provides an excellent estimate of the allowed difference. Figure~\ref{fig:normalizedDifferences} shows the distribution of the measured differences between all fit results normalized to their estimated allowed statistical differences.  The distribution agrees well with a unit Gaussian distribution, given the correlations observed when comparing all pairwise differences.

\paragraph{Averaging of the Results}
Because of the high statistical overlap in the data samples used in the different reconstruction and analysis methods, the \oam values obtained from those methods are highly correlated.  While a combination of results based on their covariance matrix is possible in principle, in practice two issues arise concerning sensitivity to correlations among the results.  First of all, for \oam values from different methods (asymmetry or unit weighted) based on the same data, the \oam values are essentially critically correlated~\cite{Lusiani:2024md} --- their correlation coefficient is close to the ratio of their uncertainties.  As a result, the optimal combination gives zero weight to the less precise value, in this case, the unit-weighted result.  For the asymmetry-weighted results, their precise average depends on the accurate determination of the correlations, which is challenging for the systematic uncertainties.  As a result, a simple unweighted average of the asymmetry-weighted results provides the optimal method for combining them~\cite {Lusiani:2024md,Cowan:1998ji}.

\subsubsection{Closure Tests and Crosschecks} \label{sec:wa-crosschecks}

During the analysis, the consistency of the \oam result across different data divisions was checked as part of the standard analysis procedure. 
These divisions discussed below all yield statistically compatible values of \oam. 
Divisions of the data during which the value of \oam is expected to change but for which the ratio of \oam to the measured field is constant, for instance across datasets with large changes in the magnetic field, are discussed later in Section~\ref{sec:consistencyOfAmu}.

\paragraph{\oam variation vs. bunch number}

\begin{figure}[tb]
    \centering
    \includegraphics[width=0.95\linewidth]{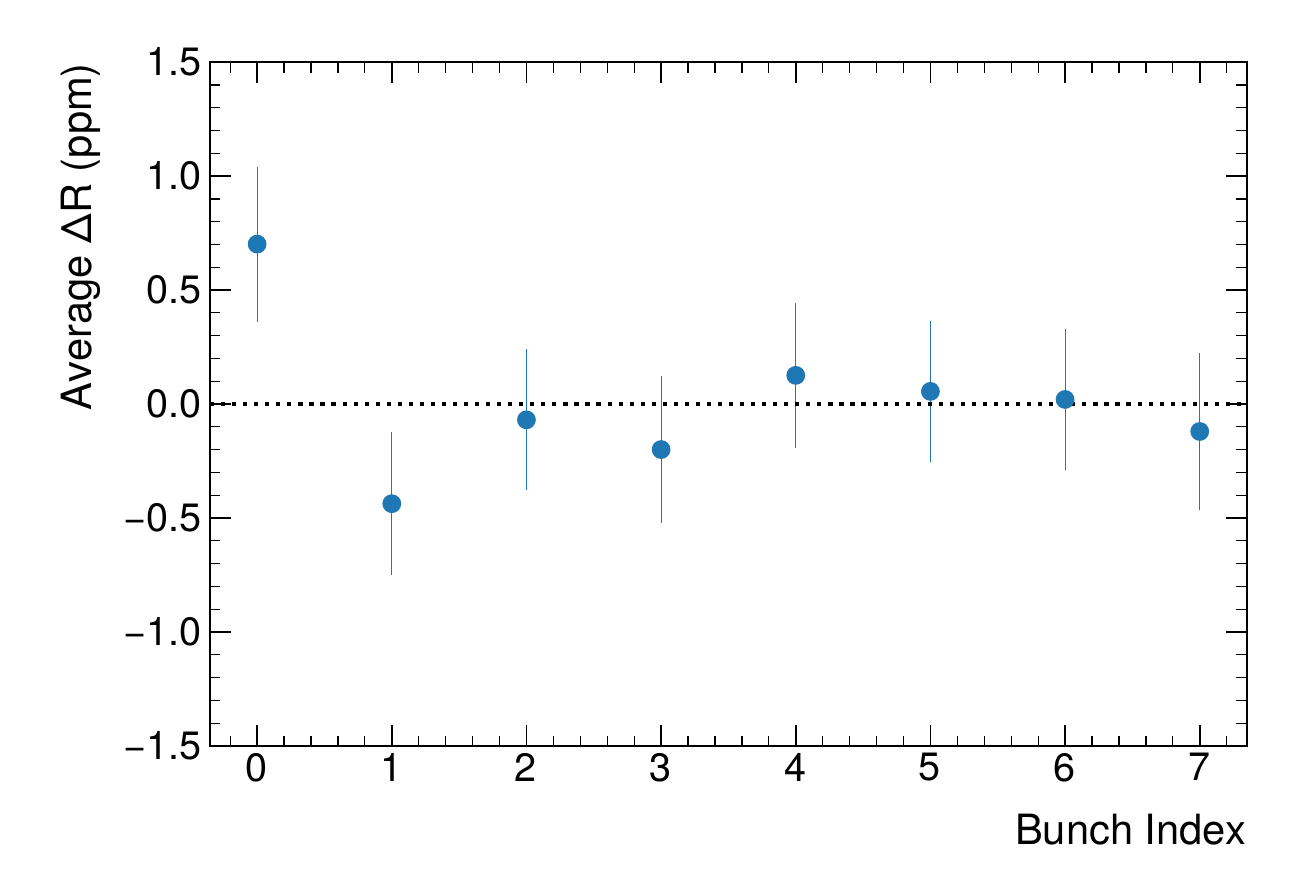}
    \caption{The difference between the \Roa values determined in each of the eight injected muon bunches and the nominal value determined using all bunches combined, averaged over the four datasets. These results are from the analysis group \wagroupTyler. The distribution has a probability of $\chi^2$ of 54\%.}
    \label{fig:wa-bunch-variation}
\end{figure}

The \ac{FNAL} accelerator complex provides two sequences of $8$ muon bunches, each with a unique pulse shape. Except for a minute difference (on average $\pm \SI{5}{mrad}$) in the starting \oa phase, this structure should have no effect on the extraction of \oam. Groups \wagroupZep, \wagroupTyler, and \wagroupEuropa performed fits to each of the injected bunches individually and compared to the summed fit result. The distributions of \oam per-bunch were all well described by a constant fit, and the average of the individual bunch fits agreed with the summed fit to within \SI{10}{ppb}. An example of this analysis is shown in Fig.~\ref{fig:wa-bunch-variation}.

\paragraph{\oam variation vs. calorimeter}

Each calorimeter images the muon beam at a slightly different phase in its cyclotron motion around the storage ring. While the phase of \oam does not vary significantly, many of the beam dynamics frequencies (as aliased combinations of the cyclotron frequency and betatron oscillations $\omega_x$ and $\omega_y$) exhibit a significant phase advance around the ring. This phase advance creates a partial cancellation of the frequency around the ring when data from all detectors are summed. 
Differences in acceptance across the calorimeters also couple to the beam dynamics frequencies. 
These effects together can be seen in the variation in \ac{CBO} envelope functions $\alpha_{CBO}(t)$ and $\beta_{CBO}(t)$ in Fig.~\ref{fig:slidingwindows}.
Fitting the calorimeters individually and confirming that the average value of \oam matches the calorimeter-summed value, therefore provides a cross-check of the handling of the beam dynamics frequencies in each of the fit models. It also provides a cross-check that the residual effect of the cyclotron oscillation has been removed (see Fig.~\ref{fig:Rwave}), and that the sag in the calorimeter gains from beam injection has been corrected well.
All groups performed this cross-check and found agreement at the \SI{10}{ppb} level between the average of the calorimeter results and the calorimeter-summed fit results.

\paragraph{Start Time Stability}

\begin{figure}[tb]
    \centering
    \includegraphics[width=0.98\linewidth]{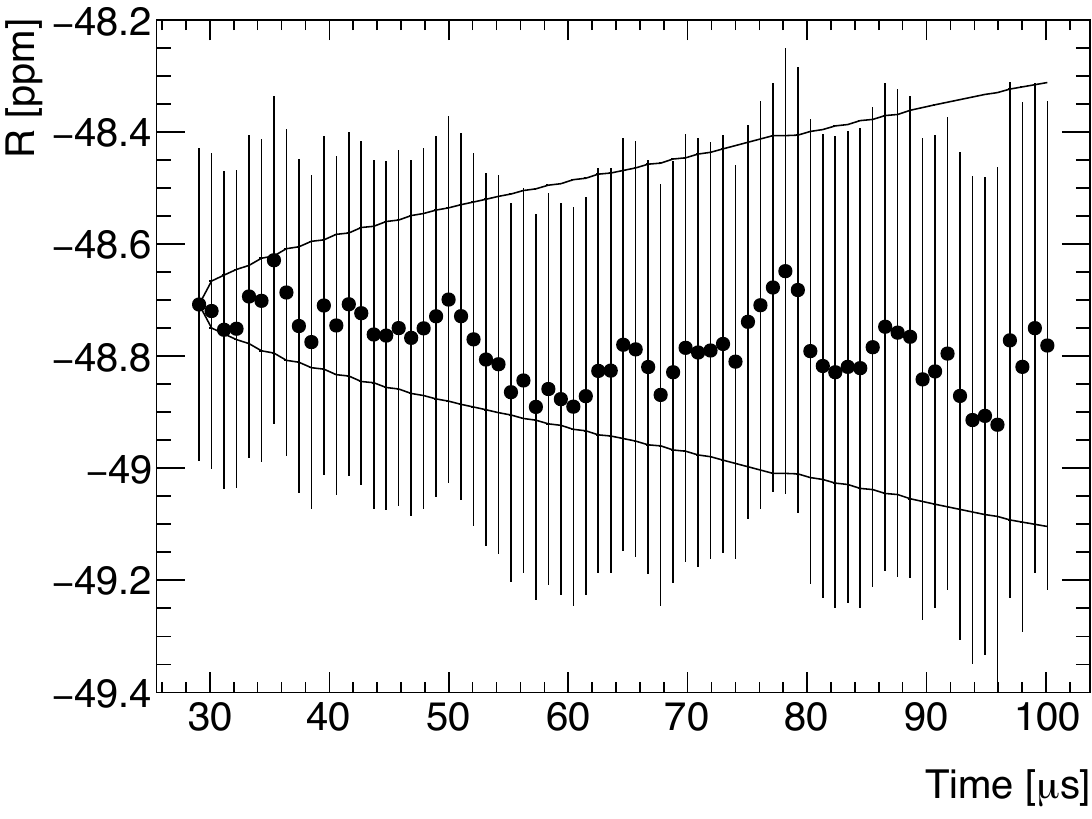} \\
    \includegraphics[width=0.98\linewidth]{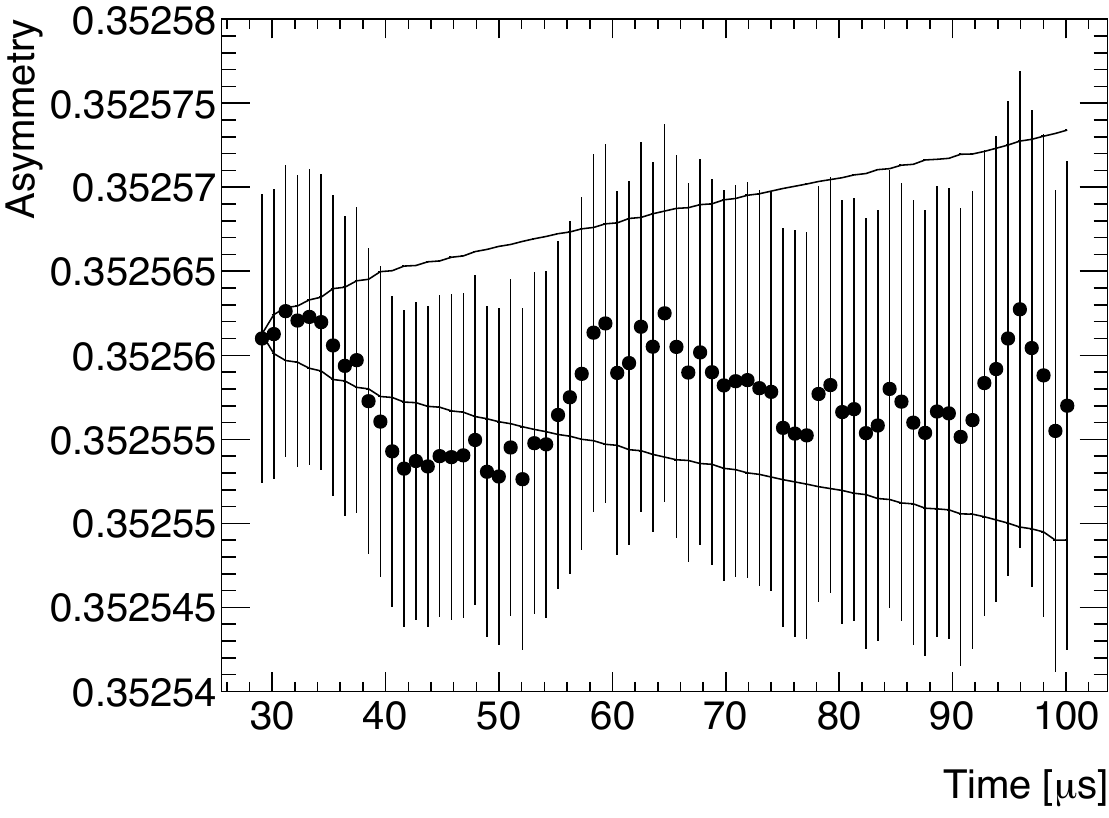} 
    \caption{Start time scans for \Roa (top) and $A_0$ (bottom) from analysis group \wagroupEuropa. The black solid lines indicate the $\pm1\sigma$ allowed variation arising from the difference in statistics between the set and the subset. Significant excursions outside this band would indicate an uncontrolled systematic effect. The error bars represent the statistical uncertainties on each individual fit.}
    \label{fig:wa-start-time-scans}
\end{figure}

If not properly accounted for, early-to-late variations within a muon fill under experimental conditions (for instance, changes in detector gains) can lead to instabilities in the fit parameters as a function of time in fill. To test for such unphysical variation, the fit start and stop times are varied, and the consistency of the result is compared to the expected variation given the difference in statistics between the full fit and the subset of the data. 
Due to an overall decrease in statistical power later in the fill, start-time scans with all parameters floating are typically restricted to no later than \SI{100}{\us}. For start-time scans extending beyond this period, higher-order terms in the fit are fixed to their nominal values.
All groups performed these cross-checks and found no statistically significant variations in the fitted parameters.
An example of such a scan can be seen in Fig.~\ref{fig:wa-start-time-scans}. 

\paragraph{Histogram Swapping}

Up to the limit of systematic uncertainties, the final fitted value of \oam should be independent of the fit method. 
After the fit methods of the six analysis groups reached a satisfactory level of sophistication (all major datasets were processed, methods for evaluating all major systematics were determined, etc.), the groups switched to a common software blinding offset (see the introduction to Sec.~\ref{sec:oam}). This ``relative unblinding'' allowed cross-comparisons between the analysis groups, while the software and hardware-based absolute clock blinding remained in place. These cross-comparisons were performed in part by swapping one or more representative histograms across groups and performing the same fit on the new input. The swaps ensured that the cross-checks occurred across groups using different reconstructions.
After an initial comparison with their standard fit functions, the groups would morph their analysis procedures from one group's to the other's to understand any differences.  
During this process, an error in the treatment of the \ac{CBO} phase-advance was discovered in one group: a fixed reference frequency used when evaluating $\phi_{\text{CBO}}(t)$ was left floating in the main analysis. Upon correction, the result used in the final average from that group changed by only a few ppb.

\subsection{Systematic Uncertainties} \label{sec:wa-systematic-uncertainties-subsection}

\begin{table}
    \centering
    \caption{The systematic uncertainties on the \oam measurements for each dataset, averaged over the seven results that were input to the final reported \oam averages.  The statistical uncertainty is provided for reference.}
    \begin{tabular}{lcccc} \toprule
Category                     & NoRF  & xRF   & xyRF5 & xyRF6  \\ 
                             & (ppm) & (ppm) & (ppm) & (ppm)  \\ \midrule
Statistical                  & 0.182 & 0.241 & 0.246 & 0.277  \\
CBO                          & 0.022 & 0.019 & 0.020 & 0.024  \\
Residual slow gain variation & 0.022 & 0.020 & 0.016 & 0.011  \\
Gain corrections             & 0.004 & 0.005 & 0.006 & 0.007  \\
Pileup                       & 0.006 & 0.006 & 0.006 & 0.005  \\
Muon loss                    & 0.002 & 0.003 & 0.003 & 0.003  \\
Ratio method inputs          & 0.001 & 0.001 & 0.000 & 0.000  \\
Randomization                & 0.001 & 0.002 & 0.002 & 0.002  \\ \midrule
Total systematic             & 0.032 & 0.029 & 0.027 & 0.028 \\ \bottomrule
    \end{tabular}
    \label{tab:wa_syst}
\end{table}

Table~\ref{tab:wa_syst} summarizes the systematic uncertainties associated with each dataset and the final result.  The uncertainties for a systematic effect that is common to all analysis groups (and across datasets) are treated as fully correlated.  The uncertainties associated with the \ac{CBO}, its decoherence envelopes, and the correction for the slow gain variation dominate. A brief description of the evaluation of each uncertainty follows.

\paragraph{Beam motion decoherence envelope}
The evaluation of the systematics arising from modeling the effects of beam motion on the data depends on the methods used to describe the envelopes.

{\it Parameterization}  Groups who used a functional parameterization of the beam envelopes replaced their nominal functional form with a variety of other forms that could reasonably capture the decreasing amplitude of a given beam motion frequency or phase shift. For the exponential lifetimes associated with the amplitude envelope for higher harmonics of a fundamental beam frequency, these groups also compared floating those lifetimes in the fit versus fixing them relative to the lifetime for the associated fundamental.  Acceptable models must satisfy the Akaike Information Criterion (AIC)~\cite{Akaike:1100705} (in the Gaussian limit) $\Delta{\rm AIC} = \Delta\chi^2 + 2\Delta k < 7$, where $\Delta\chi^2$ is the change in the \oam fit $\chi^2$ relative to the nominal fit model, and $\Delta k$ is the change in the number of floating parameters.  This criterion corresponds to including the AIC categories of ``substantial support'' and ``considerably less support''. 

\begin{figure*}
    \centering
    \includegraphics[width=1\linewidth]{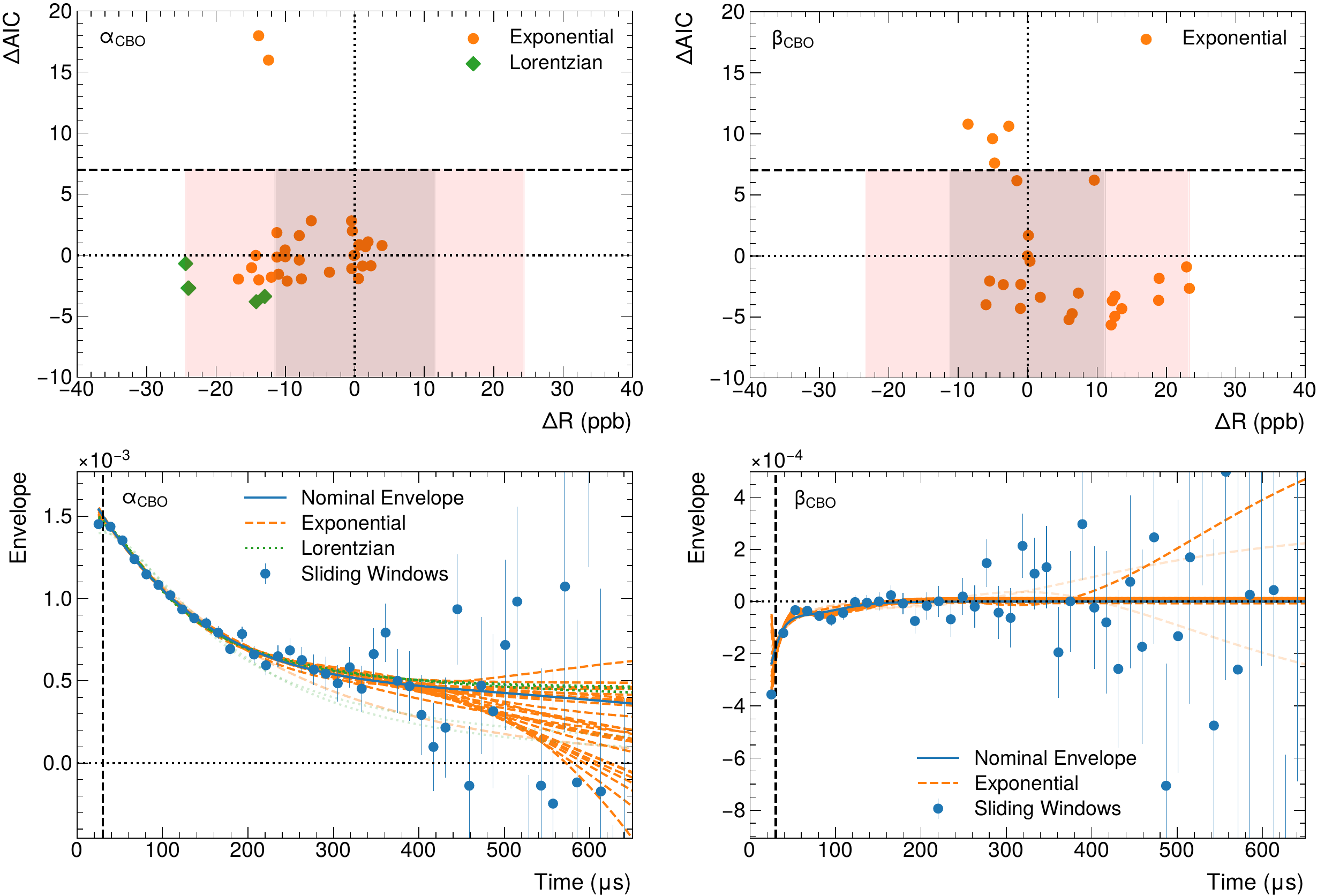}
    \caption{Top: The Akaike Information Criterion (AIC) versus $\Delta R$ for the $\alpha(t)$ envelope (left) and $\beta(t)$ envelope (right) for the horizontal betatron oscillation in the \wagroupTyler fit model. The horizontal lines indicate the AIC acceptance cutoff --- larger values are discarded.  The outer pink bands show the systematic uncertainty contributions to \oam for those envelopes, while the inner gray bands show the root mean square width of the points. Bottom: The best fit curve for each model variation used in the top plot.  Note that some curves correspond to points with large AIC values off the plot. These points represent the values obtained from a corresponding sliding window study (see text) of those envelopes.}
    \label{fig:aic}
\end{figure*}

Figure~\ref{fig:aic} illustrates this process for the $\alpha(t)$ and $\beta(t)$ envelopes for the dominant horizontal betatron oscillation term in the \wagroupTyler fit model.  

{\it Data-driven non-parametric envelopes} The data-driven methods determine the beam-related oscillation amplitudes of Eq.~\eqref{eqn:alpha_beta} from fits to many small, contiguous time windows, which are then interpolated.  Assessment of the uncertainties for these methods uses $100$ trials, in which each individual amplitude is randomly shifted by its fit uncertainty to produce a new set of envelopes. The width of the distribution of \Roa from fits using these fluctuated envelopes determines the systematic uncertainty.  Additionally, the size of the time windows was varied from the minimal size that provided sensitivity to the beam oscillation to the largest size at which the window fits were not significantly affected by decoherence itself.

Studies showed that the systematic uncertainty for a given beam frequency has little correlation with the uncertainties at other frequencies, and similarly, that the ``sin'' and ``cosine'' envelopes have little correlation. Their contributions to the final uncertainty were added in quadrature.

\begin{figure}
     \centering
     \includegraphics[width=0.95\linewidth]{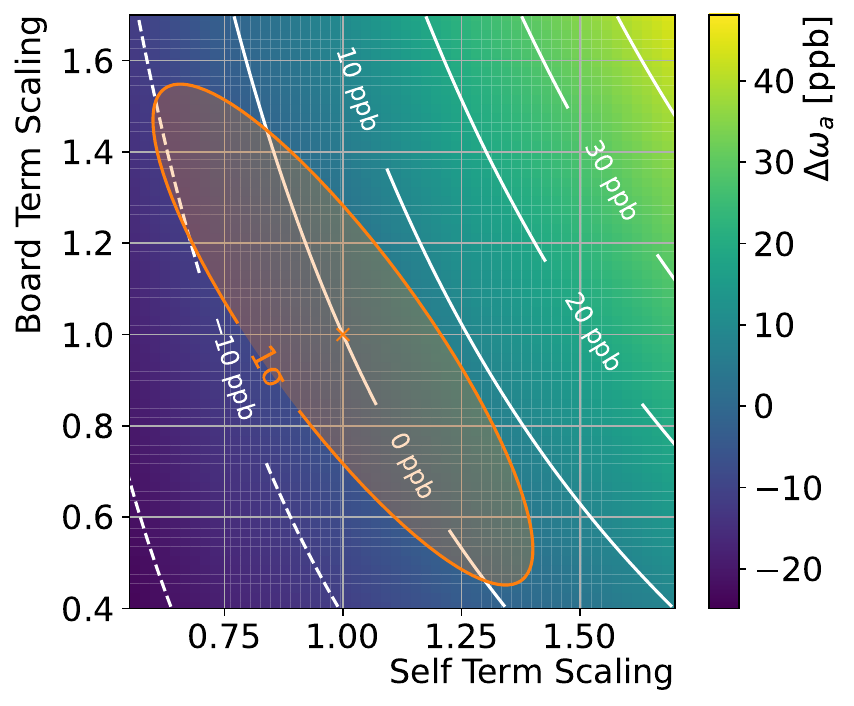}
     \caption{The variation in \oam as the scale factors for the amplitude of the gain sag related to activity within a given crystal and within crystals that share the same power distribution board with that crystal.  The ellipse represents the one standard deviation uncertainty ellipse on those amplitudes,  which determines the systematic uncertainty contribution to the \oam measurement.}
     \label{fig:sagEllipse}
 \end{figure} 
\paragraph{Residual slow gain variation}
Two contributions dominate the uncertainties in the post-reconstruction correction for the residual slow gain variation.  The largest arises from uncertainties in the amplitude of the correction, which has a direct contribution from the uncertainties in the amplitude itself in the standalone laser studies. That amplitude has two contributions: sag arising from activity within a crystal channel itself (self term), and sag arising from activity in crystals that share the power distribution board with that crystal (board term).  Figure~\ref{fig:sagEllipse} shows the variation in \oam for a scan over the scale factors for those two amplitudes. The one-standard-deviation error ellipse for those two parameters then provides the uncertainty scale for \oam.
A second contribution arises from uncertainties in the energy scale, which feeds back into the amplitude determination.  The linear sum of these effects provides the amplitude uncertainty.  The second arises from evidence for a residual slow effect in both the global and local reconstructions, though with somewhat different characteristics in the two cases. Variation of the correction parameters to cover the remaining effect, with a phase shift relative to the \oa phase that conservatively maximizes the effect.  This contribution, about 1/2 the size of the first, gets combined in quadrature.  A handful of other effects, found to be much smaller, have also been included, but don't contribute significantly.

\paragraph{Calorimeter gain corrections}
The systematic uncertainty arising from the two gain corrections applied during reconstruction stems from the uncertainties in the in-fill and standalone laser data measurements. A systematic sweep over scale factors applied to the corrections, followed by fitting for \oa with the modified data, determines the sensitivity of \oa to these corrections. As in previous publications~\cite{Run1PRDomegaa,PhysRevD.110.032009}, the product of the measurement uncertainty with the observed sensitivity provides the final systematic uncertainty.  The contributions from the two corrections are combined in quadrature.

\paragraph{Pileup}
The pileup estimation algorithm requires above-threshold activity in a calorimeter one cyclotron period after observed activity in that calorimeter, so that the DAQ system has saved windows that can be overlaid.  However, a pileup event with one of the showers below threshold may still affect the precession data, and the empirical correction misses this contribution.  A prescaled subset of the data always recorded a snapshot of activity in a calorimeter one cyclotron period following observed activity.  Analysis of that subset provided a correction to the standard algorithm that shifted \oam by \SI{3}{ppb}, the largest contribution to the systematic.  The uncertainty scale from this analysis confirmed the simulation-based estimates used in the earlier publications~\cite{Run1PRDomegaa,PhysRevD.110.032009}. Previous estimates~\cite{Run1PRDomegaa,PhysRevD.110.032009} of the uncertainties in the time and energy associated with empirical pileup candidates, as well as in a small effect from the neglect of the spin precession itself, both at the \SI{1}{ppb} level, carry over conservatively to this lower-intensity dataset.  Finally, the overall scale uncertainty contributes at the sub-ppb level. These contributions have been combined linearly.

\paragraph{Muon loss}
Muon loss from the stored beam occurred at very low rates in \RunFourFiveSix because of the improvements in the position and preparation of the stored beam.  As a slow effect, the muon-loss contribution is difficult for the fit to disentangle from other slow processes, such as muon decay and the residual slow gain variation.  The assigned systematic receives a contribution from the difference in \oam between floating and zeroing the loss rate scale factor.
A second contribution arises from systematic differences in the shapes of the measured muon loss rate versus time, depending on the selection criteria used by different analysis groups to identify lost muons.

\paragraph{Ratio method inputs}
The ratio method requires an initial estimate of the muon anomalous precession frequency for the initial sorting and time-shifting of the positron data, as well as the muon lifetime for the small rate correction in the shifts (see Ref.~\cite{Run1PRDomegaa,PhysRevD.110.032009}).  The analyses vary that period by $\pm 50$~ppm, which conservatively covers the variation in the magnetic field, and the resulting uncertainties are under \SI{2}{ppb}.

\paragraph{Randomization}
Each analysis performed the time randomization of their data to suppress the muon cyclotron frequency in their fit data at least $100$ times (more for the ratio method analyses) with different random seeds each time.  The results reported for each group are the average of the fits obtained using each seed.  The randomization systematic uncertainty corresponds to the residual uncertainty on that average.

\section{Beam Dynamics Corrections} \label{sec:bd}
The deviations of the stored beam from the ideal magic momentum orbit and the presence of vertical betatron oscillations require two beam dynamics corrections to the measured precession frequency: the electric field correction, $C_e$, and the pitch correction, $C_p$, introduced in Eq.~\eqref{eq:R}. The rigorous derivations of these effects from the Thomas-BMT equation and their evaluation methodologies are detailed in~\cite{PRAB_On_Pitch,Run1PRAB,PhysRevD.110.032009}.

\subsection{Electric field correction: $C_{e}$}
\label{sec:efield}

The electric field correction $C_e$ accounts for the spin precession induced by the momentum spread of the stored muon beam. Although the radial electric field effect vanishes at the magic momentum $p_0=\SI{3.094}{GeV/c}$, deviations from this nominal value within the momentum acceptance necessitate an additional correction to the measured $\omega_a^m$. Following Ref.~\cite{PhysRevD.110.032009}, the electric field correction is
\begin{equation} \label{eq:Ce}
    C_e = \frac{2n \beta_0^2}{1-n} \langle\delta^2\rangle = 2n(1-n) \beta_0^2 \frac{\langle x_e^2 \rangle}{R_0^2},
\end{equation}
with $\langle x_e^2 \rangle = \langle x_e \rangle^2 + \sigma_{x_e}^2$,
where $\delta = (p - p_0) / p_0$ is the relative momentum, $\beta_0$ is the magic momentum velocity, $n \approx 0.108$ is the effective field index, $R_0 = \SI{7.112}{m}$ is the magic momentum orbit radius, and $x_e$ is the horizontal equilibrium position of the detected positrons with respect to the central orbit.
Muons with $p\neq p_0$ orbit at equilibrium radii $x_e \neq 0$
 experience a net electric field from the focusing quadrupoles that, in the muon rest frame, transforms to an additional magnetic field altering their spin precession; the correction scales with the variance of this radial distribution.

The more optimized kicker strength and additional RF system in \RunFourFiveSix result in a more centered radial beam distribution and momentum spectrum closer to magic momentum, reducing $C_e$ by approximately \SI{30}{\percent} relative to \RunTwoThree.

Three complementary methods for extracting the momentum distribution required for the electric field correction are described below: the Fast-Rotation analysis, the tracker-based positron-tracking analysis, and direct beam-measurement cross-checks. 

\subsubsection{Fast-Rotation analysis}
\label{sec:fast_rotation}

The momentum distribution of the stored beam is accessible through the cyclotron frequency spectrum. For a muon orbiting at radius $r = R_0 + x_e$ with tangential velocity $v$, the cyclotron frequency is
\begin{equation} \label{eq:fastRotationBasic}
\omega_c = \frac{v}{R_0 + x_e}.
\end{equation}
Since the stored muons are highly relativistic, their tangential velocity $v \approx \beta_0 c$ remains nearly constant (stable at the ppm level) even as their momenta deviate from $p_0$. Therefore, variations in the measured cyclotron frequency $\omega_c$ directly reflect variations in the equilibrium radius $x_e$, providing a precise probe of the radial (and hence momentum) distribution needed to determine $\langle x_e^2 \rangle$.

\begin{figure}[htbp!]
\centering
\includegraphics[width=0.5\textwidth]{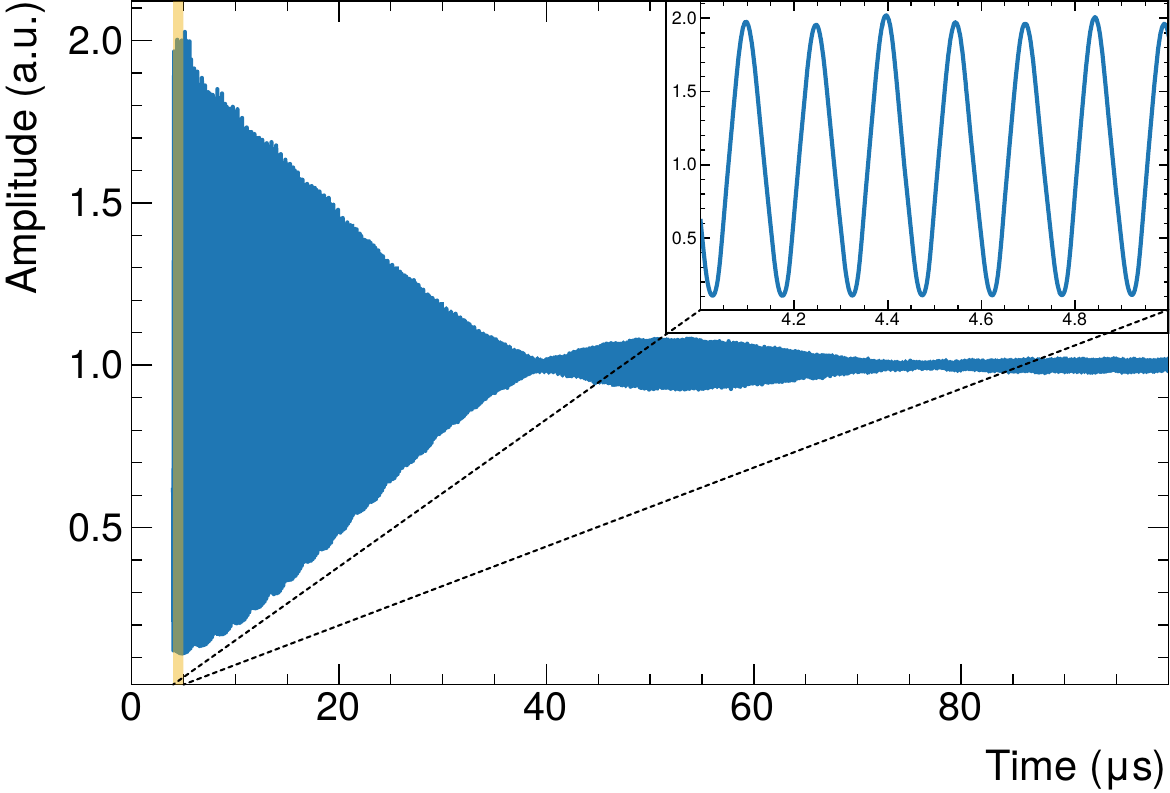} 
\caption{Fast-Rotation signal from \RunFournoRF dataset showing individual turns around the storage ring over short timescales (insert) and decoherence envelope over long time scales.}
\label{fig:fastRotation}
\end{figure}

The decay positron spectrum measured by the calorimeters is modulated by the cyclotron frequency and other beam dynamics frequencies. This modulation, referred to as the Fast-Rotation signal, is used to reconstruct the momentum distribution of the stored muon beam, which is crucial for the determination of $C_e$. When injected into the storage ring, muons are tightly bunched. Over time, they spread longitudinally due to their momentum distribution, leading to the decoherence of the Fast-Rotation signal, which fades away after approximately \SI{40}{\micro\second}, as illustrated in Fig.~\ref{fig:fastRotation}.

A filtering procedure, the {\it smearing method}, extracts the Fast-Rotation component from the decay positron time spectrum.

The procedure has two steps:
In the first step, the time of individual positron hits is shifted by a multiple, the calorimeter number, of the cyclotron period divided by $24$ (the number of detectors), effectively aligning all hits as if they originated from a single detector.
The numerator histogram is filled at the shifted hit time, while the denominator histogram is filled at the hit time shifted by a random offset drawn from a triangular distribution over $[-T_c/2, T_c/2]$, where $T_c$ is the cyclotron period.

This time randomization suppresses the Fast-Rotation signal in the denominator, while preserving slower beam dynamics effects. Taking the ratio of the two histograms effectively cancels the slow features, isolating the Fast-Rotation component in the numerator. 

This preparation marks a significant improvement over the \RunTwoThree analysis. In particular, the previously used uniformly time-smeared method did not fully suppress the $f_c \pm f_a$ sidebands in the denominator, corresponding to the beat frequency between the cyclotron and anomalous precession frequencies, thereby degrading the stability of subsequent fits. A better suppression of these sidebands was achieved by convolving two box kernels of width $T_c = 1/f_c$, yielding a triangular kernel of support $2T_c$ that reduces the sideband amplitude from 3.5\% to $\sim$0.1\% of the input signal.

The Fast-Rotation signal $S(t)$ can be modeled as a weighted sum of periodic intensity modulations with frequencies $\omega$ and time offsets $\tau$, representing the periodic detection of circulating muon bunches
\begin{equation} \label{eq:fast_rot_signal}
    S(t) = \int_{-\infty}^{\infty}\int_{-\infty}^{\infty}\sum_m\delta\bigg[t - \bigg(\frac{2\pi m}{\omega}+\tau\bigg)\bigg]\rho(\omega,\tau)d\omega d\tau,
\end{equation}
where $m$ is the turn index around the ring, and $\rho(\omega,\tau)$ is the joint distribution of revolution frequencies and injection times for stored muons. 
This integral represents the superposition of contributions from all stored muons, each characterized by its revolution frequency $\omega$ and injection time $\tau$; their collective interference pattern creates the observed modulation in the calorimeter signal.

Standard approaches based on Fourier analysis~\cite{ORLOV2002767} aim to extract the frequency distribution by assuming that $\rho(\omega,\tau)$ is separable into independent frequency and time offset components, $\rho(\omega,\tau) = \rho_1(\omega) \cdot \rho_2(\tau)$.
However, this assumption is not strictly true, since the kicker's magnetic field is not uniform over the injection time, and the stored beam's momentum distribution depends on the impulse given to each muon. The resulting {\it time-momentum correlation} will distort the momentum distribution obtained by Fourier analysis.

To address this effect, a new chi-squared method, inspired by techniques developed for the CERN muon storage ring experiments, was introduced in the \RunTwoThree analysis~\cite{PhysRevD.110.032009}. In this method, the signal in Eq.~(\ref{eq:fast_rot_signal}) is discretized over narrow bins in $\omega$ and $\tau$, with the weights denoted as $\beta_{ijk}$, where $i$ and $k$ index the $\omega$ and $\tau$ bins, and $j$ labels the time bins of the signal $S_j$. The total signal is expressed as
\begin{equation}
S_j = \sum_{i,k} \beta_{ijk} \rho_{ik},
\end{equation}
where $\rho_{ik}$ represents the discretized joint distribution $\rho(\omega,\tau)$, with  $25$ bins in each of $\omega$ and $\tau$, and approximately $10^5$ time bins for the Fast-Rotation signal. Rather than treating all $25\times25=625$ entries of $\rho_{ik}$ as independent free parameters, the distribution is assumed to be separable into a reference frequency distribution $\rho_i$ at a central time slice and an injection time profile $\rho_k$, each with $25$ free bin contents determined through an iterative linear chi-squared minimization.
To account for the time-momentum correlation, the shape of the frequency distribution is assumed to be consistent across time slices $\tau_k$, with only its mean frequency, width, skewness, and intensity varying between slices. Each of these three characteristics is modeled as a fourth-order polynomial in $\tau_k$ with no constant term, introducing $3\times4=12$ additional correlation parameters. These parameters are optimized through a non-linear minimization that wraps around the iterative linear solution for $\rho_i$ and $\rho_k$.

As mentioned above, the improved randomization significantly reduced the $f_c \pm f_a$ sidebands in the Fast-Rotation signal; however, they were not fully eliminated. This residual structure originates from the intrinsic properties of the muon distribution. In particular, the time offset $\tau$ is widely spread, introducing variations in the $\omega_a$ phase across different injection times. For each $\tau_k$, the Fast-Rotation signal contributes with a different phase in the $\omega_a$ oscillation.

To account for this phase variation across injection times, the propagator functions $\beta_{ijk}$ must be modified as follows
\begin{equation}
\beta_{ijk} \rightarrow \beta_{ijk}\frac{1 + A \cos[\omega_a(t_j + \tau_k) + \phi]}{1 + A \cos(\omega_a t_j + \phi)},
\end{equation}
where $A$ and $\phi$ are the asymmetry and phase parameters, respectively, obtained from the standard 5-parameter fit to the positron time spectrum.\\

Each signal constructed through this procedure was fitted with a set of initial parameters. Although the chi-squared values of the fits are remarkably good, they tend to converge to local minima. This occasionally results in variations of several ppb in the correction. To mitigate this issue, the fits are iterated multiple times using different initial parameter values.
This effect was more pronounced in \RunTwoThree, but has been reduced thanks to the implementation of new procedures described above. 

Due to systematic shape variations in the beam pulses, the fits were performed separately for the time spectra of each individual bunch and for the combined (summed) spectrum as a consistency check.

\begin{figure}[htbp!]
\centering
\includegraphics[width=0.5\textwidth]{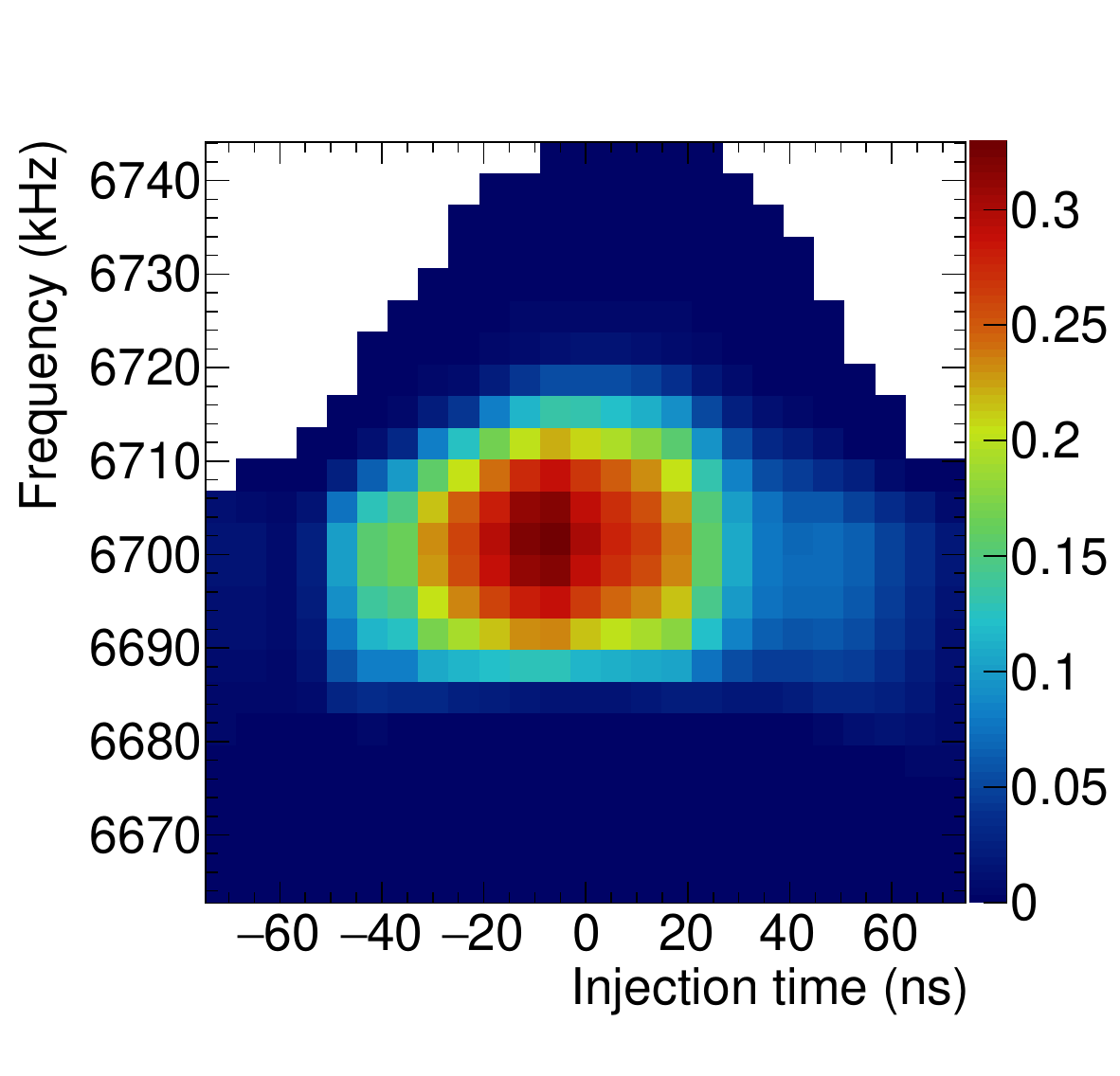} 
\caption{Joint distribution from the Fast-Rotation $\chi^2$ method of revolution frequency and injection time determined for the data subset 6A for the first bunch in the beam pulse sequence. The color represents intensity in arbitrary units.}
\label{fig:pt_distribution}
\end{figure}

Figure~\ref{fig:pt_distribution} shows a joint distribution obtained for data subset 6A.

To assess the sensitivity of the result to the fit procedure, the set of best-fit parameters was varied within the parameter space around the minimum.
To quantify the impact of each parameter on the reduced $\chi^2$, we varied each parameter above and below its optimal value. These variations were then used to guide the sampling of the parameter space around the minimum. From this procedure, 2500 random sets of 12 parameters were extracted from the allowed variation determined in the parameter scan. Each new set of parameters was used to fit the data and to extract the electric field correction. To obtain a conservative estimate of the total error, only fits with a reduced $\chi^2$ within 2.5~$\sigma$ of the minimum were retained.

In addition, to account for possible local minima, a wider scan of the parameter space was executed. For each bunch, a new series of fits was performed, using as initial conditions the best-fit parameters obtained from the other bunches of the same dataset. The corresponding error was then estimated as the maximum deviation between the new fits and the initial best fit. The final error on the correction was taken as the larger of the results from the two procedures described above and is reported in Table \ref{tab:ce_all_combined}.

Once the joint distribution $\rho(\omega,\tau)$ is obtained from the $\chi^2$ fit, the momentum distribution is extracted by marginalizing over the injection time $\tau$. The cyclotron frequency distribution is then converted to the radial distribution following Eq.~\eqref{eq:fastRotationBasic}. From this distribution, the variance is computed and used in Eq.~\eqref{eq:Ce} to determine the electric field correction $C_e$. This procedure accounts for the time-momentum correlation through the fitted $\tau$-dependent parameters, providing a more accurate determination than methods assuming separability.

Finally, the following systematic uncertainties associated with the Fast-Rotation analysis were investigated: the choice of the signal start time, incomplete suppression of residual frequencies, and modifications to the muon distribution due to scraping. These effects have a negligible, sub-ppb impact on the final correction.

Using the correlated distribution $\rho(\omega,\tau)$ obtained from the $\chi^2$ method, the distorted momentum distribution obtained from the Fourier method can then be corrected using an analytic description of the distortion. With this correction, the Fourier method is not fully independent of the $\chi^2$ method, but it provides a cross-check that shows good consistency within \SI{10}{ppb}, an improvement over \RunTwoThree. This demonstrates the robustness of the improvements in the $\chi^2$ analysis in \RunFourFiveSix.

\subsubsection{Positron tracking analysis}

An independent determination of $C_e$ is obtained through tracking of decay positrons.
The straw tracker system measures the spatial distribution of the stored muon beam by reconstructing decay positron trajectories. Since the equilibrium radius of the muon directly relates to its momentum through the dispersion function $D$ ($x_e=D \delta$), the radial position distribution measured by the trackers encodes the underlying momentum distribution.
However, muons undergo \acf{CBO} around their equilibrium orbits, which must be disentangled from the dispersion-driven offsets to accurately extract $\langle\delta^2\rangle$.
From the momentum-dependent equilibrium radii and the reconstructed radial betatron amplitude distribution, the muon beam's momentum variance is determined, yielding the electric field correction $C_e$ using Eq.~\eqref{eq:Ce}.

\begin{figure}[ht!]
\centering
\includegraphics[width=\linewidth]{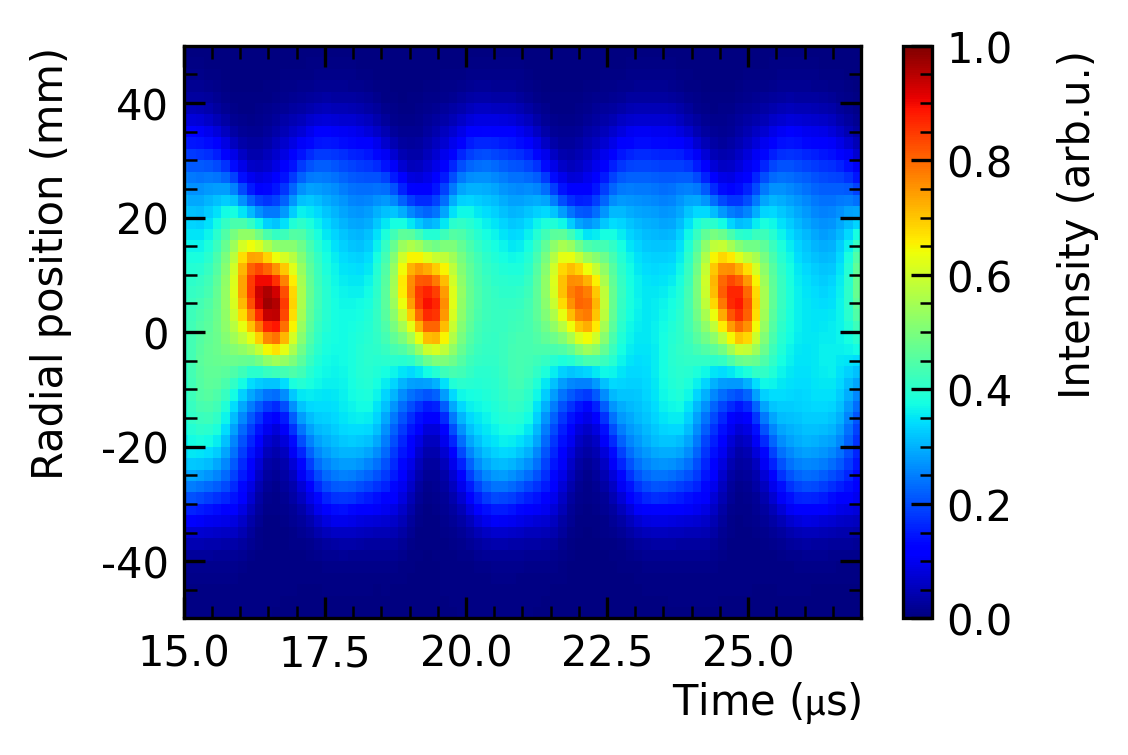}
\caption{Example of  muon beam intensity as a function of radial position, $x$, and time, $\tau$, coordinates. The tracker-based method takes the $x-\tau$ beam distribution as input.}
\label{fig:CEinputs}
\end{figure}

The radial motion of a stored muon over time, measured relative to the nominal radius $R_0$ of the storage ring, is modeled as
\begin{equation}
    x=D\delta+A\cos{\left( \omega_{\mathrm{CBO}}t+\phi \right)},
\label{eq:x_motion}
\end{equation}
where $D$ is the radial dispersion function, $\delta$ the relative momentum defined in Eq.~\eqref{eq:Ce}, $A$ the radial betatron amplitude, and $\omega_{\mathrm{CBO}}$ the CBO frequency~\cite{PhysRevD.110.032009}. The first term describes the equilibrium radial offset due to the muon's momentum deviation, while the second term describes the transverse betatron oscillation around that equilibrium orbit.
Considering two times, $t_0$ and $t_f=t_0+T_{\mathrm{CBO}}/2$, separated by half a CBO period, $T_{\mathrm{CBO}}$, the relative momentum offset's variance in terms of the radial motion is expressed as
\begin{align}
\left\langle\delta^2\right\rangle &= \frac{1}{2 D^2}\left[\left\langle x_0^2\right\rangle+\left\langle x_f^2\right\rangle - \left\langle A^2\right\rangle\right.\nonumber\\
  &\quad\left.-\left\langle A^2 \cos \left(2 \omega_{\mathrm{CBO}} t_0+2 \phi\right)\right\rangle\right].
\label{eq:dpvar_from_x}
\end{align}
All terms on the right-hand side of Eq.~\eqref{eq:dpvar_from_x} can be determined from tracker data. The $\langle x^2 \rangle$ terms are straightforwardly obtained from $x-t$ histograms (see Fig.~\ref{fig:CEinputs}) by taking the muon-ensemble average of $x^2$ over all hits within a time bin of width equal to one cyclotron period.

The radial betatron amplitude ($A$) distribution of the stored muons is not directly measured but reconstructed through non-negative least squares (NNLS) minimization. The reconstruction precision was quantified using simulations.

The NNLS minimizer provides a reconstruction of the probability density function $f_x(x_e,\phi,A)$ that best reproduces the $x-t$ histogram. In the algorithm, the dynamics of a single muon follow Eq.~\ref{eq:x_motion}. The projection of $f_x(x_e,\phi,A)$ onto the $A
$-axis yields the $A$-distribution, reconstructed from the $x
-t$ data within a time window of one \ac{CBO} period centered at $t_f = t_0 + T_{\text{CBO}}/2$.

The initial $A$-distribution used as input to the NNLS minimizer for the first iteration follows a Gaussian beam with a width of $\sigma_A=\SI{18.7}{\milli\meter}$, 
which approaches the underlying, not perfectly Gaussian distribution, as validated by the \texttt{GM2RINGSIM} simulation.
We also considered the impact of different initial amplitude distributions on the reconstructed central values to assess NNLS performance when initial conditions deviate from simulation-based expectations.

To determine $\langle \delta^2 \rangle$, we fit the quantity 
\begin{equation} \label{eq:trk:fit}
\frac{\left( \langle x_0^2 \rangle + \langle x_f^2 \rangle - \langle A^2 \rangle \right)}{2D^2}
\end{equation} from tracker data and the reconstructed $A$-distribution as a function of time using the function $f(t)=c_1+c_2\cos{\left( 2\omega_{\mathrm{CBO}}t+c_3 \right)}$ over \SI{2}{\micro\second} intervals starting at $t_0=\SI{30}{\micro\second}$, which matches the functional form of Eq.~\eqref{eq:dpvar_from_x}, rearranged to solve for Eq.~\eqref{eq:trk:fit}. The relative momentum offset distribution variances are then extracted from the time-independent fit parameter, $\langle \delta^2 \rangle\equiv c_1$.

Since this method calculates the electric field correction $C_{e,i}$ independently in each time bin $i$ of the fill, the total electric field correction is then computed as the muon-lifetime-weighted average
\begin{equation}
C_e=\frac{\sum_i C_{e,i}\exp{(-t_i/\gamma_0\tau_\mu)}}{\sum_i \exp{(-t_i/\gamma_0\tau_\mu)}}
\label{eq:CeTrack}
\end{equation}
for $t>\SI{30}{\micro\second}$, where $C_{e,i}$ follows Eq.~\eqref{eq:Ce} for each $\langle \delta^2 \rangle_i$, and $\langle \delta^2 \rangle_i$ is the momentum variance determined in each bin.

The statistical uncertainties from the positron tracking analysis are negligible, while the systematic uncertainties originate from the $A$-distribution reconstruction and detector acceptance effects. Cross-calibration with independent vertical MiniSciFi detector measurements provides an additional constraint on these systematics, as described below.

The determination of the variance of the radial amplitude distribution, $\langle A^2 \rangle$, is an intrinsic uncertainty of the method used for these results. Although the NNLS minimizer converges to a solution after less than $400$ iterations (with a tolerance of $10^{-3}$), the value of the reconstructed amplitude variance depends on the initial guess of the $A$-distribution. We estimate these fluctuations to contribute a systematic uncertainty of \SI{18}{ppb}.

To constrain the systematic uncertainties in the tracker-based $C_e$ determination, we applied the method to eight dedicated systematic run datasets in which the vertical MiniSciFi was inserted to provide an independent, unbiased measurement of the beam's radial distribution. The vertical MiniSciFi-based extraction of $\langle A^2 \rangle$ yields smaller and less correlated uncertainties than tracker measurements. However, vertical MiniSciFi measurements require correction for energy loss from scintillating fibers and deuteron contamination~\footnote{Some deuterons do get stored; because they do not decay, their profile which is also observed by the MiniSciFi is increasingly important at late times in the fill.}, contributing a \SI{12}{ppb} systematic uncertainty. Additional systematics from radial resolution and alignment are present, while acceptance effects are negligible. By correcting for the difference between tracker and vertical MiniSciFi determinations, we obtain a net correction of \SI{-26\pm13}{ppb} that improves the precision of the tracker-based $C_e$, driven by the MiniSciFi's more directly measurable acceptance properties compared to the straw tracker.

This new positron tracking analysis method was validated with the \texttt{GM2RINGSIM} framework and achieves $C_e$ determination to \SI{26}{ppb} precision, competitive to the Fast-Rotation analysis. Unlike the previous tracker-based analyses in \RunTwoThree, this method eliminates dependence on radial width motion approximations and correlations between radial and momentum coordinates. Direct comparisons with contemporaneous vertical MiniSciFi measurements allow cancellation of the method's leading systematic uncertainty associated with tracker acceptance, reducing the method's uncertainty. Table~\ref{tab:ce_all_combined} provides the final results. 

\begin{figure}
    \centering
    \includegraphics[width=0.9\linewidth]{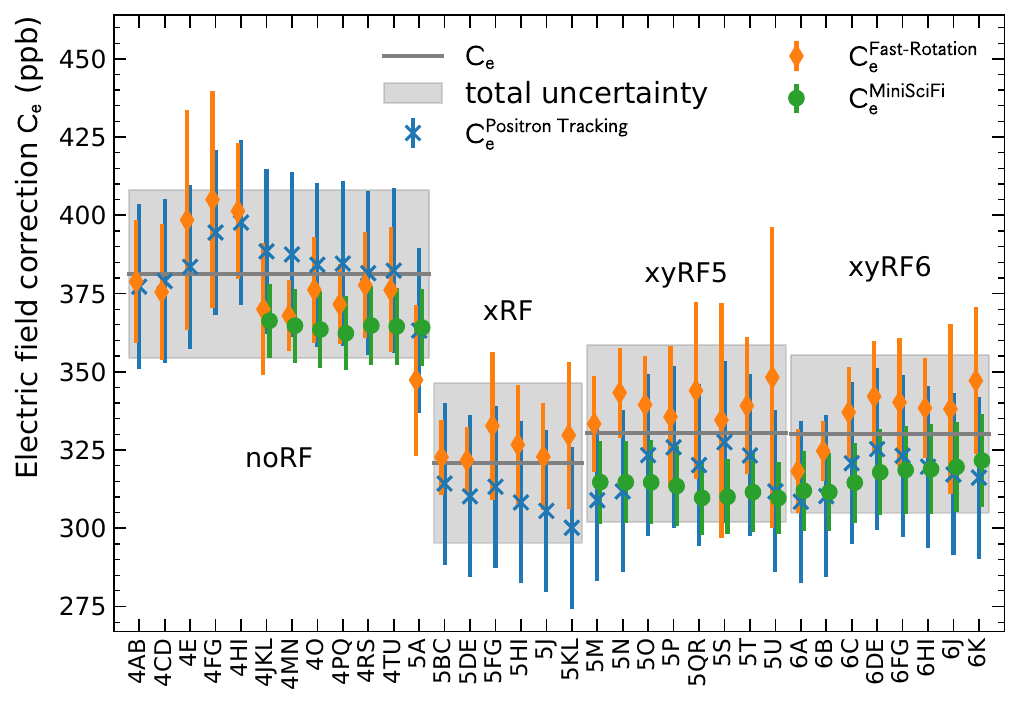}
    \caption{ Electric field correction $C_e$ determined from tracker (blue crosses), calorimeter Fast-Rotation (orange diamonds), and MiniSciFi (green circles) methods for individual lettered datasets in \RunFourFiveSix. The gray bands indicate the combined $C_e$ values with total uncertainties for each main dataset group (\RunFournoRF, \RunFiveX, \RunFiveXY, \RunSixXY). The three methods show consistent agreement within uncertainties across all datasets, with the MiniSciFi providing independent validation of the indirect positron-based measurements.}
    \label{fig:Ce}
\end{figure}

\begin{table*}[htbp]
    \centering
    \caption{Summary of electric field corrections $C_e$ for all run periods. For \RunFourFiveSix sub-datasets, both Fast-Rotation (calorimeter) and Positron tracking measurements are shown alongside the combined correction. The total correction for \RunOne and \RunTwoThree is listed as reference.}
    \label{tab:ce_all_combined}
    \begin{tabular}{lcccccc}
        \toprule
        \multirow{2}{*}{Dataset} & \multicolumn{2}{c}{Fast-Rotation} & \multicolumn{2}{c}{Positron tracking} & \multicolumn{2}{c}{Combined Result} \\
        \cmidrule(lr){2-3} \cmidrule(lr){4-5} \cmidrule(lr){6-7}
        & Corr. (ppb) & Unc. (ppb) & Corr. (ppb) & Unc. (ppb) & Corr. (ppb) & Unc. (ppb) \\
        \midrule
        \RunOne     & \textemdash & \textemdash & \textemdash & \textemdash & 489   & 53  \\
        \RunTwoThree   & \textemdash & \textemdash & \textemdash & \textemdash & 451   & 32  \\
        \midrule
        noRF      & 379.1       & 19.8        & 385.0       & 26.3        & 382.1 & 26.8 \\
        xRF       & 326.0       & 17.1        & 309.1       & 25.9        & 317.6 & 25.5 \\ 
        xyRF5     & 339.2       & 23.6        & 319.7       & 25.9        & 329.5 & 28.3 \\ 
        xyRF6     & 335.2       & 16.5        & 317.9       & 25.9        & 326.6 & 25.2 \\ 
        \bottomrule
    \end{tabular}
\end{table*}

\subsubsection{Direct beam measurement cross-check}
\label{sec:miniSciFi}

\begin{figure}
    \centering
    \includegraphics[width=0.98\columnwidth]{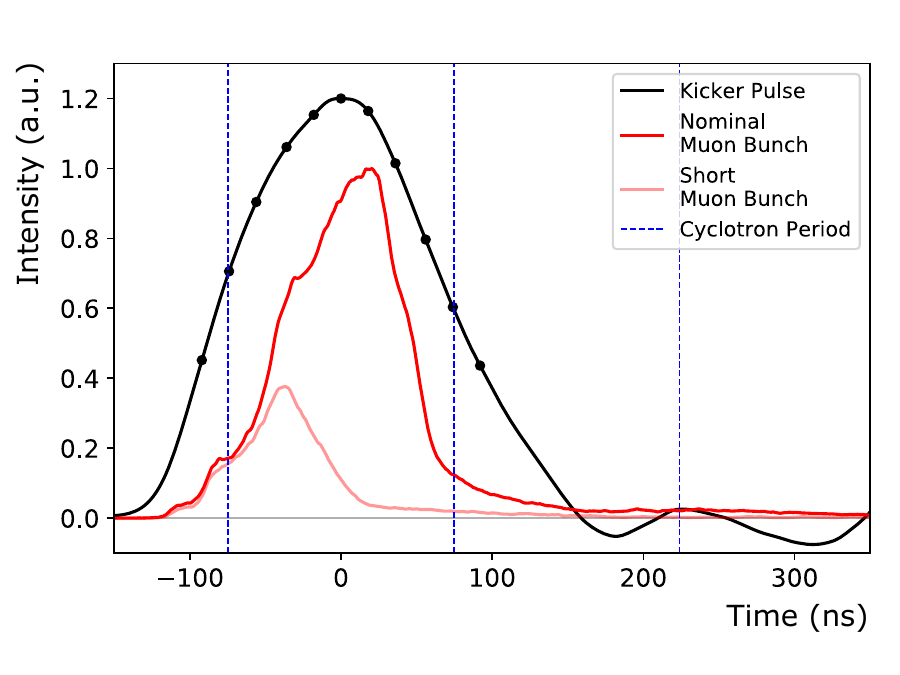}
    \caption{Time distributions of the kicker pulse (black), a nominal muon bunch (red), and a shortened muon bunch (light red) used for the $\tau-p$ correlation measurement, that were created with an upstream abort kicker. 
    MiniSciFi data was recorded at each relative delay (black circles) between the kicker and the narrow bunch. The cyclotron period (blue dashed) is shown as a reference. }
    \label{fig:miniSciFiBunches}
\end{figure}

Direct measurements of the beam with the MiniSciFi detectors provided cross checks of the $C_e$ methods, which rely on positron detections. In addition to reducing the tracker-based $C_e$ systematics with the vertical MiniSciFi, the horizontal MiniSciFi provided an independent determination of $C_e$. The novel approach to map the time-momentum ($\tau\text{--}p$) correlation in time slices entailed scanning the delay time of the kicker relative to the injected muons, and reconstructing the momentum at each step via Fourier Fast-Rotation analysis. 
A weaker effective kick preferentially stores higher-momentum muons, which sit at larger equilibrium radii and require a smaller inward deflection; this momentum dependence is directly visible across the delay scan.
Specially shortened muon bunches, created with an upstream abort kicker, enabled sharper time slices at the expense of reduced intensity, as illustrated in Fig.~\ref{fig:miniSciFiBunches}. 
The direct-beam measurement provides substantially higher statistical power per stored muon than positron-based methods, making the reduced intensity tolerable.
The $\tau\text{--}p$-dependent muon storage efficiency is recovered from the measurement by unfolding the residual temporal width of the shortened bunch, which limits the sharpness of the time slices.

For a given running period, the $\tau\text{--}p$ distribution of stored muons is determined by weighting this intrinsic, universal efficiency with the recorded muon injection time distribution.
$C_e$ is then calculated from the reconstructed momentum distribution using Eq.~\eqref{eq:Ce}. Because the muon storage efficiency is specific to the kicker, \ac{ESQ}, and \ac{RF} configuration, $C_e$ was calculated with this method only for datasets corresponding to the efficiency measurement configurations. Systematics from the unfolding procedure, time alignment between muon storage efficiency and muon injection time distribution, and $\tau\text{--}p$ correlation in the Fourier Fast-Rotation analysis result in total uncertainties of \SIrange{12}{15}{ppb}. This approach agreed with the Fast-Rotation fit and positron tracking analyses within \SI{13}{ppb}, well within their combined uncertainty, confirming the indirect positron detection methods with a direct beam measurement. 
Figure~\ref{fig:Ce} provides an overview of the individual results per lettered dataset. 

\subsubsection{Combined electric field correction}
The final electric field correction values for \RunFourFiveSix combine the calorimeter-based Fast-Rotation determination and the straw tracker-based positron tracking method.
Since the measurements from the MiniSciFi detectors covered only part of the data-taking configurations, they were used as a cross-check to assess systematics in the positron-tracking analysis rather than entering the combination directly.
The two methods were combined using equal weighting, with a conservative treatment assuming full correlation between datasets and methods. The combination incorporates both correlated systematic uncertainties and an additional uncorrelated component of \SI{12.3}{ppb} to account for observed inter-dataset variations. Independent cross-checks using the MiniSciFi detector system provided strong validation of these results, with agreement within \SI{13}{ppb} for the bunch-averaged $C_e$ values and consistent trends across datasets. The robustness of the combined approach is further supported by the improved consistency between the two methods and the corrected Fourier method compared to previous runs. The final $C_e$ values, ranging from \SI{318}{ppb} for the \RunFiveX dataset to \SI{382}{ppb} for the \RunFournoRF dataset with total uncertainties of \SIrange{26}{28}{ppb}, summarized in Table~\ref{tab:ce_all_combined}, represent an improvement in precision over \RunTwoThree while simultaneously improving confidence with the additional independent direct measurements.
The improved beam centering from optimized kicker and \ac{RF} settings reduced the $C_e$ correction amplitude from $\sim$\SI{450}{ppb} in \RunTwoThree to $\sim$\SI{325}{ppb} with \ac{RF}.
In summary, the combined $C_e$ correction achieves a precision of \SIrange{26}{28}{ppb}, validated by the consistency between independent calorimeter and tracker-based methods, and novel direct measurements.

\subsection{Pitch correction: $C_{p}$} \label{sec:pitch}
A pitch correction, $C_{p}$, compensates for the small reduction in the measured anomalous precession frequency $\omega_{a}^{m}$ that arises from vertical betatron oscillations.
As muons oscillate vertically within the storage ring, their trajectories acquire a pitch angle $\psi$ with respect to the horizontal plane. 
This pitch angle causes the muon to sample a vertical electric field, which introduces a radial component to the anomalous spin-precession vector, leading to a systematic shift in the observed precession frequency~\cite{PRAB_On_Pitch}.

The tracker measures the vertical positions of decay positrons as a function of time and azimuthal location. A positron's vertical position reflects its parent muon's vertical betatron amplitude and phase at the time of decay. By measuring the distribution of vertical positions across many decays, the underlying amplitude, $A$, distribution can be reconstructed, allowing the determination of the pitch correction
\begin{equation} \label{eq:pitch_correction}
    C_p = \frac{1}{2} \langle \psi^2 \rangle = \frac{n}{4R_0^2}\langle A^2 \rangle,
\end{equation}
where $\psi$ is the pitch angle, $n$ is the effective field index, and $R_{0}$ is the magic momentum radius. 
The pitch angle is directly related to the vertical betatron amplitude through $\psi = A k \cos(kz + \phi) / R_0$, where $k$ is the vertical betatron wave number and $z$ is the longitudinal coordinate. We use $\langle A^{2} \rangle$ rather than $\langle \psi^{2} \rangle$, because calorimeter acceptance biases the sampling of vertical positions, while the amplitude distribution remains a robust observable unaffected by position-dependent acceptance effects. The method used is described in more detail in Ref.~\cite{PhysRevD.110.032009}.

Two independent analyses, Method~1 and Method~2, determine $C_{p}$ by transforming straw-tracker decay-position spectra into amplitude space and evaluating Eq.~\eqref{eq:pitch_correction}. Both methods largely follow the methods used in \RunTwoThree. These analyses start from the same raw tracker dataset and apply similar selection criteria, yet they differ in the details of the resolution correction and the mapping from position to amplitude. Their shared workflow consists of tracker preselection, subsequent corrections for resolution and geometric acceptance, conversion of vertical-position spectra to amplitude space, a calorimeter acceptance correction, an azimuthal reweighting to account for variations in the vertical beam envelope, and an azimuthally averaged $C_{p}$ for each dataset.

\begin{figure}
    \centering
    \includegraphics[width=1\linewidth]{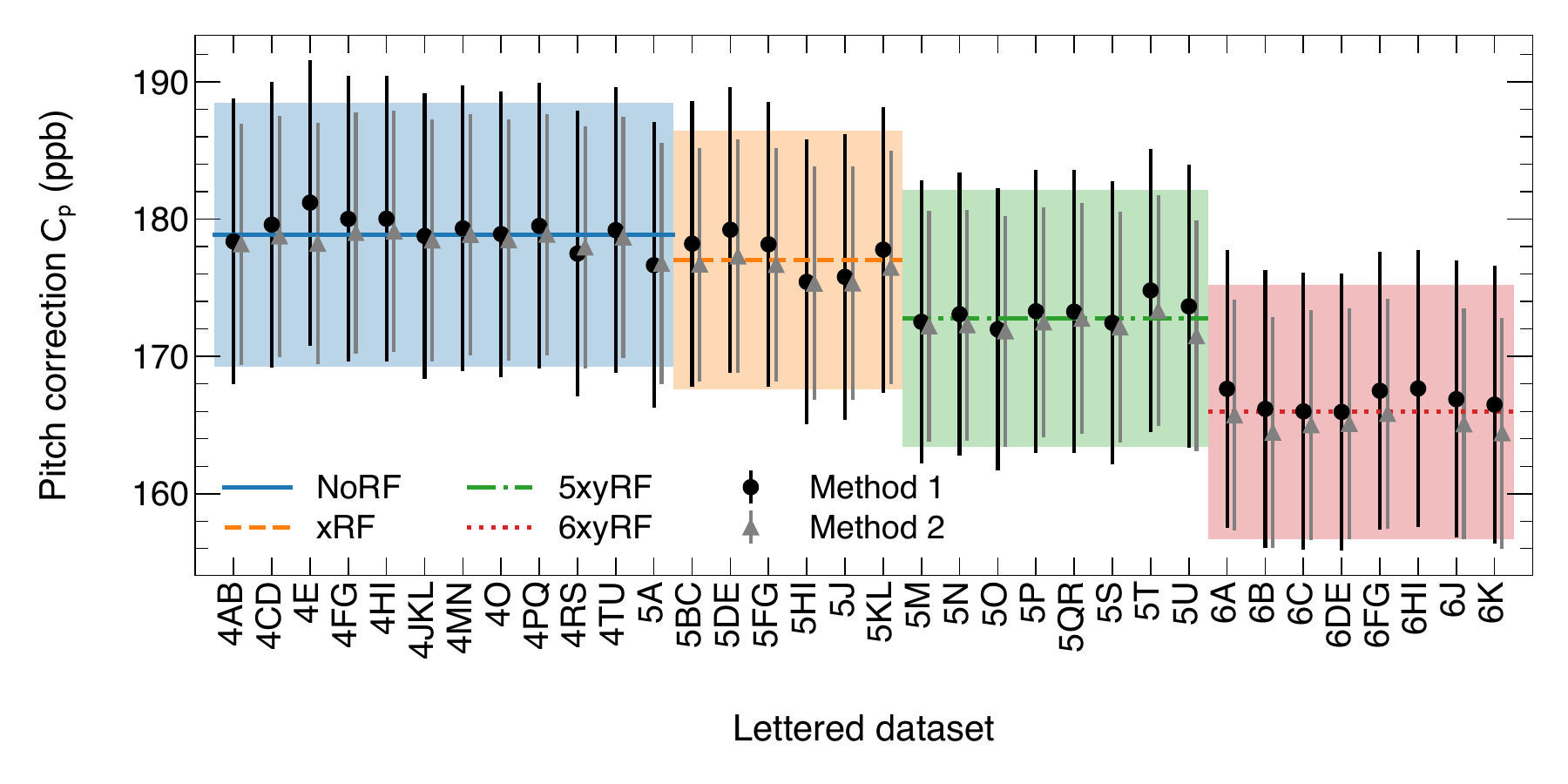}
    \caption{Pitch correction $C_{p}$ for the lettered datasets, as obtained by Method 1 and Method 2.}
    \label{fig:cp_vs_datasets}
\end{figure}

Method~1 assigns each positional bin a range of allowable amplitudes via an analytic mapping, whereas Method~2 employs a data-driven procedure that iteratively fits the measured position spectra using templates generated at discrete amplitude values. 
With respect to \RunTwoThree, Method~2 adopts a refined fitting approach with improved parameter handling, numerical stability, and convergence efficiency, and replaces the previously random sampling with deterministic generation of fit templates. 
Each method ultimately yields an average $C_{p}$ for every dataset, and the final result is the unweighted mean of the two.
The two methods agree within ${\sim}$\SI{3}{ppb}, well inside their combined statistical and systematic uncertainties. 
The dominant systematic contribution of ${\sim}$\SI{8}{ppb} arises from tracker hardware and reconstruction effects. 
Figure~\ref{fig:cp_vs_datasets} compares the two determinations for all lettered datasets, and Table~\ref{tab:summary_cp} lists the final $C_{p}$ values for \RunFourFiveSix. 
In summary, the pitch correction remains consistent with previous runs, with precision limited primarily by tracker systematics.

\begin{table}[htbp]
    \centering
    \caption{Summary of $C_{p}$ for all Run periods.}
    \label{tab:summary_cp}
    \begin{tabular}{lrr}
        \toprule
        Dataset & Correction & Uncertainty \\
        & (ppb) & (ppb) \\
        \midrule
        \RunOne & 180 & 13 \\
        \RunTwoThree & 170 & 10 \\
        \midrule
        noRF & 178.9 & 9.6 \\
        xRF & 176.8 & 9.5 \\
        xyRF5 & 172.8 & 9.4 \\
        xyRF6 & 166.0 & 9.3 \\
        \bottomrule
    \end{tabular}
\end{table}
 
\section{Time Varying Muon Ensemble} \label{sec:timeVaryingEnsemble}
Any time dependence of the measured \gm phase, $\phi_{0}$, over a muon fill introduces a bias to the measured anomalous precession frequency \oam that requires a correction of the form
\begin{equation}
    C_{\phi_0-t} = -\frac{\Delta \omega_{a}}{\omega_{a}} = \left( \frac{1}{\omega_{a}} \right) \left( \frac{d \phi_0}{dt}\right).
\end{equation}
Note that $\phi_0$ is defined as used in Eq.~\eqref{eq:omega_fiveparameterfit} with a negative sign. The measured phase depends on the ensemble's transverse decay position $\mathbf{t}=(x,y)$ and momentum $p$, allowing the source of $\frac{d\phi_{0}}{dt}$ to be split into distinct contributions. The potential dependence of $\phi_{0}$ on $\tau$ is averaged in the analysis by binning with $\sim T_{c}$.
The first contribution arises from acceptance effects: the observed phase of an ensemble depends on the transverse decay position and couples with beam motion $\mathbf{t}(t)$.
A second contribution arises from phase-momentum correlations combined with time-dependent changes in the ensemble's momentum spectrum. These momentum changes originate from lost muons ($ml$) and differential decay ($dd$).
The total time derivative can be expressed as
\begin{equation} \label{eq:dpi-dt}
\frac{d\phi_{0}}{dt} = \frac{\partial\phi_{0}}{\partial\mathbf{t}} \frac{d\mathbf{t}}{dt} + \frac{\partial\phi_{0}}{\partial p} \left(\left( \frac{dp}{dt} \right)_{dd} + \left( \frac{dp}{dt} \right)_{ml}\right).
\end{equation}
The differential decay term $(dp/dt)_{dd}$ describes how the muon beam's average momentum changes from higher-momentum muons having longer lifetimes due to time dilation, $\gamma(p)\tau_{\mu}$. 
The muon loss term $(dp/dt)_{ml}$ describes the effect of time-dependent changes in the momentum spectrum from lost muons.
Following Eq.~\eqref{eq:dpi-dt}, the correction due to the time-varying ensemble is split into three parts
\begin{equation} \label{eq:C-dphi-t0}
    C_{\phi_0-t} = C_{pa} + C_{dd} + C_{ml},
\end{equation}
with $C_{pa}$, $C_{dd}$, and $C_{ml}$ introduced in Eq.~\eqref{eq:R}.

\subsection{Phase acceptance correction: $C_{pa}$}\label{subsec:phaseacceptance}
\label{sec:Cpa}
\begin{figure}[!htbp]
\centering
\includegraphics[width=0.35\textwidth]{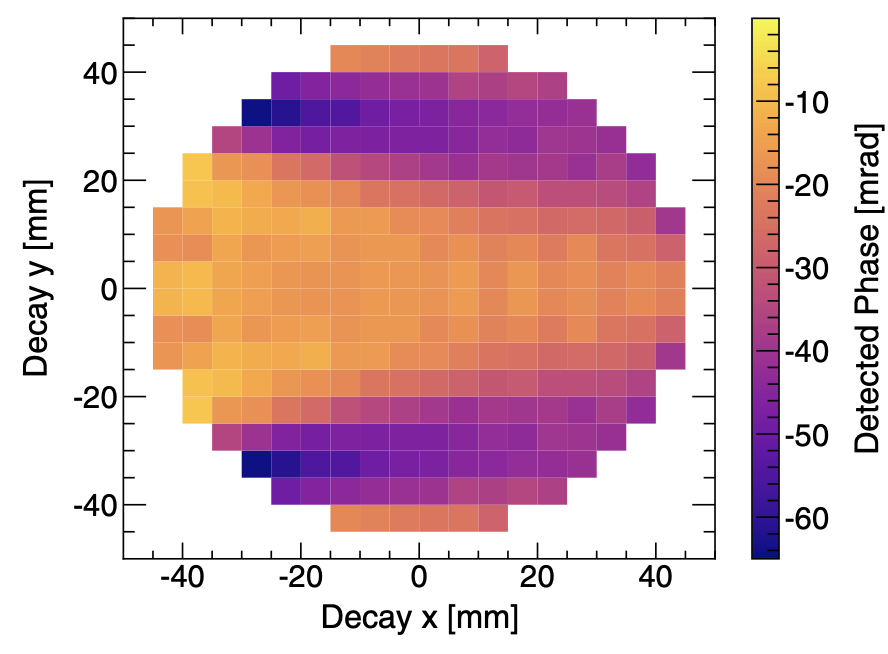} 
\caption{The azimuthally averaged phase map for the asymmetry-weighted analysis, showing the detected phase as a function of transverse muon decay position $(x,y)$.}
\label{fig:phase_map}
\end{figure}

Following Eq.~\eqref{eq:C-dphi-t0}, the phase acceptance contribution
\begin{equation} \label{eq:Cpa}
C_{pa}=\left(\frac{1}{\omega^{}_{a}}\right)\frac{\partial\phi_0}{\partial\mathbf{t}} \frac{d\mathbf{t}}{dt}
\end{equation}
arises from the coupling between the detector acceptance's muon decay position-dependent phase response $\phi^{m}_{0}(\mathbf{t}) = \phi^{m}_{0}(x,y)$ and time-dependent changes in the muon beam distribution, $\mathbf{t}(t)$, where the superscript $m$ denotes measured quantities.
The time-dependent phase arising from acceptance-phase effects is denoted $\phi_{pa}(t)$. It is computed by averaging over time the phase response as a function of transverse coordinates ($x$, $y$), obtained from a \texttt{GEANT4}-based simulation of the experiment (\texttt{GM2RINGSIM}), with the time evolution of the beam transverse coordinates measured by straw trackers, $M(x,y,t)$. Figure~\ref{fig:phase_map} shows the ``phase map" averaged over azimuth and weighted by the asymmetry method used to extract $\omega_a^m$.

The tracker stations measure the $M(x,y,t)$ distribution at two locations around the ring, but the extraction of $\omega_a^m$ is performed by calorimeters at 24 azimuthal locations. To transport the tracker data to the calorimeter locations, both \texttt{GM2RINGSIM} and \texttt{COSY INFINITY} simulation programs are used to predict the behavior of the beam around the ring as detailed in Ref.~\cite{PhysRevD.110.032009}.
Evolving the $M(x,y,t)$ distribution, including both vertical and radial manipulations, a complete spatial and time distribution of the muons $M^c(x,y,t)$ in the vicinity of the calorimeter locations is obtained. Combining the simulated maps with the muon distributions, a time-dependent phase can be computed for each calorimeter, $\varphi^c_{pa}(t)$, using the following weighted sum

\begin{figure}[htbp!]
\centering
\includegraphics[width=0.45\textwidth]{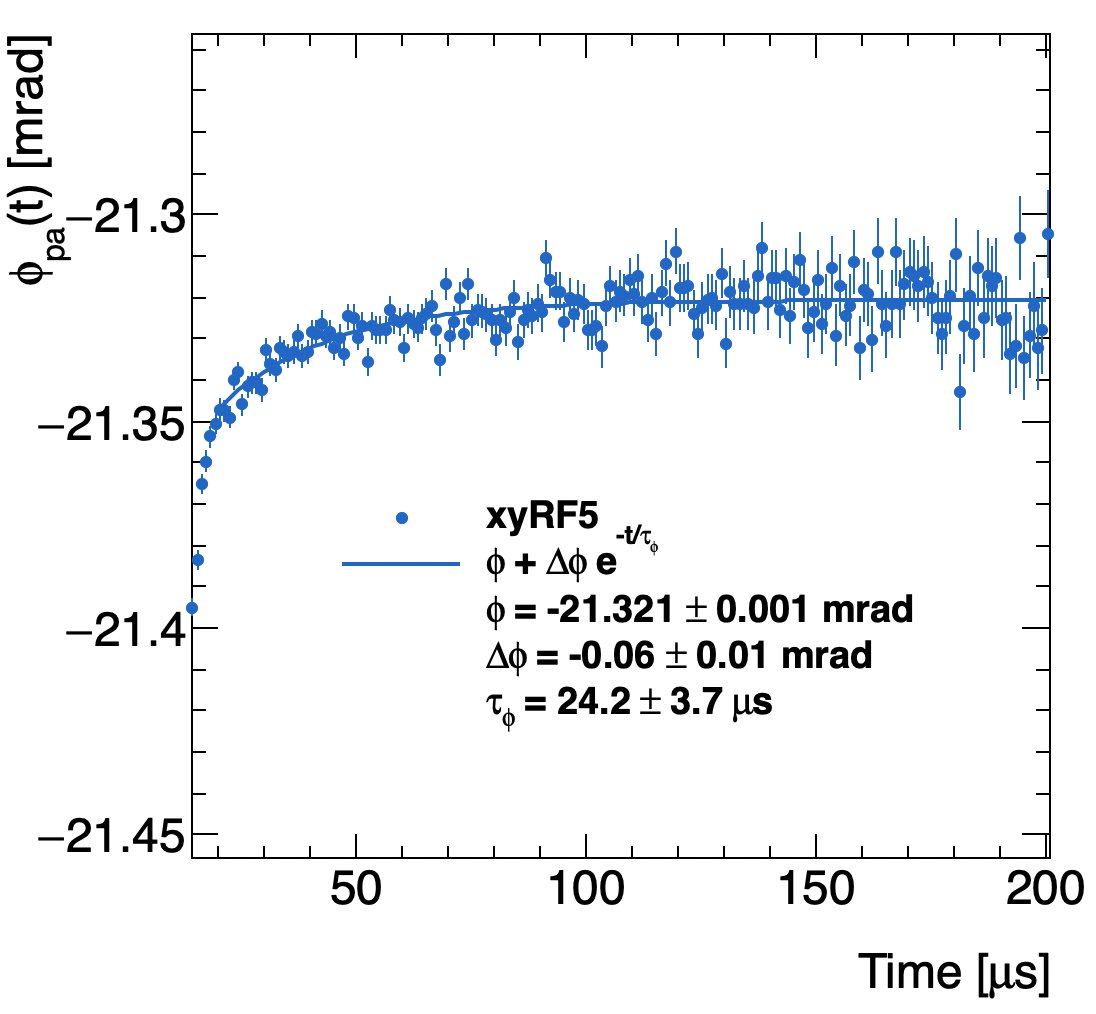} 
\caption{Calculated $\phi_{pa}(t)$ in data subset \RunFiveXY using data from the tracker station at 180$^{\circ}$. The fitted function is of the form $\phi_0 + \Delta\phi\cdot e^{-t/\tau_{\phi}}$.}
\label{fig:phi_t_5xy}
\end{figure}

\begin{widetext}
\begin{equation} \label{eq:master_formula}
\phi_{\mathrm{pa}}^c(t) = \mathrm{arctan}\frac{\sum_{ij}M^c(x_i,y_j,t)\cdot \varepsilon^c(x_i,y_j)\cdot A^c(x_i,y_j)\cdot \sin[\phi^c_{\mathrm{pa}}(x_i,y_j)]}{\sum_{ij}M^c(x_i,y_j,t)\cdot \varepsilon^c(x_i,y_j)\cdot A^c(x_i,y_j)\cdot \cos[\phi^c_{\mathrm{pa}}(x_i,y_j)]},
\end{equation}
\end{widetext}
where $\varepsilon^c$, $A^c$, and $\phi^c_{\mathrm{pa}}$ represent 
the acceptance, asymmetry, and phase maps, respectively, evaluated at t=0 for calorimeter $c$, which are then convoluted with the beam temporal evolution $M^c(x,y,t)$. 
Figure~\ref{fig:phi_t_5xy} shows the calculated $\phi_{pa}(t)$ time evolution for the \RunFiveXY dataset.

The size of the bias of $\omega^m_a$ due to this time-dependent phase is evaluated with a toy Monte Carlo procedure. For each calorimeter, a histogram is generated which includes all beam dynamics terms with the time-varying phase. The relative difference between the injected $\omega_a$ frequency and the fitted value gives the size of the correction for that calorimeter. The total correction is computed as the average difference among the calorimeters.

The evaluations of the statistical and systematic uncertainties are reported in Table~\ref{tab:final_value}.
The statistical uncertainty, which ranges between $2.0$ and \SI{3.0}{ppb}, originates from the number of tracks from $M(x,y,t)$ collected by tracker stations. The sources of systematic uncertainty can be categorized into three main groups. First, uncertainties arise from incomplete knowledge of the tracker detector, including alignment, resolution, and acceptance corrections, which directly impact the measured $M(x,y,t)$. Second, uncertainties stem from the estimation of the phase maps in Eq.~\eqref{eq:master_formula}, which are derived using \texttt{GM2RINGSIM}. Finally, the procedure relies on the beam dynamics functions obtained through simulation to derive the calorimeter $M^c(x,y,t)$ distribution from the tracker-measured $M(x,y,t)$.
Uncertainty contributions in the asymmetry and acceptance maps are negligible at the sub-ppb level.

The analysis methodology for \RunFourFiveSix follows the established procedure from \RunTwoThree~\cite{PhysRevD.110.032009}, with additional validation checks informed by accumulated experience.
Given the sensitivity of this correction to detector acceptance, both tracker and calorimeter acceptance models were re-evaluated.
The tracker acceptance was validated using the MiniSciFi detector, with differences well below systematic uncertainties.
For the calorimeter acceptance, azimuthal dependence was incorporated, extending previous analyses that considered only radial and vertical position.
These studies revealed systematic effects of about \SIrange{1}{2}{ppb}, increasing confidence in the robustness of the correction.

Finally, Table~\ref{tab:summary_pa} summarizes the $C_{pa}$ results, comparing the values from \RunFourFiveSix with those from previous runs.
In \RunOne, the correction was significantly larger due to
the suboptimal state of
the \ac{ESQ} system
that manifested in a slower \ac{ESQ} electric field ramp-up and unstable beam conditions at early times in the muon fill~\cite{Run1PRAB}. The correction in \RunFourFiveSix is slightly larger than in \RunTwoThree, primarily due to the use of the additional \ac{RF}. While it actively dampens beam oscillations, it induces greater time-dependent variations in the beam's vertical width and radial position than under the beam conditions of \RunTwoThree.

\begin{table}[!htbp]
\centering
\caption{Overview of the uncertainty contributions to the phase-acceptance correction $C_{\mathrm{pa}}$ for the \RunFourFiveSix dataset.}
\begin{tabular}{lrrrr}
\toprule
Dataset           & \RunFournoRF & \RunFiveX & \RunFiveXY & \RunSixXY\\ \midrule
Stat. uncertainty (ppb) & 2.4   & 2.9  &  2.7 & 3.4\\
   Tracker and \ac{CBO} (ppb)   & 10.5    &   10.0    &  10.9 & 9.9 \\
   Phase maps (ppb)       & 10.2    &   10.2   &  9.8 &  9.8 \\
   Beam dynamics (ppb)    & 1.3    &   1.3    &   1.3  & 1.3  \\ 
\midrule
Total uncertainty (ppb) & 15.0    & 14.7  & 15.0  &  14.4 \\ \bottomrule
\end{tabular}
\label{tab:final_value}
\end{table}

\begin{table}[htbp]
    \centering
    \caption{Summary of phase acceptance corrections $C_{pa}$ for all Run periods.}
    \label{tab:summary_pa}
    \begin{tabular}{lrr}
        \toprule
        Dataset & Correction & Uncertainty \\
        & (ppb) & (ppb) \\
        \midrule
        \RunOne & -158 & 75 \\
        \RunTwoThree & -27 & 13 \\
        \midrule
        \RunFournoRF & -32.8 & 15.0 \\
        \RunFiveX    & -34.3 & 14.7 \\
        \RunFiveXY   & -32.2 & 15.0 \\
        \RunSixXY    & -33.7 & 14.4 \\
        \bottomrule
    \end{tabular}
\end{table}

\subsection{Differential decay correction: $C_{dd}$} \label{subsec:dd}
\label{sec:DiffDec}
Following Eq.~\eqref{eq:C-dphi-t0}, the differential decay contribution 
\begin{equation} \label{eq:Cdd}
    C_{dd}=\left(\frac{1}{\omega^{}_{a}}\right)\left(\frac{d\phi_0}{dp}\right)\left(\frac{dp}{dt}\right)_{dd}
\end{equation}
arises from the time evolution of the muon ensemble's average momentum due to relativistic time dilation. Higher momentum muons have slightly longer lifetimes in the lab frame because their proper decay time is dilated by the slightly larger $\gamma$. This differential decay causes high-momentum muons to survive preferentially, shifting the ensemble's average momentum over time. When combined with any existing phase-momentum correlations in the initial distribution, this momentum shift produces a time-dependent phase evolution that needs to be corrected for.

The rate of momentum shift is
 \begin{equation} \label{eq:dd:dpdt}
 \left(\frac{dp}{dt}\right)_{dd} \approx \frac{p_0}{\gamma_0\tau_{\mu}}\sigma^2_{\delta},
  \end{equation} 
where $\sigma^2_{\delta}$ is the variance of the fractional-momentum distribution, $p_0$ is the magic momentum and $\gamma_0$ is the corresponding Lorentz factor. $\delta$ is the fractional momentum as introduced for Eq.~\eqref{eq:Ce}.
This rate is proportional to $\sigma^2_{\delta}$: a wider momentum distribution leads to faster evolution of the ensemble average, as the preferential survival of high-momentum muons (due to time dilation) more rapidly shifts the mean when the distribution is broader.

The phase-momentum correlation, 
\begin{equation}
    \frac{d\phi_{0}}{dp}=\frac{1}{p_0}\frac{d\phi_0}{d\delta},
\end{equation} 
is a full derivative with respect to the  orbital coordinates
at a reference longitudinal position $s$ along the muon trajectories,
where $x$, $x'$ are the radial position and angle, $y$, $y'$ are the vertical coordinates and $\tau$ is the relative arrival time within a bunch:
\begin{align} 
\frac{d\phi_0}{d\delta} &= \underbrace{\left.\frac{\partial\phi_0}{\partial \delta}\right|_{s}}_{\text{direct}} 
+ \underbrace{\left.\frac{\partial\phi_0}{\partial x}\right|_{s}\frac{dx(s)}{d\delta} + \left.\frac{\partial\phi_0}{\partial x'}\right|_{s}\frac{dx'(s)}{d\delta}}_{\text{radial}} \nonumber\\
&\quad + \underbrace{\left.\frac{\partial\phi_0}{\partial y}\right|_{s}\frac{dy(s)}{d\delta} + \left.\frac{\partial\phi_0}{\partial y'}\right|_{s}\frac{dy'(s)}{d\delta}}_{\text{vertical}}
+ \underbrace{\frac{\partial\phi_0}{\partial \tau}\frac{d\tau}{d\delta}}_{\text{longitudinal}}. \label{eq:dphidp}
\end{align}
 This decomposition, for specific $s$, for example at the $T_0$-detector or at the end of the inflector, was validated through extensive beam tracking studies using the \texttt{GM2RINGSIM} framework.

\paragraph{Direct Contribution}
The direct contribution $\partial\phi_0/\partial\delta$ originates primarily from the upstream beamline, where the muon spin phase advance scales with momentum. After four revolutions through the delivery ring, assuming the idealized case, this produces a momentum-dependent phase advance with an expected gradient of approximately -8.6$~\text{mrad}/(\%\delta)$. End-to-end simulations confirm a slope of $\sim$-9$~\text{mrad}/(\%\delta)$ in the stored muon distribution, with negligible sensitivity to the presence of cooling wedges. Small vertical bends ($\mathcal{O}(\SI{15}{\degree})$) in the upstream beamline are included in these simulation studies.

\paragraph{Transverse Contributions}
Correlations between spin phase and transverse coordinates ($\partial\phi_0/\partial x$, $\partial\phi_0/\partial x'$, etc.) originate from the fundamental physics of pion decay~\cite{PhysRevD.110.032009}. When a $\pi^{+}$ decays to a muon and neutrino, the muon spin and momentum are anti-aligned in the pion rest frame due to conservation of angular momentum and the left-handedness of neutrinos. When transformed to the lab frame, this property creates specific correlations between the spin angle and transverse coordinates at birth that persist through beamline transport, though modified by the betatron phase advance.

The momentum acceptance properties of the storage ring introduce correlations between transverse coordinates and momentum ($dx/d\delta$, $dx'/d\delta$) during injection. The finite aperture and acceptance constraints result in storage probabilities that depend on both momentum and transverse phase space coordinates, creating complex phase space-momentum correlations. The dispersive injection process through the inflector and kickers introduces additional mixing between these variables.

\paragraph{Longitudinal Contribution}
The longitudinal contribution arises from the temporal spread of muon arrival times in the storage ring. Muons arrive at different relative times $\tau$ within each bunch. Since earlier muons enter the magnetic field sooner, their spins begin precessing while later muons are still approaching, creating a relative phase shift between them. This time-dependent phase evolution is characterized by $\partial\phi_0/\partial \tau = \omega_a$.

The correlation between arrival time and momentum ($d\tau/d\delta$) emerges from the combined acceptance efficiency of the storage ring and kicker system for muons arriving at different relative times. The novel MiniSciFi detectors enabled the mapping of these underlying distributions as described in more detail in Sec.~\ref{sec:miniSciFi}. This coupling between injection time and momentum, combined with the intrinsic time-phase relationship, produces the longitudinal contribution to the differential decay correction.

The analysis approach for these contributions has evolved with an improved understanding of the injection process. In \RunTwoThree, the direct and transverse components were treated separately: the direct component was determined from dedicated measurements of the $\partial\phi_0/\partial\delta$ correlation, while transverse components were evaluated using beam tracking simulations. 
However, in \RunFourFiveSix, the direct and transverse components are combined into a single \textit{injection} component due to their complex mixing during the injection process and the difficulty in reliably decoupling them position independently. This entire injection component is evaluated from simulation, reflecting a more conservative approach that better accounts for our understanding of injection correlations.

    To determine the \textit{injection} component, a set of dedicated simulations constrains the possible range of phase-momentum correlations. \texttt{BMAD}-based simulations of the muon production and delivery were used to constrain the input phase space before the injection into the storage ring. A set of \texttt{GM2RINGSIM}-based simulations that span feasible ranges of injection and beam storage configurations constrained the total effect. These configurations included different injection angles, injector fields, radial field configurations, kicker, and \ac{RF} strengths and timing settings. These simulations yield a range of possible phase momentum correlation $d\phi_0/d\delta$ of $-26$~\text{mrad}$/(\%\delta)$ to 0~mrad$/(\%\delta)$. The momentum spread term in Eq.~\eqref{eq:dd:dpdt} remains data-driven. These constraints are broadly consistent with direct measurements of $-13.5\pm1.4$~\text{mrad}$/(\%\delta)$ detailed in Ref.~\cite{PhysRevD.110.032009}, which are used as the central value with the full simulation range providing a conservative uncertainty that accounts for potential cancellations between different injection-related contributions that cannot be reliably separated.

The evaluation of the \textit{longitudinal} ($p-\tau$) contribution follows the same data-driven method used in \RunTwoThree, utilizing momentum-injection time distributions driven by the electric field correction analysis. The analysis is based on 2D $\tau$-$x_e$ distributions from the electric field analysis, which are transformed to $\phi_0$-$\delta$ distributions using the relationships $x_e = D\delta$ (where $D\sim$\SI{8}{m} is the dispersion relation) and $\phi_0 = -\omega_a \cdot \tau$. 
To quantify the differential decay effect, toy muon ensembles are generated following these distributions and used to create synthetic wiggle plots. Each muon's contribution includes the momentum-dependent lifetime $\tau(\delta) \approx \tau_{0}(1+\delta)$. Five-parameter fits to these synthetic datasets extract the bias $\Delta\omega_a$, yielding the correction $C_{dd} = -\Delta\omega_a/\omega_a$. 

\begin{figure}
    \centering
    \includegraphics[width=0.98\linewidth]{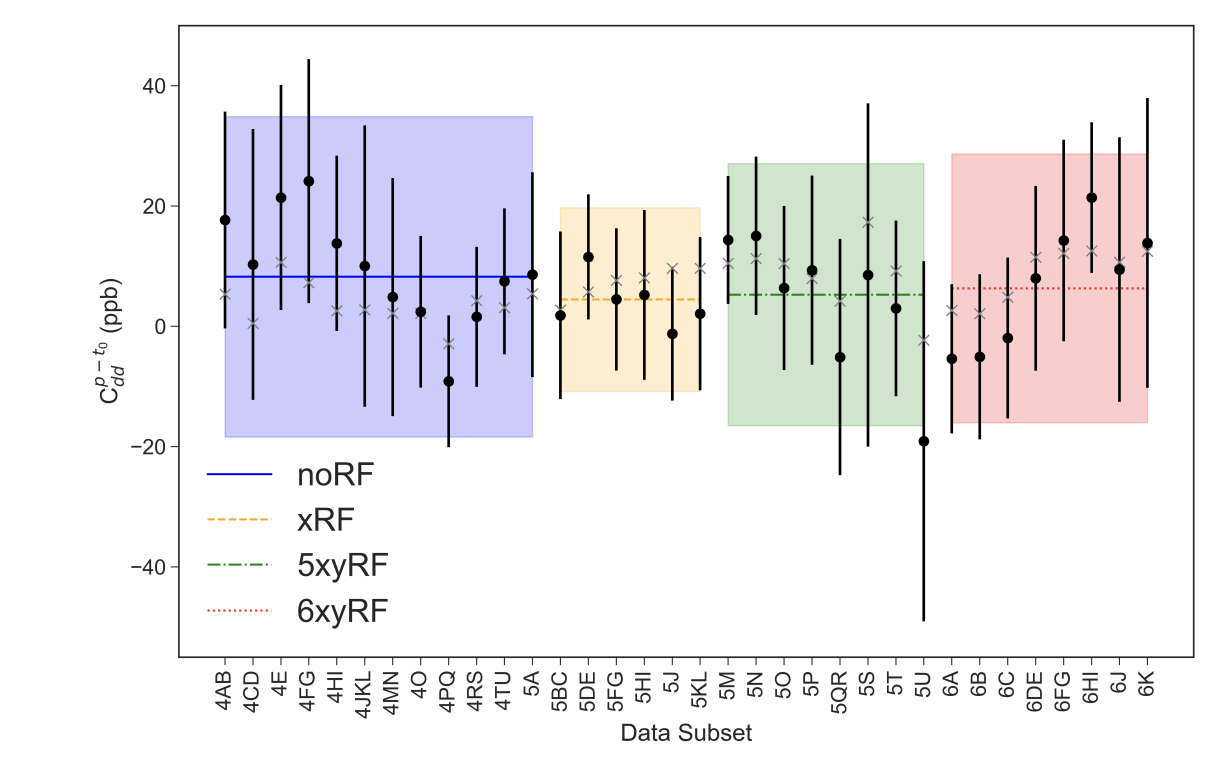}
    \caption{Longitudinal differential decay contributions bunch-averaged values for each lettered dataset in black and the weighted averages of the main datasets in color.}
    \label{fig:dd:longitudinal}
\end{figure}

The longitudinal correction contributions are calculated per lettered dataset per muon bunch. 
Each muon bunch has its own characteristic time distribution, leading to significant bunch-to-bunch differences spanning $\pm\SI{30}{ppb}$ and above. Results are combined, taking into account the different number of stored muons per bunch and lettered dataset. Statistical uncertainties are evaluated using the same spread of different input distributions that are also used for the electric field correction. In addition, the bunch-to-bunch spread leads to an additional conservative systematic uncertainty. Figure~\ref{fig:dd:longitudinal} shows the bunch-averaged values per lettered dataset, along with the weighted averages and corresponding uncertainties, where the weights are the number of stored muons.

The Fast-Rotation-based extraction of the relative muon timing within the bunch, the injection time $\tau$, can only resolve one cyclotron period of about \SI{149}{\nano\second}. This limitation causes muons in the very tails of the timing distributions to be wrapped around, leading to an incorrect $\phi_0$ assignment.
To evaluate the worst-case impact of this wrapping effect, toy simulations were performed using MiniSciFi-based distributions, which do not suffer from this cyclotron period limitation. This yielded an additional conservative uncertainty of \SI{12}{ppb}. The total correction value and uncertainties are summarized in Table~\ref{tab:dd:run456}.

The additional scrutiny of the \RunFourFiveSix analysis revealed a sign error in the \RunTwoThree direct (beamline) contribution which was corrected in the combination of results, leading to a \SI{+32}{ppb} shift, with additional uncertainty increases applied to reflect the improved understanding of injection correlations, increasing the \RunTwoThree total uncertainty from \SI{17}{ppb} to \SI{22}{ppb}.

The straw tracking detectors provide a complementary approach to determining the differential decay correction, though systematic limitations prevent their use in the final combination. The analysis extracts radial-dependent \gm phases of detected positrons within the momentum range \SIrange{1.8}{2.5}{GeV/c} to obtain the gradient $d\phi_0/dx$. After removing the radial phase spread from betatron motion, the phase-momentum correlation $d\phi_0/d\delta$ is determined using the dispersion relation. While reconstructed phase uncertainties of $\approx\SI{5}{mrad}$ yield statistical precision at the $\mathcal{O}(10~\text{ppb})$ level, detector-driven systematic phase shifts at the \SI{100}{ppb} level limit the utility of this method. It served as a cross-check of the sign and magnitude of the differential decay correction.

\begin{table}
\centering
\caption{\RunFourFiveSix differential decay correction $C_{dd}$ values with their injection and longitudinal contributions. The total, corrected (corr.) \RunTwoThree values are shown for reference.}

\begin{tabular}{lrrrrrr}
\toprule
& \multicolumn{2}{c}{Injection} & \multicolumn{2}{c}{Longitudinal} & \multicolumn{2}{c}{Total} \\
\cmidrule(lr){2-3} \cmidrule(lr){4-5} \cmidrule(lr){6-7}
Dataset & Corr. & Unc. & Corr. & Unc. & Corr. & Unc. \\
& (ppb) & (ppb) & (ppb) & (ppb) & (ppb) & (ppb) \\
\midrule
\RunFournoRF & 20.1 & 20.3 & 8.2 & 19.3 & 28.3 & 28.0 \\
\RunFiveX & 19.1 & 19.3 & 4.4 & 16.6 & 23.5 & 25.5 \\
\RunFiveXY & 19.6 & 19.8 & 5.3 & 19.4 & 24.9 & 27.7 \\
\RunSixXY & 19.7 & 19.9 & 6.3 & 18.1 & 26.0 & 26.9 \\
\midrule
\RunTwoThree & -- & -- & -- & -- & 17 & 22 \\ 
\bottomrule
\end{tabular}

\label{tab:dd:run456}
\end{table}

\subsection{Muon loss correction: $C_{ml}$}\label{subsec:muonloss}
\label{sec:MuonLoss}
Following Eq.~\eqref{eq:C-dphi-t0}, the muon loss contribution 
\begin{equation} \label{eq:Cml}
    C_{ml}=\left(\frac{1}{\omega^{}_{a}}\right)\left(\frac{d\phi_0}{dp}\right)\left(\frac{dp}{dt}\right)_{ml}
\end{equation}
arises from the time-dependent momentum spectrum of lost muons in the presence of any phase-momentum correlation.
$C_{ml}$ is estimated by measuring the shape of the muon loss function $L(t)$ and the measured phase-momentum correlation explained above in Sec.~\ref{sec:DiffDec}. In addition, the cumulative measured muon loss spectrum
is used as a scale factor. 
The implementation of the \ac{RF} system in three of the four datasets not only reduces the \ac{CBO} amplitude but also the cumulative muon losses by roughly a factor of five.
Table~\ref{tab:Cml} summarizes the values of $C_{ml}$.

Under worst-case assumptions, treating all absent muons as lost exclusively from one side of the momentum distribution, the maximum possible bias is \SI{2.6}{ppb} for the \RunFiveXY dataset.

\begin{table}[]
    \centering
    \caption{Muon loss corrections $C_{ml}$ and uncertainties.}
    \label{tab:Cml}
    \begin{tabular}{lrr}
        \toprule
         Dataset & Correction & Uncertainty \\
         & (ppb) & (ppb) \\
         \midrule
         \RunFournoRF & 0 & 3 \\
         \RunFiveX, \RunFiveXY, \RunSixXY & 0 & 2 \\ 
         \bottomrule
    \end{tabular}
\end{table}

\section{Magnetic field measurement: \op}\label{sec:op}
The magnetic field within the storage ring is determined using a combination of mobile and fixed \ac{NMR} probes.
A mobile magnetic field camera, the trolley, comprising $17$ \ac{NMR} probes arranged in a transverse plane, provides detailed spatial maps of the magnetic field in the muon storage volume.
The trolley is effectively calibrated against shielded protons in a spherical water sample at a reference temperature $T_r$.
This calibration consists of two steps: first, the trolley probes are calibrated against a cylindrical water-based calibration probe by measuring magnetic-field differences in the same environment within the muon storage volume. Second, the calibration probe measurements are converted to absolute field values by correcting for the probe's magnetic environment.
Every three to seven days, the trolley maps the magnetic field inside the storage ring over approximately \SI{70}{\minute} at approximately $9000$ azimuthal positions (every \SI{5}{mm}). 
Fixed \ac{NMR} probes installed at $72$ stations around the ring, above and below the muon storage volume outside the vacuum chamber, track the field over time, sampling asynchronously with respect to muon injection at approximately \SI{1.2}{\second} intervals. More details can be found in Ref.~\cite{Run1PRAField}.
The combination of field maps from the trolley and continuous fixed probe measurements allows reconstruction of the magnetic field inside the storage volume as $\omega_{p}'(x,y,\varphi,t)$.
These time-tracked field maps are weighted with the reconstructed muon distributions $M(x,y,\varphi,t)$ from the straw-trackers and corrected for synchronous transient magnetic fields to obtain the magnetic field experienced by the muons during storage, \opprimetildeofT, as used in Eq.~\eqref{eq:Rmu}.
The analysis presented here follows the same approach used in \RunOne~\cite{Run1PRAField} and \RunTwoThree~\cite{PhysRevD.110.032009}.
The following sections detail each step of this process.

\subsection{Calibration} \label{sec:wp:calibration}

All measurements by the trolley probe $n$, $\omega_{n}^{\mathrm{tr}}$, taken at a temperature $T_{m}$, are corrected by probe-specific ($n$) constants, $\delta_{n}^{\mathrm{calib}}$, with respect to shielded protons in a spherical water sample at a reference temperature $T_r$
\begin{equation} \label{eq:wp:calib}
    \omega'(T_r)_n = \omega_{n}^{\mathrm{tr}}(1+\delta_{n}^{\mathrm{calib}}).
\end{equation}

The calibration constants are split into four parts,
\begin{equation}
    \delta_{n}^{\mathrm{calib}} = (\delta^{\mathrm{abs}} + \delta^\text{x-check} + \delta_{n}^{\mathrm{env}} + \delta_{n}^{\mathrm{trans}})
\end{equation}
corresponding to the two-step calibration chain, where $\delta^{\mathrm{abs}}$ corresponds to an absolute calibration of measurements from the calibration probe at $T_r$ with respect to shielded protons in a spherical water sample and $\delta^\text{x-check}$ accounts for differences in the cross-calibration measurements with other absolute probes. $\delta_n^{\mathrm{trans}}$ represents the probe-specific transfer of this absolute calibration to the individual trolley probes. $\delta_n^{\mathrm{env}}$ accounts for differences in the environments of the first two terms.
These differences include the calibration probe's temperature during the transfer measurements relative to $T_r$.
They also account for effects from the in-situ calibration's magnetic environment, such as magnetic image effects on the calibration probe, which are only present when placed inside the experiment's storage volume. 

In contrast to previous publications~\cite{Run1PRL, PhysRevD.110.032009}, this analysis uses $T_{r}=\SI{25}{\celsius}$.
This change is motivated by the modified CODATA 2022~\cite{mohr2024codatarecommendedvaluesfundamental} recommendations for $\gamma_{p}'$ to use averaged results from Refs.~\cite{flowers_measurement_1993} and~\cite{schneider_3he+_2022}, and hence the need to use CODATA's temperature reference. Previously, the analysis used Ref.~\cite{flowers_measurement_1993} directly and the corresponding reference temperature of that measurement.

Since the largest correction is $\sim$\SI{1.5}{ppm}, $\ll$1, only first-order corrections are applied when multiple terms of the form $(1+\delta_a)(1+\delta_b)\approx (1+\delta_a+\delta_b+\mathcal{O}(\delta^2))$ are combined.

\subsubsection{Absolute Calibration: $\delta^{\mathrm{abs}}$} \label{sec:wp:absCalib}
The pulsed \ac{NMR} probe that is used to transfer the absolute calibration {\it in situ} to the individual trolley probes is a cylindrical water probe designed to minimize field perturbations.
The probe used for the \RunFourFiveSix data presented in this publication is the same as that used for \RunTwoThree, upgraded from but nearly identical to that used in \RunOne, which was described in detail in Ref.~\cite{FlayPP}.

The corrections $\delta^{\mathrm{abs}}$ correct raw measurements from this probe to the frequency that would be measured in the same field by shielded protons in an ideal spherical sample. These corrections are given by
\begin{equation}
\delta^{\mathrm{abs}} = \delta^{\mathrm{b}}(T_r) + \delta^{\mathrm{s}} + \delta^{\mathrm{P}} + \delta^{\mathrm{RD}} + \delta^{\mathrm{d}}
\end{equation}
and are summarized in Table~\ref{tb:wp:absCalib}. The individual terms account for:

\paragraph{Sample-shape correction $\delta^{\mathrm{b}}$} The magnetic shielding produced by the individual magnetic moments within the sample is influenced by the sample’s geometry and is governed by its magnetic susceptibility together with a shape-dependent factor. The temperature-dependent correction is 
\begin{equation}
    \delta^{\mathrm{b}}(T)=\chi_{V}(T)(\epsilon-1/3),
\end{equation}
where $\chi_{V}(T)$ is the volume magnetic susceptibility of the water sample at temperature $T$, and
    $\epsilon=0.4994(6)$
    for the finite cylindrical sample, which was calculated from Ref.~\cite{OsbornShapeCorrection} using the approximation that the sample is a long prolate spheroid with semimajor axis half its length and semiminor axis half its inner diameter. A conservative uncertainty was chosen, covering up to the infinite-cylinder case ($\epsilon=1/2$). Numerical simulation of the sample, approximating small volume elements as magnetic dipoles created by the water's magnetization, agrees well with this choice and also concludes that $0.499<\epsilon<0.500$.

    The temperature-dependent volume susceptibility of water is 
\begin{align}
    \chi_{V}(T) &= \chi_{V}(\SI{22}{\celsius}) \times  \left[ \frac{\rho(T)}{\rho(\SI{22}{\celsius})} \right] \times \left[ \frac{\chi_{m}(T)}{\chi_{m}(\SI{20}{\celsius})} \right] \nonumber \\
    &\quad \times \left[ \frac{\chi_{m}(\SI{20}{\celsius})}{\chi_{m}(\SI{22}{\celsius})} \right],
\end{align}
where $\chi_V(T)$ is the volume susceptibility of water, $\rho(T)$ is the mass density of water and $\chi_m(T)$ is the mass susceptibility of water at temperature $T$.
Here $\chi_{V}(\SI{22}{\celsius})=-9.056\times  10^{-6}$ is the value recommended by CODATA~\cite{CODATA1998}, using a $3\times 10^{-8}$ uncertainty (compared to CODATA's $6\times 10^{-9}$) due to additional measurements at unspecified temperatures~\cite{Blott1993}.
The ratio of mass susceptibilities from Ref.~\cite{philo1980} is used, which directly provides the term $\chi_m(T)/\chi_m(\SI{20}{\celsius})$, and the temperature dependence of the mass density of water from \cite{Kell1967}. For more details, see Ref.~\cite{PhysRevD.110.032009}.

\paragraph{Material effects $\delta^{\mathrm{s}}$}\label{subsubsubsec:pp_material_effect}
The material effect refers to field perturbations caused by the nonzero magnetic susceptibility of the calibration probe's components (glass tubes, aluminum and copper coil wires, aluminum shell, etc.) when magnetized by the external field. These perturbations are directly measured in a dedicated stable, homogeneous, large-bore \SI{1.45}{T} solenoid, accounting for the probe's orientation relative to the respective magnetic field vectors.
The measurement is performed by removing the calibration probe's water sample and repeatedly placing the rest of the calibration probe on and off a small stationary test probe approximately the same size as the water sample, and observing a $\mathcal{O}(10)$~ppb field shift.

Table~\ref{tab:delta_s} summarizes the contributions to $\delta^{\mathrm{s}}$, which include the measurement in the test solenoid and a few additional systematic uncertainties. 
Systematic uncertainties arise from the alignment of the calibration and test probes relative to each other and to the external field in the test solenoid, and from the range of possible magnetic susceptibilities in the mock holder stick. The mock stick, a mechanical support structure used for these measurements, was made of an identical-cross-section but shorter-length piece of U-Channel compared to the real stick, and cannot be guaranteed to have identical magnetic properties.

\begin{table}[htbp]
\centering
\caption{Contributions of the material effects $\delta^{\mathrm{s}}$ of the calibration probe.} \label{tab:delta_s}
\begin{tabular}{lrr}
\toprule
Description & Correction & Uncertainty \\
            & (ppb) & (ppb) \\
\midrule
Measurement & -11.6 & 1.8 \\
Alignment & 0 & 1.0 \\
Tilt w.r.t. field & 0 & 2.0 \\
Roll around its axis & 0 & 1.0 \\
Mock stick alloy & 0 & 3.8 \\
\midrule
Total & -11.6 & 4.9 \\
\bottomrule
\end{tabular}

\end{table}

\paragraph{Additional uncertainties $\delta^{\mathrm{P}} + \delta^{\mathrm{RD}} + \delta^{\mathrm{d}}$}
These additional uncertainties account for possible water sample imperfections~\footnote{Imperfections such as curvature of the cylindrical water sample or trace impurities.} and dissolved oxygen in the water ($\delta^{\mathrm{P}}$), radiation damping effects from the dynamics of the \ac{NMR} measurement ($\delta^{\mathrm{RD}}$), and distant dipolar field proton-proton spin interactions in the water sample ($\delta^{\mathrm{d}}$). Table~\ref{tb:wp:absCalib} summarizes the full set of calibration probe intrinsic corrections, including these uncertainties.

\begin{table}[ht]
    \centering
    \caption{The calibration probe intrinsic corrections and uncertainties summing up to $\delta^{\mathrm{abs}}$}
    \label{tb:wp:absCalib}
    \begin{tabular}{llrr}
        \toprule
        Description & & Corrections & Uncertainties \\
                    & & (ppb) & (ppb) \\
        \midrule
        Shape, susceptibility  & $\delta^{\mathrm{b}}(T_r)$ &  -1503.4 & 5.0 \\
Material effects & $\delta^{\mathrm{s}}$     & 11.6 & 4.9 \\
        Radiation damping & $\delta^{\mathrm{RD}}$   & 0 & 3.0 \\
        Proton dipolar field & $\delta^{\mathrm{d}}$ & 0 & 2.5 \\
        Sample purity & $\delta^{\mathrm{P}}$        & 0 & 2.0 \\
        \midrule
        Subtotal & $\delta^{\mathrm{abs}}$ & -1491.8 & 8.3 \\
        \bottomrule
    \end{tabular}
\end{table}

\subsubsection{Cross-checks and uncertainty inflation}
Different subsets of the corrections within $\delta^{\mathrm{abs}}$ were cross-checked in independent, dedicated cross-calibration efforts. The material effects were checked in a cross-calibration with respect to a water-based calibration probe of the MuSEUM~\cite{Strasser2025} and upcoming Muon \gm/EDM~\cite{10.1093/ptep/ptz030} experiments at J-PARC. The magnetic susceptibility of the water sample in the exact setting used in the calibration probe was cross-checked in dedicated measurements. The full absolute calibration term, including the sample-shape correction, was measured independently in a cross-calibration with respect to a $^3$He-based pulsed \ac{NMR} probe~\cite{Farooq}.

\paragraph{Material Effect $\delta^{\mathrm{s}}$} \label{sec:wp:USJP}
The absolute calibration constant $\delta^{\mathrm{abs}}$, excluding the bulk magnetic susceptibility and shape factor contribution $\delta^{\mathrm{b}}$, was cross-checked against the J-PARC MuSEUM experiment's \ac{NMR} calibration probe~\cite{yamaguchi_development_2019}. 
This comparison was performed in the same solenoid used to measure the material effect. We define the remaining contributions as $\delta^{\mathrm{t}}=\delta^{\mathrm{s}} + \delta^{\mathrm{P}}+\delta^{\mathrm{RD}}+\delta^{\mathrm{d}}$, in which the material effect $\delta^{\mathrm{s}}$ dominates.
 
Both probes are water-based, but while the \ac{FNAL} probe uses pulsed \ac{NMR}, the Japanese counterpart employs a continuous-wave technique.
Comparisons were performed over a four-year time span at \SI{1.45}{T} (twice), \SI{1.7}{T}, and \SI{3}{T}. Over the course of the different measurements, the Japanese probe was improved. 
This cross-calibration directly probes $\delta^{\mathrm{t}}$ by measuring the difference:
\begin{eqnarray} \label{eq:delta}
  &\Delta^{\text{JP-US}} &\opprime = \left[ \opprimeJPatTexp - \opprimeUSatTexp \right] \times (1 + \deltasetup) \nonumber \\
  & = & \left(\opmeasUS(\TUS) \times \left[ 1 + \deltaT(\TUS) + \deltab(\TUS) + \deltatUS \right]\right.  \nonumber \\
  &  & \left. - \opmeasJP(\TJP) \times \left[ 1 + \deltaT(\TJP) + \deltab(\TJP) + \deltatJP \right]\right) \nonumber \\
  & & \times (1 + \deltasetup),
\end{eqnarray}
where $US$ denotes the \ac{FNAL} and $JP$ the J-PARC probes, and $\delta^{\mathrm{setup}}$ are differences from the cross-calibration setup.
Due to the very similar form factors, the $\delta^{\mathrm{b}}$-terms cancel at the \SI{3}{ppb}-level. The same is true for the $\delta^{\mathrm{T}}$ terms due to the very similar temperatures at the 5~ppb-level, and any remaining effects from the calibration setup, $\delta^{\mathrm{setup}}$, cancel out by averaging two measurements with swapped probes within an uncertainty of $\sim$\SI{2}{ppb}.

\begin{table}[htbp]
    \centering
    \caption{Cross-calibration differences $\Delta^{\text{JP-US}}\opprime$ between the J-PARC and \ac{FNAL} calibration probes.}
    \begin{tabular}{lrr}
        \toprule
        Description & Difference & Uncertainty \\
        &  (ppb) & (ppb) \\
        \midrule
        2019, \SI{1.45}{T} & -63.0 & 17.0 \\
        2020, \SI{1.7}{T}  & -58.3 & 16.6 \\
        2022, \SI{3.00}{T} & -3.6 & 9.1 \\
        2024, \SI{1.45}{T} & -11.5 & 11.7 \\
        \bottomrule
    \end{tabular} \label{tab:crosscall}
\end{table}

Table~\ref{tab:crosscall} shows the results from the four cross-calibration campaigns.
The expectation is that $\Delta^{\text{US-JP}}\opprime$ is consistent with zero within uncertainties. 
However, the first two of the four campaigns showed tensions that remained unresolved.
The two more recent results are consistent with expectations.
Therefore, the uncertainty on $\delta^{\mathrm{s}}$ is inflated, equally distributed between the two probes, such that the reduced $\chi^2$ of the four measurements becomes unity. 
This leads to an additional uncertainty on $\delta^{\mathrm{t}}$ of \SI{25}{ppb}, which is used for $\delta^\text{x-check}$.

\paragraph{Susceptibility of water and shape factor $\delta^{\mathrm{b}}$}
The bulk magnetic susceptibility of water was cross-checked using the same homogeneous solenoid used for the material-effect measurements. 
In a dedicated measurement, the identical cylindrical water sample used in the calibration probe was used to measure the volume susceptibility of water by exploiting the $\sin(2\theta)$ angular dependence of the form factor $\epsilon$ when rotating the cylindrical sample from parallel to perpendicular with respect to the field.
The measurements were performed in a nitrogen atmosphere to eliminate corrections due to paramagnetic oxygen, under well-controlled pressure and temperature. 
The measured value of water's susceptibility is consistent with the CODATA value within uncertainties.

\paragraph{Absolute calibration  $\delta^{\mathrm{abs}}$ with respect to $^{3}$He}

The absolute calibration constant $\delta^{\mathrm{abs}}$ was also measured with respect to a spherical $^{3}$He-based pulsed \ac{NMR} probe in the same test solenoid.
Cross-calibrating that probe with the standard calibration probe allows an independent, simultaneous determination of the form factor and susceptibility terms ($\delta^{\mathrm{b}}$) and the material effects (dominated by $\delta^{\mathrm{s}}$).
The negligible temperature dependence of $^{3}$He provides an independent handle on the temperature dependence of the water-based results.
The difference between the standard calibration probe, $CP$, and the probe based on helium, $He$, can be expressed as
\begin{equation} \label{eq:he3}
\Delta^{\text{CP-He}}\omega_p' = \omega_{p,CP}' - \frac{\gamma_{p'}}{\gamma_{He}}\omega_{He}' 
\end{equation}
where $\frac{\gamma_{p'}}{\gamma_{He}}$ is the ratio of the gyromagnetic moments of shielded protons and $^{3}$He from literature.

The ratio of the gyromagnetic ratios, $\frac{\gamma_{p'}}{\gamma_{He}}$, was directly measured by Ref.~\cite{flowers_measurement_1993} in 1993 with an uncertainty of \SI{4.3}{ppb}, but can also be derived from a combination of the measurement of $\frac{\mu_{p}'}{\mu_e}$ by~\cite{phillips_magnetic_1977} in 1977 and $\frac{\mu_{^{3}He}}{\mu_{N}}$ by~\cite{schneider_3he+_2022} in 2022 in combination with the well known ratio $\frac{\mu_{e}}{\mu_{N}}$ with a total uncertainty of \SI{11}{ppb} driven by the first term. 
However, these two determinations show a difference of \SI{15.5}{ppb} ($1.3\sigma$).

The cross-calibration measurement yields $\Delta^{\text{CP-He}}\omega_p^\prime = \SI[separate-uncertainty=true]{30 \pm17.6}{ppb}$ which is consistent with zero at a $1.7\sigma$-level.

\subsubsection{Environment Corrections: $\delta^{\mathrm{env}}$}
The following terms contribute corrections arising from the environment in which the transfer measurements take place; these corrections are evaluated for each probe $n$ from the sum 
\begin{equation} \label{eq:delta-env}
\delta_{n}^{\mathrm{env}} = \delta^\mathrm{stick} + \delta^\mathrm{oxygen} + \delta^\mathrm{fp,trly} + \delta_{n}^\mathrm{fp,cp} + \delta_{n}^{img}
\end{equation} and are summarized in Table~\ref{tb:wp:envCalib}. The individual terms account for:

\paragraph{Holding stick field environment $\delta^\mathrm{stick}$:}
A correction due to the field perturbations produced at the location of the calibration probe by the long aluminum u-channel stick used to hold the probe in the storage volume. The stick extended from the storage volume radially inwards into the fringe field of the magnet. The effects of a much shorter stick were measured in the test solenoid and used with measurements of the fringe field to estimate the field perturbation of the longer stick. The effect is determined to \SI{-0.7+-0.2}{ppb}.

\paragraph{Oxygen Effect $\delta^\mathrm{oxygen}$}
A correction accounting for the fact that the measurement in the test solenoid is performed in air and thus moving the probe causes a displacement of (magnetized) gaseous oxygen, whereas the calibration probe is used in vacuum in the storage ring. The effect is determined to \SI{+2.0+-1.0}{ppb}.

\paragraph{Magnetic footprints of probes on each other $\delta^\mathrm{fp,trly}$ and $\delta_n^\mathrm{fp,cp}$ }
The magnetic footprint of the retracted trolley reduces the field seen by the calibration probe,
hence a positive correction of $\delta^\mathrm{fp,trly} = \SI{14.3+-8.0}{ppb}$ is needed. The magnetic footprint of the retracted calibration probe influences the field seen by the trolley and depends on the position of the calibration transfer, and thus is probe-specific. Corrections range from \SIrange{-4.4}{-8.7}{ppb} and uncertainties range from \SIrange{7.0}{7.7}{ppb}.

\paragraph{Magnetic Images $\delta_{n}^{\mathrm{img}}$}

Boundary conditions for the magnetic flux density at the iron pole faces of the storage ring slightly complicate the material effect considerations described in Sec.~\ref{subsubsubsec:pp_material_effect}. While $\delta^{\mathrm{abs}}$ provides the full correction to find the absolute magnetic field measured by the probe at the reference temperature and in vacuum far from other materials, the probe is in fact used fairly close to the storage ring's iron poles. The boundary conditions for the magnetic field at the interface between low- and high-permeability materials create an effect analogous to ``image charges'' in electrostatics near the boundary between materials of differing permittivity. The magnetic images appear as a reflection of the perturbation of the material magnetization across the pole face, so their size was measured similarly to the material effect but by placing the calibration probe at an appropriate distance next to the test probe. Simulations translate this measurement to the various distances to the pole face at which the calibration probe was used and extrapolate the infinite series of images-of-images, since the probe is in fact between the two poles, each of which contributes image perturbations.

\begin{table}[ht]
    \centering
    \caption{Corrections and uncertainties caused by different magnetic field environments in the test magnet and in the storage ring magnet sum up to $\delta^{\mathrm{env}}$. $^\dagger$ probe specific corrections and uncertainties have been averaged, accounting for the beam shape distribution.}
    \label{tb:wp:envCalib}
    \begin{tabular}{llrr}
        \toprule
        Description & & Corrections & Uncertainties \\
                    & & (ppb) & (ppb) \\
        \midrule
        Oxygen effect & $\delta^\mathrm{oxygen} $ & -2.0 & 1.0 \\
        Stick's field environment & $\delta^\mathrm{stick}$ & 0.7 & 0.2 \\
        Footprint trolley & $\delta^\mathrm{fp,trly}$ & 14.3 & 8.0 \\
Footprint CP$^\dagger$ & $\delta^\mathrm{fp,CP}$ & -5.8 & 7.7 \\
        Mag. images$^\dagger$ & $\delta^\mathrm{img}$ & -18.3 & 8.9 \\
        \midrule
        Subtotal & $\delta^{\mathrm{env}}$ & -11.1 & 14.3 \\
        \bottomrule
    \end{tabular}
\end{table}

\subsubsection{Trolley calibration: $\delta^{\mathrm{trans}}$} \label{sec:wp:trlyCalib}

Before and after each running period, the absolute calibration is transferred from the water-based cylindrical calibration probe, which is mounted {\it in-situ} on a 3D stage, to the trolley's $17$ \ac{NMR} probes.
This transfer is achieved by positioning both probe types at identical locations within the storage ring, ensuring they experience the same external magnetic field. 
The measured frequency differences between the calibration probe and each trolley probe under these identical field conditions yield the probe-specific ($n$) transfer constants $\delta_{n,k}^{\mathrm{trans}}$ in several campaigns $k$, given by
\begin{eqnarray} \label{eq:delta-trans}
    \delta_{n,k}^{\mathrm{trans}} & = & \left.\frac{\omega_{n,k}^{\mathrm{cp}}(T_{n,k}^{\mathrm{cp}}) - \omega_{n,k}^{\mathrm{tr}}(T_{n,k}^{\mathrm{tr}})}{\omega_{n,k}^{\mathrm{cp}}(T_{n,k}^{\mathrm{cp}})}\right|_{o.c.} + \delta_{n,k}^\mathrm{align} + \delta^\mathrm{av} \nonumber \\
    & & + \delta^\mathrm{freq,cp}  + \delta^{\mathrm{T}}(T_n) + (\delta^{\mathrm{b}}(T_n)-\delta^{\mathrm{b}}(T_r)) \nonumber \\
    & & + \delta_{n,k}^{T,\mathrm{trly}}(T^{\mathrm{tr}}_{n,k}) + \delta^\mathrm{var},
\end{eqnarray}
where the subscript ``o.c.'' indicates that oscillation corrections, accounting for field drifts during the measurement, have been applied to the frequency difference.

\paragraph*{Probe alignment: } 
The calibration procedure uses $200$ surface coils that span the full ring and are located at the top and bottom of the storage volume between the vacuum chambers and the magnet pole pieces. These coils allow the superposition of arbitrary transverse magnetic field gradients inside the muon storage volume. During data-taking periods, they are used to shim the azimuthally averaged magnetic field. At the dedicated calibration region, where the calibration probe is located, additional calibration-specific coils that are normally inactive allow linear gradients to be superimposed along the azimuthal direction.

These coils serve two functions for the calibration. First, they locate the \ac{NMR} probes precisely in space by superimposing first-order linear gradients. By toggling such gradients on and off for each direction, the difference in the observed field (\ac{NMR} frequency) uniquely defines the probe location along the gradient axis. Typically, gradients around \SIrange{0.5}{0.7}{ppm/mm} in the transverse plane and \SIrange{0.1}{0.2}{ppm/mm} in the azimuthal direction are used. This allows the precise location of the trolley probes to be determined and the calibration probe to be aligned with them at the sub-millimeter level.

\paragraph*{Field shimming}
Second, the same coils are used to locally shim the magnetic field around each calibration location, as much as possible, thereby minimizing systematic effects from imperfect alignment between the calibration probe and trolley probes. This local shimming is achieved by mapping the local field with the calibration probe itself. Typical linear gradients of approximately \SI{10}{ppb/mm} or below are achieved, with some outliers exhibiting larger local gradients. Combined, this procedure results in per-probe corrections and uncertainties from alignment of \SIrange{-20}{24}{ppb} and \SIrange{0}{30}{ppb}, respectively.

\paragraph*{Probe swapping}
The frequency difference between the two probes is measured in multiple iterations, with one probe moved to the calibration location and the other retracted far enough to minimize cross-talk contributions. 
The differences are extracted from ABA-scheme~\cite{Swanson_2010} measurements that correct for linear field drift. One such swap typically took around \SIrange{9}{12}{minutes}, while the trolley measures for around \SI{30}{\second} at its nominal \SI{2}{Hz} sampling frequency, and the calibration probe takes between $6$ and $10$ measurements separated by a few seconds. The position repeatability of the 3D stage was \SI{0.2}{\milli\meter}, and the trolley's positioning accuracy was \SI{0.6}{\milli\meter}. Common-mode field drifts are suppressed by a calibration-specific setting of the active feedback system that excludes measurements in the vicinity of the calibration region, and the remaining effects are tracked by the fixed installed \ac{NMR} probes and corrected for in the analysis.

Corrections and uncertainties are summarized in Table~\ref{tb:wp:trlyCalib}, and the individual terms account for:

\paragraph{Alignment and drift correction $\delta^\mathrm{align}$}
Regular fiducial marks on the bottom of the vacuum chambers allow the trolley's azimuthal position to be determined at the \SI{0.3}{\milli\meter} level and corrected for accordingly.
In addition, monitoring data from fixed \ac{NMR} probes, which have good resolution and are positioned far enough from the calibration region to avoid being influenced by local probe motion, are used to correct for higher-order common-mode field drifts within measurement series. The typical per-probe spread from different corrected swaps was \SIrange{2}{9}{ppb}.

\paragraph{Active volume correction $\delta^\mathrm{av}$}
The pulsed \ac{NMR} probes pick up the \ac{FID} from spins precessing in the sample material. The spin-flip and pick-up efficiency depend on the geometry of the probe coils. Hence, the smaller trolley \ac{NMR} probes with 32-turn coils of diameter \SI{5}{\milli\meter} and length \SI{15}{\milli\meter} surrounding petroleum jelly samples of diameter \SI{2.5}{\milli\meter} sample slightly different field regions than the larger calibration probe with a 5.5-turn coil of diameter \SI{15}{\milli\meter} and a length of \SI{14}{\milli\meter} surrounding the \SI{4.2}{\milli\meter} diameter, \SI{229}{\milli\meter} long water sample. While the two types of probes would measure exactly the same frequency in a perfectly uniform field (after the set of corrections described in the previous sections), and linear gradients cancel out due to symmetry, higher-order gradients can lead to different frequency shifts even with perfect alignment of probe centers. This occurs because the probes weight slightly different volumes differently.

For this publication, this effect was studied in greater detail using dedicated measurements in the test solenoid, with careful shimming and a well-controlled setup capable of applying linear and quadratic field gradients along the probe. Due to the probes' coils and the sample's long extent in the azimuthal direction and comparably small transverse dimensions, they are most sensitive to gradients along their cylinder axis. By applying quadratic gradients of the form $B_y = \frac{\partial^2 B}{\partial z^2}z^2 + B_0$ to both probes separately in these gradient-on--gradient-off studies, with $\frac{\partial^2 B}{\partial z^2}\in[-3.6, 3.6]\text{~ppb/mm}^2$, it was possible to calculate the necessary correction to account for the difference in probe volume and sensitivity needed for the gradients measured during any given trolley probe calibration.

The difference between the two probe types is characterized by a correction of the form
\begin{eqnarray}
    \delta^{\text{active-volume}} = \SI{8.5(3.6)}{\milli\meter\squared} \times \frac{\partial^2 B}{\partial z^2}, \end{eqnarray} where $\frac{\delta^2 B}{\delta z^2}$ is the corresponding second-order gradient along the azimuthal direction. 
This leads to per-probe corrections and uncertainties of \SIrange{-21}{100}{ppb} and \SIrange{0}{74}{ppb}, respectively.   

\paragraph{Frequency extraction and cross-check $\delta^\mathrm{freq,cp}$}
The frequency extraction of the calibration probe is based on a zero-crossing algorithm, and that of the trolley on a Hilbert transform algorithm. The uncertainties in the frequency extraction depend on the actual magnetic field gradients during rapid swapping.

\begin{figure}
    \centering
\includegraphics[width=1.0\linewidth]{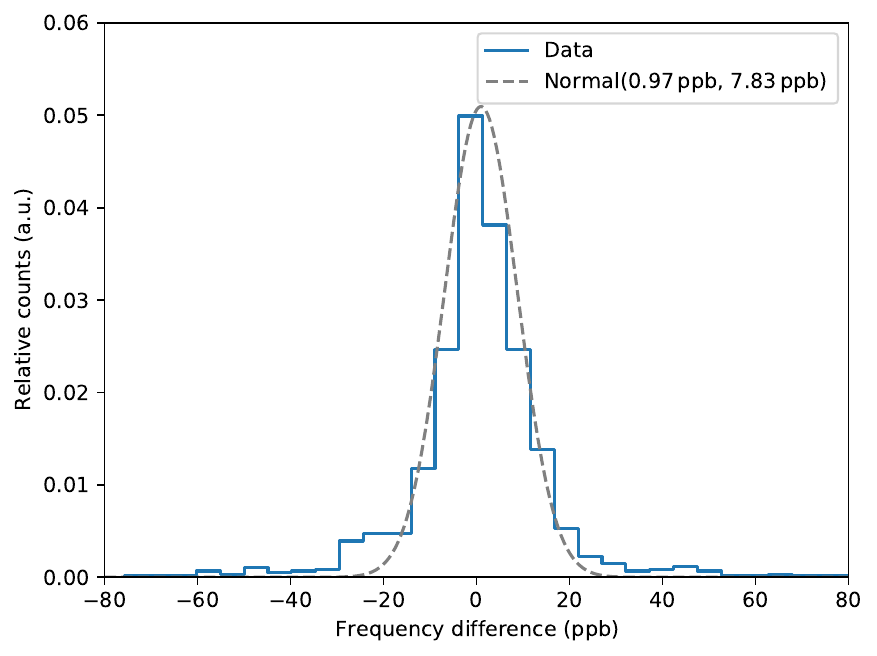}
    \caption{Comparison of absolute probe frequencies in the spin echo pulsing scheme. The distribution of frequency difference between the echo and $f_0$ (the standard frequency extraction) is centered within \SI{1}{ppb} of 0, with a standard deviation of $<$\SI{15}{ppb}, indicating no observable bias in our standard method. The significant width of this distribution is expected due to the short fits used to extract $f_0$ in these signals.}
    \label{fig:f_echo}
\end{figure}

In the post-\RunSix calibration, a second \ac{RF} pulsing scheme was implemented for the absolute calibration probe to be compared to the standard pulsing scheme. In this new scheme, a second \ac{RF} pulse sent \SI{14}{ms} after the initial pulse induces a spin-flip of $\pi$, leading to a re-coherence of the \ac{FID} signal called the spin echo. The frequency extracted from the standard \ac{FID} signal and that extracted at the spin echo are consistent at the \SI{1}{\hertz} level, verifying the standard frequency extraction on the same scale as the overall precision of the calibration measurements. The primary limitation of this comparison was field drift during the calibration campaign. The comparison was therefore performed by extracting the frequency from both the first \SI{13}{ms} of each signal, which behaves like a normal \ac{FID}, and from the spin echo within the same measurement. The distribution of the difference between these frequencies is shown in Fig.~\ref{fig:f_echo}. The short \SI{13}{ms} fit window leads to a larger uncertainty in the extracted frequency than the usual fits to \ac{FID}s, which are generally $\mathcal{O}$(\SI{100}{ms}). Over a set of 1140 normal \ac{FID}s taken in this campaign, the frequencies extracted using a \SI{13}{ms} fit and those using a fit to half of the signal's characteristic relaxation time show a spread with standard deviation \SI{14}{ppb} but well centered on 0, similar to the comparison of echo frequencies to standard frequencies in echo runs with a standard deviation of \SI{15}{ppb}.

\paragraph{Temperature dependence of diamagnetic shielding $\delta^{\mathrm{T}}$}
The diamagnetic shielding of protons in water has a significant temperature dependence~\cite{PetleyH2Otemp}. 
The corresponding correction from $T_n$, the calibration-probe temperature during the transfer measurement for trolley probe $n$, to the reference temperature $T_r$, has the form 
\begin{equation}
    \delta^{\mathrm{T}}(T_n)=(+10.36\pm 0.30)\times \frac{10^{-9}}{\SI{}{\celsius}} (T_n-T_r).
\end{equation}
The calibration probe's temperature was measured with a platinum resistive temperature device with an accuracy of \SI{0.5}{\celsius} during each \ac{NMR} measurement. In our previous work, a reference temperature of \SI{34.7}{\celsius} was used, resulting in corrections of order \SI{100}{ppb} for this term. However, by using the newest CODATA standard reference temperature of \SI{25}{\celsius}, these corrections are now generally an order of magnitude smaller.

\paragraph{Temperature dependence of bulk magnetic susceptibility ($\delta^{\mathrm{b}}(T_n)-\delta^{\mathrm{b}}(T_r))$}
Analogous to the diamagnetic shielding, the temperature dependence of the bulk susceptibility must be corrected for temperature differences between the probe and the reference. The temperature dependence of the magnetic susceptibility of water is much smaller than that of the diamagnetic shielding, so these corrections are $\mathcal{O}(1)$~ppb.

\paragraph{Temperature dependence of trolley probes $\delta^{\mathrm{T},\mathrm{trly}}$}
Analogous to the temperature dependence of the calibration probe, the trolley probes need to be corrected for the probe's temperature differences relative to the reference. The majority of the temperature difference of $\mathcal{O}(\SI{10}{\kelvin})$ is corrected in the magnetic field maps. Here we correct for a $\mathcal{O}(\SI{1}{\kelvin})$ temperature difference to the mean temperature during calibration. Since the major effect is corrected in the magnetic field maps, we do not assign an uncertainty here.

\paragraph{Calibration campaign combination $\delta^\mathrm{var}$}
No significant time dependence was observed in $\delta^{\mathrm{trans}}$ across all calibration measurements, which were performed before and after each run period. 
However, as in earlier analyses, the spread between values obtained in different calibration campaigns is larger than expected from uncertainty estimates alone. 
An additional correlated uncertainty of \SI{11}{ppb} was introduced to achieve a reduced chi-squared of unity, a value consistent with earlier analyses.
The $\delta_{n}^{\mathrm{trans}}$ values from different campaigns are combined into a single value per probe using an uncertainty-weighted average. The post-\RunSix data were not included in this average and were only used as cross-checks. 
Two independent analyses performed on a subset of the data were used to cross-check the analysis chain.

\begin{table}[!htbp]
    \centering
    \caption{Uncertainties from the calibration procedure on the muon-weighted field. Each uncertainty is labeled as uncorrelated (Uncorr.) or correlated (Corr.) across probes.}
    \label{tb:wp:trlyCalib}
    \begin{tabular}{llr}
    \toprule
    Description & &  Uncertainty\\
    & & (ppb) \\
    \midrule
    Swapping and misalignment & $\delta^{\mathrm{align}}$ (Uncorr.) & 0.7 \\
    Frequency extraction CP & $\delta^{\mathrm{freq,cp}}$ (Uncorr.) & 1.2\\
    Active volume & $\delta^{\mathrm{av}}$ (Corr.) & 11.0 \\
    Temp. of diamag. shielding & $\delta^{\mathrm{T}}$ (Corr.) & 5.2 \\
Variance & $\delta^{\mathrm{var}}$ (Corr.) & 11.4 \\
    \midrule
    Subtotal & $\delta^{\mathrm{trans}}$ (Tot.) & 16.2 \\
    \bottomrule
    \end{tabular}
\end{table}

\begin{figure}
    \centering \includegraphics[width=1.0\linewidth]{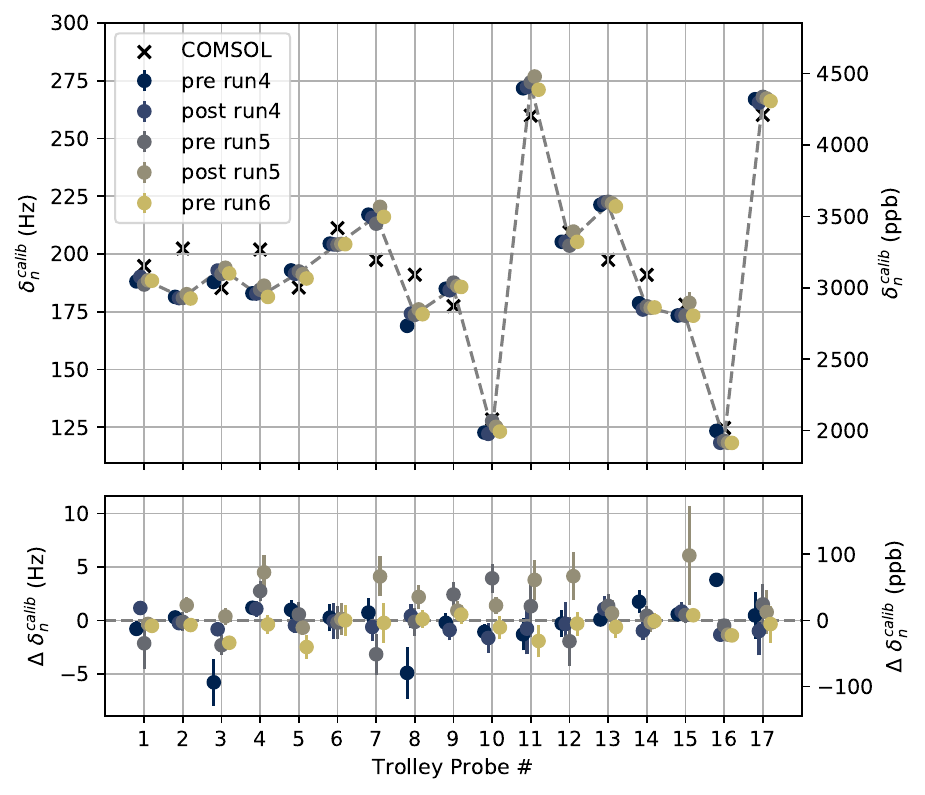}
    \caption{Calibration constants of each trolley probe determined in different calibration campaigns. COMSOL simulations of the trolley shell material effect are shown for comparison. The bottom plot shows the residuals with respect to the combined \RunFourFiveSix calibration constants.}
    \label{fig:wp:calibration}
\end{figure}

Figure~\ref{fig:wp:calibration} gives an overview of all $\delta_{n}^{\mathrm{calib}}$ and shows a comparison to the simulated effect in COMSOL from a simplified trolley geometry. 
While Table~\ref{tb:wp:trlyCalib} summarizes the uncertainty contributions from the individual effects on the muon-weighted field, Table~\ref{tb:PPCorrections} in the appendix shows the probe-specific corrections and uncertainties.
The individual probe corrections due to temperature dependence of the diamagnetic shielding range from \SI{-126.3}{ppb} to \SI{-59.1}{ppb}.

The uncertainty on the calibration constants is dominated by the additional \SI{25}{ppb} uncertainties introduced by the cross-checks of the absolute calibration. The calibration constants are directly applied to the raw trolley frequency measurements as defined in Eq.~\eqref{eq:wp:calib}, used to construct spatial magnetic field maps in the following section.
In summary, the calibration procedure yields probe-specific corrections of approximately \SI{1.5}{ppm}, dominated by sample-shape correction, with a total systematic uncertainty dominated by the \SI{25}{ppb} inflation derived from cross-check measurements, and \SI{16.2}{ppb} from the calibration procedure itself.

\subsection{Interpolation and muon weighting} \label{sec:wp:interpol}

\subsubsection{Spatial field maps}
This section describes the detailed extraction of the field maps $\omega_p^{\prime}(x,y,\varphi,t)$ from measurements of a trolley run. The field map measurement consists of the determination of the trolley probe's position and the determination of the \ac{NMR} frequency at each time.

To determine the trolley probe's position at each time, the transverse position and the azimuthal position are reconstructed.
The transverse positions inside the trolley are fixed by the probe locations, while the radial position is constrained by the trolley rails. 
In this analysis, variations in the rail's transverse position are directly accounted for. The trolley azimuthal position is determined by reading the barcodes etched into the bottom of the vacuum chamber. A detailed reimplementation of the position extraction algorithm, cross-checked with a machine learning method, validated the position extraction approach. Encoders that measure the length of the trolley cables are a backup; however, their precision is inferior to that of the barcode due to cable tension variations. In total, $195$ trolley runs were conducted for the data included in the $R_{\mu}$ measurement, using the field map acquired during \ac{CCW} motion for all except one run where \ac{CW} motion was used.

The trolley-probe \ac{NMR} active sample is affected by temperature variations over the course of the measurement. Measured frequencies are corrected to the mean temperature during the calibration (\SI{33.1}{\celsius}) using a correction of $\SI{-0.8\pm0.2}{ppb\per\celsius}$ as found in previous studies~\cite{PhysRevD.110.032009}.
The corrected frequencies are interpolated to a grid of azimuthal positions $\varphi_k(t)$. Different interpolation schemes were tested and agreed within \SI{1}{ppb}.

\begin{figure}[th]
    \centering
\includegraphics[width=1.\linewidth]{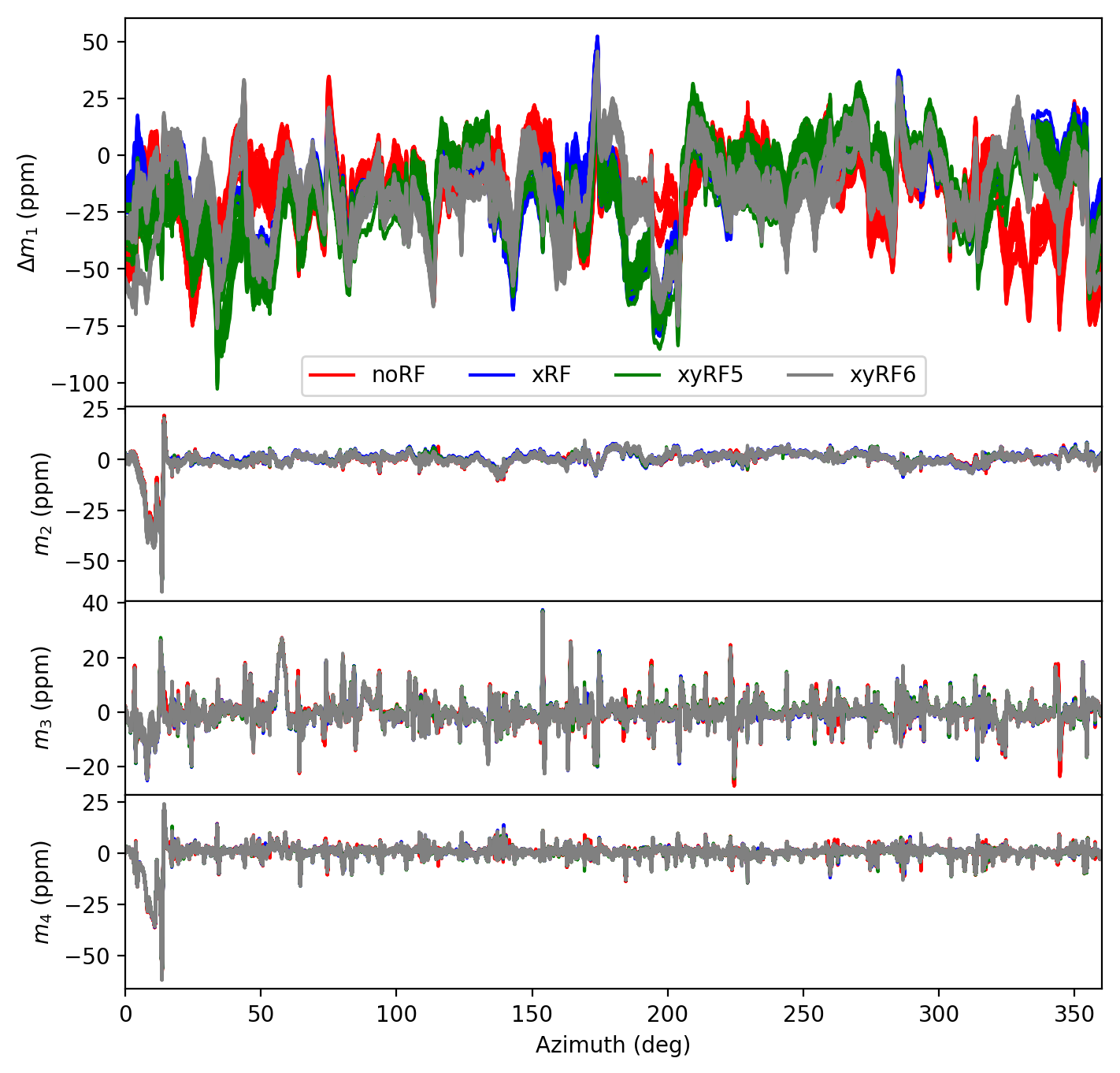}
    \caption{Representative field maps of $m_1, m_2, m_3, m_4$ as function of azimuth $\varphi$. $m_1$ is shown w.r.t. an arbitrary offset.}
    \label{fig:wp:field_maps}
\end{figure}

The time-dependent trolley maps are parametrized as 
\begin{equation}
    \opprime(x,y,\varphi,t) = \sum_{i=1}^{N_\mathrm{max}} m_i(\varphi,t) f_i(r,\theta),
    \label{eq:multipole_decomposition}
\end{equation}
where
\begin{equation}
    f_i(r,\theta) = \left\{\begin{array}{ll}1 & \mathrm{for}\,\,i=1 \\
    \left(\frac{r}{r_0}\right)^{\tfrac{i}{2}}\cos\left(\frac{i}{2}\theta\right) & \mathrm{for\,\,even}\,\,i>1 \\
    \left(\frac{r}{r_0}\right)^{\tfrac{i-1}{2}}\sin\left(\frac{i-1}{2}\theta\right) & \mathrm{for\,\,odd}\,\,i>1 \end{array}\right. ,\label{eq:multipole_definition}
\end{equation}
is the source-free 2D multipole expansion around the muon storage center.
Here, $r_0 = \SI{4.5}{\centi\meter}$ is a reference radius, and $(r,\theta)$ the radial coordinates in the transverse plane: $x=r\cos(\theta)$ and $y=r\sin(\theta)$.
The multipole coefficients $m_i$ are extracted for each azimuthal slice $\varphi_k(t_\mathrm{tr})$ by fitting Eq.~\eqref{eq:multipole_decomposition} to the temperature-corrected frequencies and rail-position-corrected probe positions. The trolley translations and rotations used for these rail-position corrections were measured using laser-based metrology prior to final installation. Multipoles are extracted up to $N_\mathrm{max}=12$ as determined in previous analysis~\cite{PhysRevD.110.032009}. Representative field maps of $m_1(\varphi)$, $m_2(\varphi)$, $m_3(\varphi)$, $m_4(\varphi)$ for different trolley runs are shown in Fig.~\ref{fig:wp:field_maps}.

\begin{table*}[t]
    \centering
    \caption{Overview of the magnetic field mapping corrections and uncertainties for all datasets.}
    \label{tb:wp:field_map_uncertainties}
    \begin{tabular}{lrrrr}
    \toprule
     & \multicolumn{4}{c}{Correction (Uncertainty) (ppb)} \\
    Dataset & \RunFournoRF & \RunFiveX & \RunFiveXY & \RunSixXY \\
    \midrule
    Frequency extraction & (16.0) & (16.0) & (16.0) & (16.0) \\
    Motion effects & -15.0 (18.0) & -15.0 (18.0) & -15.0 (18.0) & -15.0 (18.0) \\
    Position & & & & \\
    ~~~Transverse & (7.9) & (8.5) & (7.4) & (8.9) \\
    ~~~Azimuthal & (4.0) & (4.0) & (4.0) & (4.0) \\
    Temperature & (14.3) & (14.1) & (13.0) & (15.9) \\
    Parametrization & & & & \\
    ~~~Residuals & (3.8) & (4.6) & (4.7) & (8.5) \\
    ~~~2D vs 3D & (1.0) & (1.0) & (1.0) & (1.0) \\
    Azimuthal averaging & (1.8) & (1.6) & (2.3) & (1.7) \\
    Configuration & & & & \\
    ~~~Ground loop effect & & & +0.1 (0.1) & +0.8 (0.4) \\
    ~~~Garage effect & -6.4 (2.3) & -6.3 (2.3) & -6.4 (2.3) & -6.3 (2.3) \\
    ~~~Collimator effect & (0.0) & (0.0) & (0.0) & (0.0) \\
    \midrule
    Subtotal & -21.4 (29.8) & -21.3 (29.9) &  -21.3 (29.2) & -20.5 (31.8) \\
\bottomrule
    \end{tabular}
\end{table*}

Corrections and uncertainties to the trolley multipole coefficients are presented in Table~\ref{tb:wp:field_map_uncertainties} and summarized below.

\paragraph{Trolley frequency extraction} 
The trolley \ac{NMR}-probe \ac{FID} analysis is described in Ref.~\cite{field_FID_frequency_extraction}. In summary, the phase function of the \ac{FID} is extracted by a Hilbert transform fit to a polynomial of varying order. The fit function is extrapolated to $t_\mathrm{FID}=0$, which yields the proton precession frequency averaged over the active volume. Detailed studies of polynomial order, fit range, and temperature dependence have been performed for the \RunTwoThree analysis~\cite{PhysRevD.110.032009}.

\paragraph{Trolley motion effects} 
The trolley motion in a nonuniform magnetic field induces eddy currents in the conducting components, most significantly the aluminum shell. We use estimates from previous publications~\cite{PhysRevD.110.032009}, obtained from comparisons of standard continuous-motion trolley runs with stop-and-go runs and of \ac{CW} and \ac{CCW} trolley runs.

\paragraph{Trolley transverse and azimuthal position} 
The trolley position is constrained in the transverse plane by the rails. A laser tracker was used to estimate the effects of rail distortions on the trolley's translations and rotations before the vacuum chambers were installed. The positions are corrected before the multipole decomposition. The corresponding uncertainties are estimated conservatively from a large set of randomly azimuthally shifted rail distortions, yielding a representative ensemble of distortion amplitudes.

The azimuthal trolley position is determined using the barcode, except for small gaps between adjacent vacuum chambers and for barcode errors, where cable-length encoders are used. A conservative upper bound of \SI{2}{\milli\meter} on the azimuthal position resolution at all locations is used to estimate the corresponding systematic uncertainty; correlations between resolution and azimuthal multipole variations are negligible.

\paragraph{Trolley temperature dependence} 
A dedicated study revealed a temperature dependence of the petroleum jelly frequency of \SI{-0.8\pm0.2}{ppb\per\celsius}. However, a conservative uncertainty of \SI{2}{ppb\per\celsius} is used, since the uncertainty is dominated by couplings with \ac{NMR} frequency extraction from temperature-induced changes in the \ac{FID}. An uncertainty of \SI{1}{\celsius} on the probe temperature is assumed, since only one probe is equipped with a temperature sensor. The temperature difference between field maps and calibration (around \SI{33.1}{\celsius}) is corrected for. The uncertainties on the muon-weighted average magnetic field for each dataset are calculated by propagating the temperature uncertainty through the full analysis chain.

\paragraph{Field parametrization} 
The truncation of the multipole parametrization in Eq.~\eqref{eq:multipole_decomposition} at $N_\mathrm{max}$ introduces additional uncertainties. The uncertainty is estimated from the residuals of the fit of Eq.~\eqref{eq:multipole_decomposition} weighted by the azimuthally averaged beam distribution within $\Delta l=\SIrange{1}{10}{\milli\meter}$ around each probe.
The use of 2D multipole expansion, which is only valid if there is no azimuthal magnetic field dependence, is found to be $<\SI{1}{ppb}$ effect compared to a full 3D expansion~\cite{PhysRevD.110.032009,bodwin_implementation_2019}.

\paragraph{Azimuthal averaging}
Uncertainties due to the interpolation between the finite number of azimuthal slices are estimated by comparing results from linear, quadratic, and cubic spline azimuthal interpolation.

\begin{figure}[ht]
    \centering
    \includegraphics[width=1.0\linewidth]{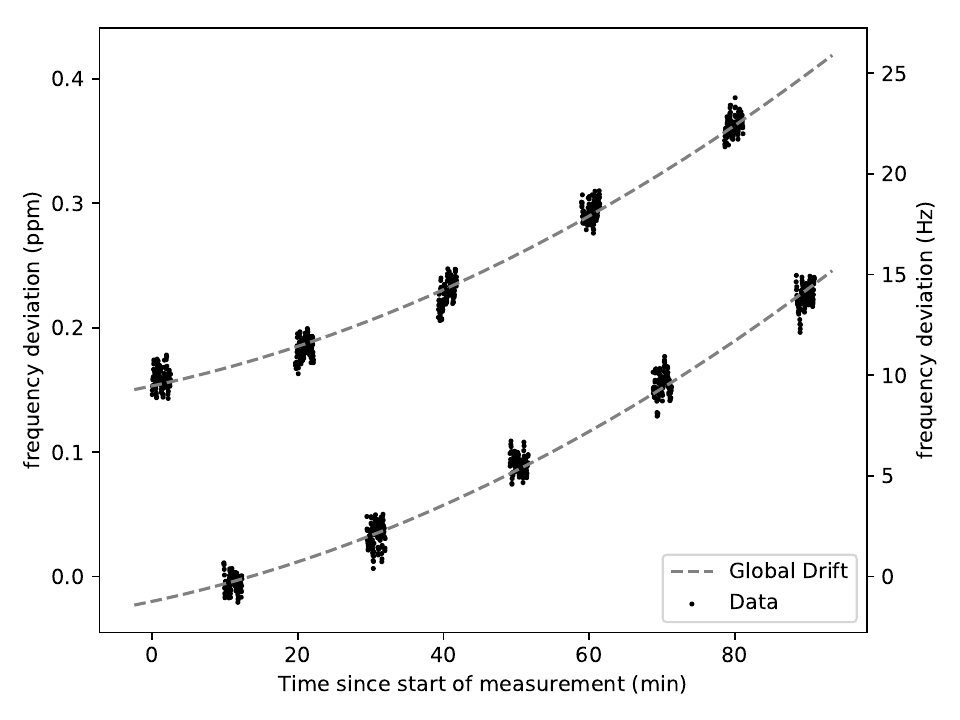}
    \caption{Measurement of magnetic field \ac{NMR} frequency with an \ac{NMR} probe fixed by an external holding structure inside the storage volume within the garage region. Data are collected while the garage rails are inserted and retracted alternately. The gray dashed line shows the magnetic field drift.}
    \label{fig:wp:garage_swapping}
\end{figure}

\paragraph{Difference in configuration}
During the field maps, the collimators that constrain the stored-muon distribution are retracted, and a section of the trolley rails that are used to insert the trolley into the storage volume is in a different position than when the muons are stored, and the trolley and corresponding piece of rails are moved radially inwards. Additional measurements significantly reduced the uncertainties on these corrections. Fig.~\ref{fig:wp:garage_swapping} shows measurements with an \ac{NMR} probe on a stiff holding structure inside the storage volume while the parking rails are inserted and retracted in the storage volume. A clear effect on the local magnetic field between the inserted and retracted states is visible. Additional measurements with 3D-printed \ac{PLA} rails enabled mapping of the entire parking region. 
These measurements reduced the uncertainty due to the parking mechanism by about a factor of $10$, since previous simulation estimates showed discrepancies relative to nearby fixed-probe data. Using a very similar setup with an \ac{NMR} probe on a stiff holding structure, it was confirmed that the effect of the collimators is negligible.

In parts of \RunFiveXY and \RunSixXY, a ground contact of a connector in the trolley pulling mechanism led to an additional ground loop whenever the trolley shell touched the grounded kicker plates. This ground loop reduced resistivity and caused a current imbalance, generating a small magnetic field that affected all fixed probe stations through which the trolley signal routes at this time. A similar effect was present in \RunOne and was studied in~\cite{Run1PRAField}. A correction and uncertainty based on the beam distribution are calculated.

\subsubsection{Time dependence}

Between two trolley field maps, the magnetic field is tracked by the fixed-probe system. The fixed probes are grouped into $72$ stations at azimuthal locations $\varphi_s$, which allows the extraction of the lowest-order gradients. The extraction of fixed probe multipole moments follows the same procedure as in earlier analysis~\cite{Run1PRAField,PhysRevD.110.032009}. This includes only tracking the lowest four multipoles by the fixed probes, linear interpolation of neighboring stations for three stations in the inflector region that have very short \ac{FID}s due to large gradients, and general data-quality cuts on \ac{FID} amplitude and power. 

The change of the magnetic field relative to a reference field map is 
\begin{equation}
    \Delta m_i^\mathrm{fp}(\varphi_s, t) = m_i^\mathrm{fp}(\varphi_s, t) - m_i^\mathrm{fp}(\varphi_s, t_s^\mathrm{tr})\,
\end{equation}
where $m_i^\mathrm{fp}(\varphi_s, t_s^\mathrm{tr})$ is the moment measured at a fixed probe station $s$ averaged around the time the trolley passes by that station. $t_s^\mathrm{tr}$ and $\varphi_s$ are obtained from the magnetic field perturbation caused by electronics and material of the trolley. The field perturbation due to the trolley is removed from fixed probe data and replaced by linear interpolation of $m_i^\mathrm{fp}(\varphi_s,t)$. 

The interpolated magnetic field moment at azimuthal location $\varphi$ and time $t$ is given by 
\begin{align}
    m_i(\varphi,t) = \sum_k W_k(t)&\left( \vphantom{\sum_j}  \right. m_i^\mathrm{tr}(\varphi,t_k) +  \nonumber \\ 
    & \left. \sum_s W_s(\varphi) \sum_j J_{ij}(\varphi_s)\Delta m_j^\mathrm{fp}(\varphi_s,t)\right)  
\end{align}
where $k$ labels the field maps; ideally, two consecutive runs are used, e.g., $k=1,2;2,3$ {\it etc.}. $W_k(t)$ is the temporal weighting of each trolley run: for linear interpolation between two consecutive runs, the weights transition from $(1,0)$ to $(0,1)$; if only one field map is available (e.g., due to an unplanned magnet ramp), $W_k(t)=1$. The azimuthal weighting factor $W_s(\varphi)$ interpolates between stations on either side of $\varphi$.
$J_{ij}(\varphi_s) = \frac{\partial m_i^\mathrm{tr}(\varphi_s)}{\partial m_j^\mathrm{fp}(\varphi_s)}$ is the Jacobian that relates small changes of the measured fixed probe moments to changes of the trolley moments for station $s$.

The multipole moments averaged over azimuth and weighted by the detected muons (see Section~\ref{sec:wp:muon_weightin}), including data quality cuts, $\langle m_i\rangle_{\varphi,t}$ are listed in Table~\ref{tb:wp:multipoles} for all datasets.
While the higher-order multipole moments such as $m_{10}/m_1$ and $m_{11}/m_1$ appear sizable, the corresponding muon-weighted sensitivity coefficients $k_i$ decrease rapidly with multipole order, ensuring convergence of the expansion. This is discussed in detail in Section~\ref{sec:wp:muon_weightin}.

\begin{table}[!htbp]
    \centering
    \caption{Azimuthal and muon weighted averaged multipole moments for each dataset in ppm with respect to the dipole $m_1$.}
    \label{tb:wp:multipoles}
\begin{tabular}{lrrrrr}
        \toprule
         Dataset & \RunFournoRF & \RunFiveX & \RunFiveXY & \RunSixXY & \RunFourFiveSix \\
         \midrule
$m_2/m_1$ & \num{    -0.0050} & \num{     0.0554} & \num{     0.0890} & \num{     0.0116} & \num{     0.0309} \\
$m_3/m_1$ & \num{    -0.0089} & \num{    -0.0393} & \num{     0.0596} & \num{    -0.0262} & \num{    -0.0041} \\
$m_4/m_1$ & \num{     0.0554} & \num{     0.0257} & \num{    -0.0195} & \num{     0.0195} & \num{     0.0270} \\
$m_5/m_1$ & \num{    -0.0212} & \num{    -0.0087} & \num{     0.0877} & \num{     0.0090} & \num{     0.0096} \\
$m_6/m_1$ & \num{     0.0214} & \num{     0.0173} & \num{     0.0490} & \num{     0.0274} & \num{     0.0273} \\
$m_7/m_1$ & \num{     0.0673} & \num{     0.0703} & \num{     0.0717} & \num{     0.0455} & \num{     0.0652} \\
$m_8/m_1$ & \num{     0.0138} & \num{     0.0120} & \num{     0.0010} & \num{     0.0016} & \num{     0.0086} \\
$m_9/m_1$ & \num{     0.0669} & \num{     0.0863} & \num{     0.0844} & \num{     0.0605} & \num{     0.0738} \\
$m_{10}/m_1$ & \num{    -0.7161} & \num{    -0.7264} & \num{    -0.7275} & \num{    -0.7277} & \num{    -0.7227} \\
$m_{11}/m_1$ & \num{    -0.1198} & \num{    -0.1269} & \num{    -0.1273} & \num{    -0.1339} & \num{    -0.1253} \\
    \bottomrule
    \end{tabular}
\end{table}

Corrections and uncertainties resulting from the interpolation using the fixed probe data are presented in Table~\ref{tb:wp:tracking_uncertainties} and summarized below.

\begin{table*}[t]
    \centering
    \caption{Overview of the magnetic field tracking uncertainties for all datasets. The subtotal lists statistical and systematic uncertainties separately. $^*$The tracking uncertainty is treated as a statistical uncertainty while all other uncertainties are treated as systematic.}
    \label{tb:wp:tracking_uncertainties}
\begin{tabular}{llrrrr}
    \toprule
     & & \multicolumn{4}{c}{Correction(Uncertainty) (ppb)} \\
    & & \RunFournoRF & \RunFiveX & \RunFiveXY & \RunSixXY \\
    \midrule
    Fixed probe resolution & & (1.0) & (1.0) & (1.0) & (1.0) \\
    Fixed probe temperature & & (0.0)  & (0.0) & (0.0) & (0.0) \\
    Tying uncertainty & & (10.2) & (9.9) & (10.0) & (9.9) \\
    Magnet ramp effect && -18.9 (2.7) & -10.5 (1.5) & -19.2 (2.7) & -4.7 (0.7) \\
    Tracking uncertainty$^*$ & & (7.6) & (12.1) & (11.2) & (14.3) \\
    Analysis choices & & (4.8) & (4.6) & (2.0) & (4.4) \\
    \midrule
    Subtotal & & -18.9 (7.6) (11.6) & -10.5 (12.1) (11.0)  & -19.2 (11.2) (10.6) & -4.7 (14.3) (10.9) \\
\bottomrule
    \end{tabular}
\end{table*}

\paragraph{Fixed probe resolution and temperature}
The relative frequency extraction of the fixed probes is robust with an uncertainty of $\sim\SI{1}{ppb}$, and uncertainties due to fixed probe temperature changes are negligible. 

\paragraph{Tying uncertainty} 
Two different strategies to replace the field perturbation caused by the trolley passing by a fixed probe station, based on local and global field drifts, give consistent results. Uncertainties are estimated by applying the replacement algorithm to regions where no trolley is present. Variation of the station position by $\sim\SI{0.25}{\degree}$ has an effect less than \SI{1}{ppb}.

\paragraph{Magnet ramp effect} 

\begin{figure}[ht]
    \centering
    \includegraphics[width=1.0\linewidth]{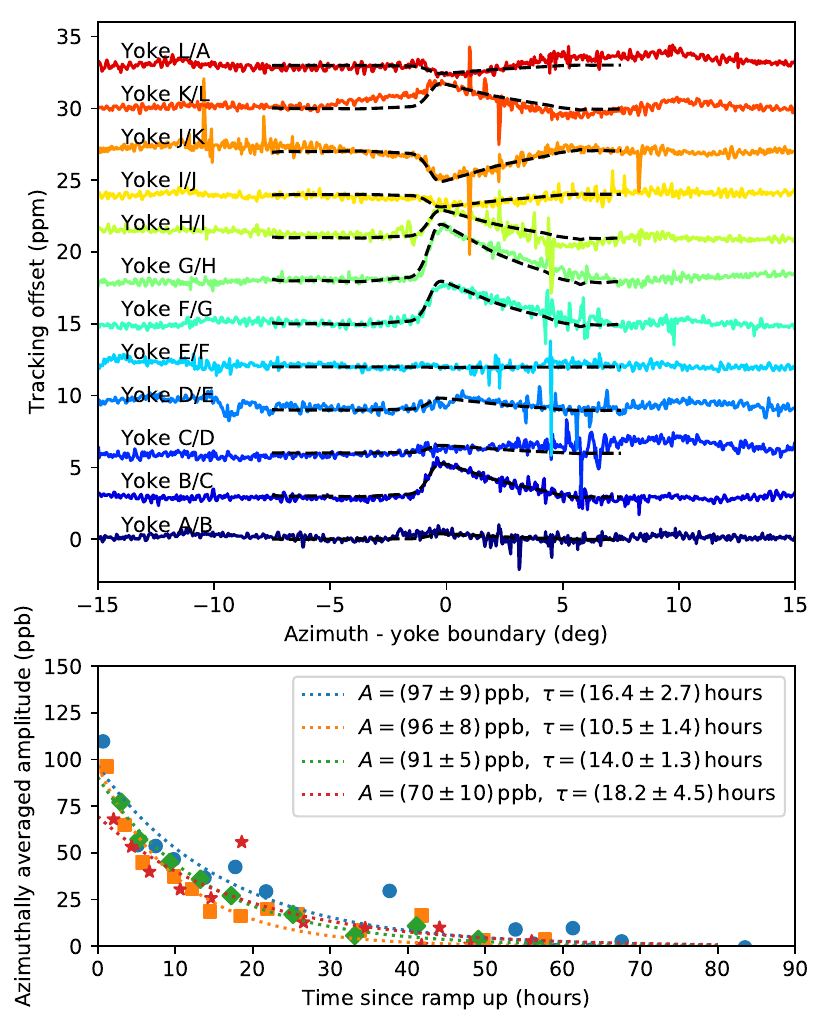}
    \caption{Untracked magnetic field changes after a magnet ramp up. The top panel shows the spatial effect at each yoke boundary fitted with a single template (black dashed line). The lower panel shows the amplitude of the azimuthally averaged effect as a function of time since ramp-up for four different magnet ramps.}
    \label{fig:wp:timeAfterRampEffect}
\end{figure}

Changes in the magnetic field that are not tracked by the fixed probe system lead to bias of $m_i(\varphi,t)$. To quantify this, predictions of the magnetic field moments from a single field map are compared with those from a subsequent field map. The difference between the prediction and the measured field moments of the second field map is called the {\it tracking offset}. 

In previous publications, an untracked change in the magnetic field was observed that depends on the time since the last ramp-up of the magnet. This effect has been studied in extensive systematic campaigns, in which the field maps were taken consecutively. The dominant part of this effect comes from reproducible magnetic field changes due to a slow mechanical relaxation of the pole positions at the boundaries between the twelve iron yokes after a magnet ramp. Systematic studies indicate that the spatial shape of this effect is the same at each yoke boundary, differing only in amplitude and sign. In the upper panel of Fig.~\ref{fig:wp:timeAfterRampEffect}, a single template has been fitted to each of the twelve yoke boundaries describing the spatial shape at all positions. 
The lower panel of Fig.~\ref{fig:wp:timeAfterRampEffect} shows the azimuthally averaged amplitude of the effect as a function of time since the last magnet ramp for four different magnet ramps. All four magnet ramps exhibit consistent amplitude and time constant for the effect. 
A \SI{210+-30}{ppb} amplitude and a \SI{15+-3}{\hour} time constant were used to estimate the correction and its uncertainty~\footnote{Note that after unblinding, we found that the amplitude was overestimated by a factor of $2$ due to a coding error. The corrected amplitudes are shown in Fig.~\ref{fig:wp:timeAfterRampEffect}.}.

\paragraph{Tracking uncertainty} 

\begin{figure}[ht]
    \centering
    \includegraphics[width=\linewidth]{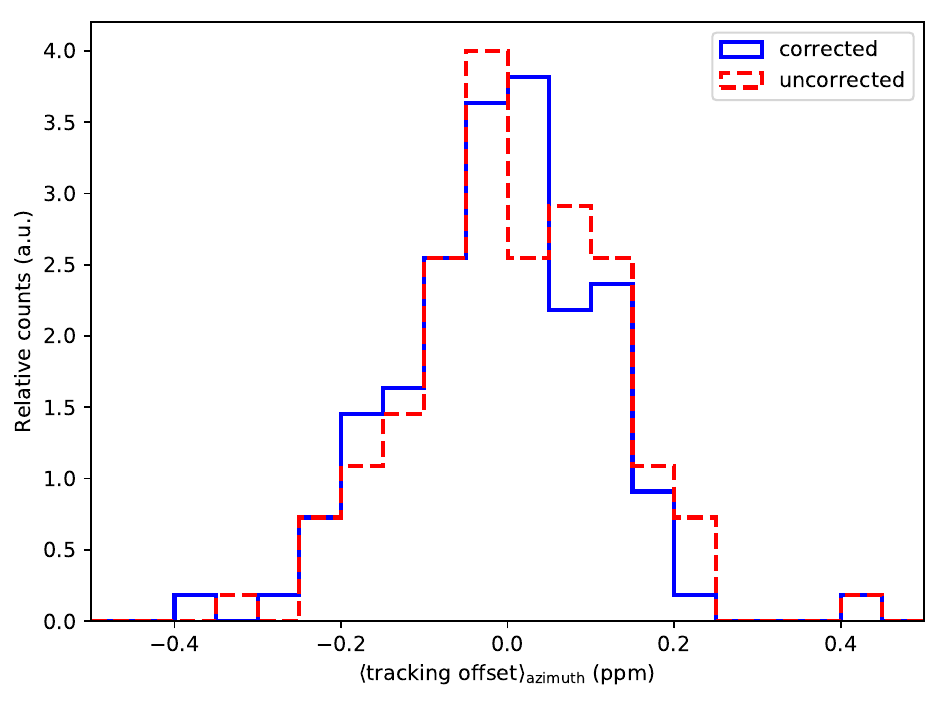}
    \caption{The azimuthally averaged tracking offsets used to determine the tracking uncertainty.}
    \label{fig:wp:syncOffsets}
\end{figure}

The azimuthally averaged synchronization offset for each field map pair, after correcting for the time-after-ramp effect, is histogrammed in Fig.~\ref{fig:wp:syncOffsets}. The synchronization offset is in agreement with zero and has a standard deviation of \SI{130}{ppb}. 

In previous publications, the remaining uncertainty was associated with a random walk arising from the limited tracking capabilities of the fixed-probe system. For a field period tied to two field maps, this process is described by a Brownian bridge model. The rate-of-change parameters for each multipole moment and their correlations are extracted from field periods that span two field maps but are tied to only one of them. The large dataset size allowed testing the Brownian bridge hypothesis, using data from two field periods that share a common field map. The prediction of a random walk model / Brownian bridge model is that the standard deviation increases with the length of the field period. However, no evidence for such behavior is observed in the data. Therefore, the result is cross-checked against an alternative model, in which the remaining difference is not attributable to a continuous random walk but to measurement uncertainties in the field maps. In that case, the width of the synchronization offset distribution is solely given by the repeatability of a field map measurement. The uncertainty in field tracking is evaluated with both the random-walk model and the constant-field-map uncertainty model. After averaging over all field periods, the results yield consistent uncertainties of \SI{5.3}{ppb} and \SI{6.0}{ppb}, respectively.

\paragraph{Analysis choices} 

The interpolation analysis was implemented by two independent teams. The two analyses were relatively unblinded in a step-by-step procedure. The analyses agree on the $\sim$\SI{4}{ppb}-level.

\subsubsection{Muon weighting}\label{sec:wp:muon_weightin}

The muon-weighted field is
\begin{equation}
    \opprimetilde = \frac{\int \opprime(x,y,\varphi,t) M(x,y,\varphi,t)\,\mathrm{d}x\,\mathrm{d}y\,\mathrm{d}\varphi\,\mathrm{d}t}{\int  M(x,y,\varphi,t)\,\mathrm{d}x\,\mathrm{d}y\,\mathrm{d}\varphi\,\mathrm{d}t}
\end{equation}
with the muon distribution $M(x,y,\varphi,t)$ determined by a combination of measurements with the straw trackers and modeling of beam dynamics. 
In principle, the magnetic field should include the positron detection efficiency rather than the muon distribution only; however, in practice the two yield consistent results within a few ppb.

\begin{figure}[ht]
    \centering
    \includegraphics[width=1.\linewidth]{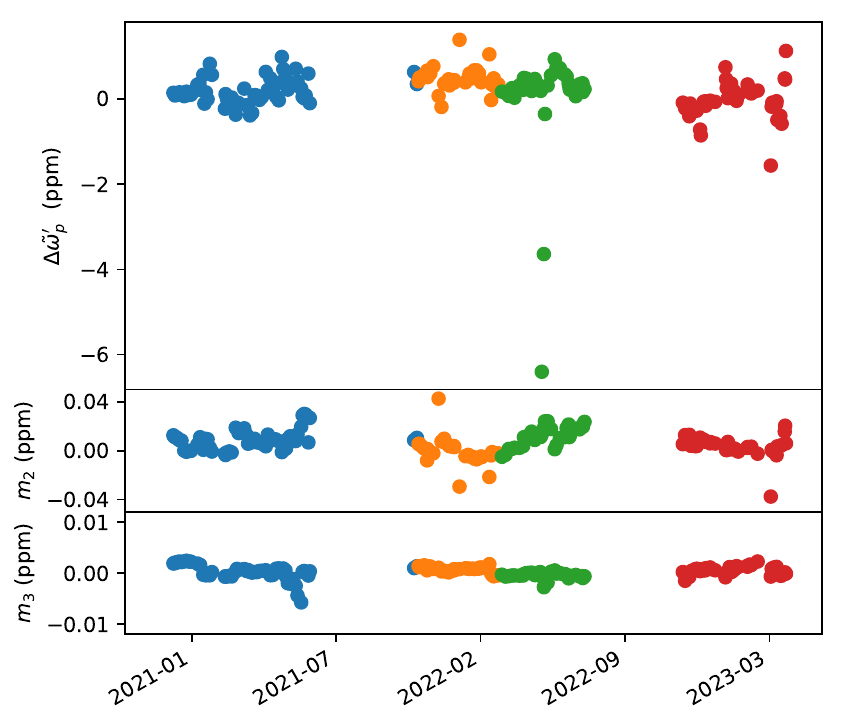}
    \caption{Muon weighted azimuthally averaged magnetic field (top) relative to an arbitrary offset and linear magnetic field gradients (middle/bottom) as a function of time over the full data taking period. 
    }
    \label{fig:wp:overview}
\end{figure}

Projecting the muon distribution in the basis introduced in Eq.~\eqref{eq:multipole_definition}, the muon beam moments $k_i$ are defined as
\begin{equation}
    k_i(\varphi,t) = \frac{\int M(x,y,\varphi,t) f_i(x,y) \,\mathrm{d}x\,\mathrm{d}y}{\int M(x,y,\varphi,t) \,\mathrm{d}x\,\mathrm{d}y}\,,
    \label{eq:beam_moments}
\end{equation}
and the muon weighted azimuthal- and time-dependent magnetic field is
\begin{equation}
    \opprimetilde(\varphi,t) = \sum_{i} m_i(\varphi,t)k_i(\varphi,t).
\end{equation}
The time-dependent azimuthally averaged field is
\begin{equation}
    \opprimetilde(t) = \frac{1}{2\pi}\int_{0}^{2\pi} \opprimetilde(\varphi,t)\,\mathrm{d}\varphi
\end{equation}
which is weighted by the number of detected muon decays and time-averaged over a few-day interval. The muon weighted, azimuthally averaged magnetic field is shown in Fig.~\ref{fig:wp:overview} as a function of time as well as the $m_2(t)$ and $m_3(t)$ components.

The spatial muon beam distribution $M(x,y,\varphi)$ is reconstructed from measured positron tracker profiles at two well-localized azimuthal positions. Tracker profiles $M^T(x, y)$ for the muon-weighted magnetic field are accumulated in time intervals $T_\mathrm{interval}$ spanning a few hours, considering only positrons within the analysis window of $\omega_a^{m}$ and corrected for detector resolution and acceptance. The time intervals $T_\mathrm{interval}$ are chosen to contain at least \num{6e5} total tracks and are optimized to minimize gaps between data taking, while staying within a trolley-run pair and containing entire $\omega_a$ DAQ runs.

The tracker profiles $M^T(x, y)$ are used to estimate the muon beam distribution at azimuthal locations where no tracker data are available by shifting the mean and scaling the width of the distribution. The beta functions and radial dispersion function are determined from \texttt{COSY INFINITY} simulation of the ring (see Sec.~\ref{sec:experiment_principle}). The mean and rms fractional momentum are extracted from the fast-rotation analysis described in Sec.~\ref{sec:fast_rotation}. Azimuthal variations in the vertical dipole component of the magnetic field cause a radial \ac{COD}, which shifts the ideal orbit away from the equilibrium position. 
The closed orbit for a given momentum is the equilibrium trajectory that a particle follows. 
The radial \ac{COD} is estimated based on the nominal radius, nominal field, effective field index, and the $N=1$ Fourier amplitudes and phase of the $m_1(\varphi)$. The Fourier components are extracted with \ac{DFT} from field maps in each $T_\mathrm{interval}$.

An azimuthally varying radial magnetic field would cause a vertical \ac{COD}, which we consider as a separate systematic. Misalignment of the \ac{ESQ} also causes radial and vertical \ac{COD}s by steering the beam. The corresponding systematic uncertainties are listed in Table~\ref{tb:field:muonWeighting}.

\begin{table*}[t]
    \centering
    \caption{Corrections and uncertainties due to spatial muon weighting of the magnetic field. }
    \label{tb:field:muonWeighting}
\begin{tabular}{l|rrrr}
    \toprule
     &  \multicolumn{4}{c}{Correction (Uncertainty) (ppb)} \\
    & \RunFournoRF & \RunFiveX & \RunFiveXY & \RunSixXY \\
    \midrule
    Detector effects  & & & & \\
    ~~Tracker acceptance  &  (0.1) & (0.1) & (0.2) & (0.0) \\
    ~~Tracker resolution  &  (0.0) & (0.1) & (0.1) & (0.1) \\ 
    ~~Tracker y-alignment & (0.6) & (0.1) & (1.4) & (0.0) \\
    ~~Tracker x-alignment &  (0.2) & (0.4) & (1.0) & (0.2) \\
    ~~Calorimeter acceptance & (0.2) & (0.0) & (0.1) & (0.1) \\
    Closed orbit distortion & & & & \\
    and azimuthal effects & & &  & \\
    ~~yCOD (radial B) & (2.4) & (2.1) & (1.6) & (2.1) \\
    ~~xCOD (quad misalig.) & +2.2 (5.7) & +1.3 (5.9) & +1.4 (6.0) & 1.9 (6.5)\\
    ~~yCOD (quad misalig.) & -0.3 (0.2) & -0.2 (0.1) & -0.2 (0.1) & -0.2 (0.1)\\
    ~~Mean momentum offset & (0.0)  & (0.0) & (0.0) & (0.0) \\
    ~~COSY & (0.0) & (0.0) & (0.0) & (0.0)\\
    \midrule
    Subtotal & +1.9(6.3) & +1.1(6.3) & +1.2(6.4) & +1.7(6.9) \\
    \bottomrule
    \end{tabular}
\end{table*}

\begin{figure*}
    \centering
    \includegraphics[width=0.38\linewidth]{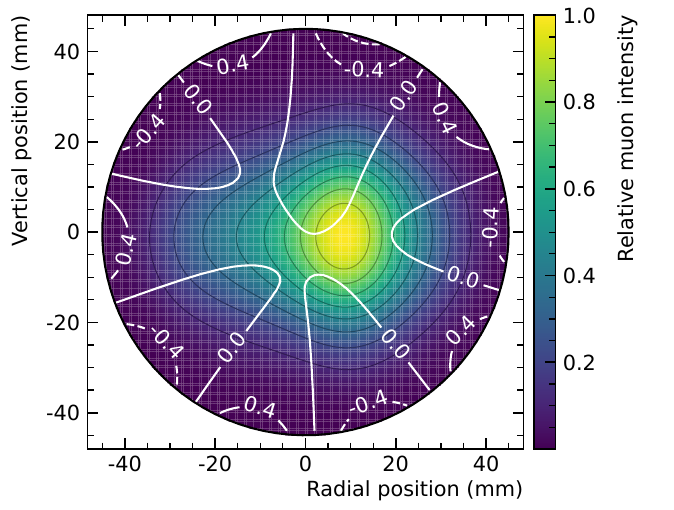}\includegraphics[width=0.38\linewidth]{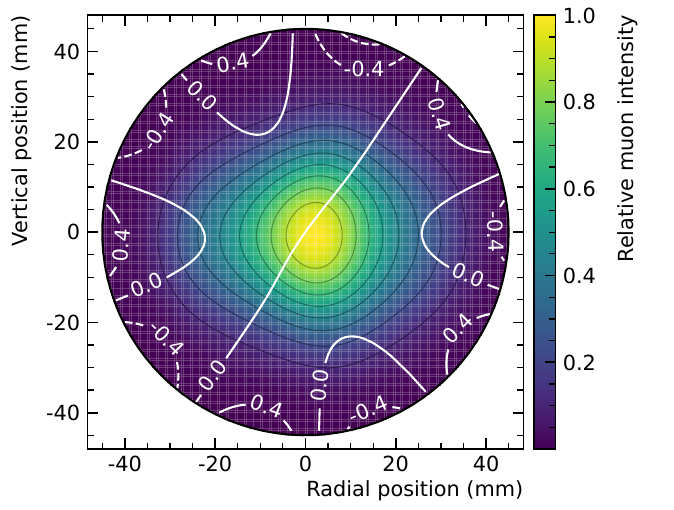}\\
    \includegraphics[width=0.38\linewidth]{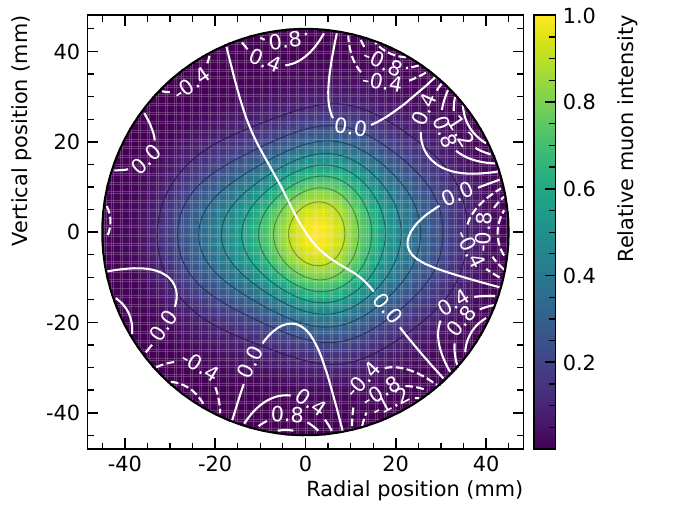}\includegraphics[width=0.38\linewidth]{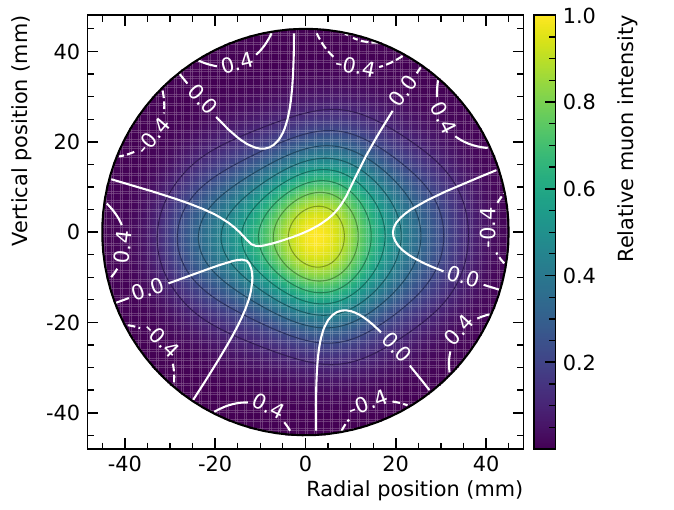}\caption{Azimuthally averaged and time averaged muon distribution and magnetic field contour lines for datasets noRF (top left), xRF (top right), xyRF5 (bottom left) and xyRF6 (bottom right). }
    \label{fig:wp:azimuthal_averaged_field_maps}
\end{figure*}

The beam moments defined in Eq.~\eqref{eq:beam_moments} averaged over azimuth and weighted by the number of muons in the storage region $N_\mu(t)$ are listed in Table~\ref{tb:wp:beamMoments}. Averaging all time intervals within a dataset, weighting by $N_\mu(t)$ and accounting for DQC cuts, yields the muon weighted magnetic field $\tilde\omega_p^\prime$ per dataset, which is listed in Table~\ref{tab:Rmu-results}. The azimuthal-averaged and time-averaged transverse muon distribution and magnetic field contours for all four datasets are shown in Fig.~\ref{fig:wp:azimuthal_averaged_field_maps}.
The muon weighting largely follows the methods used in our earlier publications~\cite{Run1PRAField,PhysRevD.110.032009}.
Combining the trolley maps with fixed-probe tracking and muon weighting provides the field $\opprimetilde$ experienced by the muons, yielding total uncertainties ranging from \SIrange{48}{51}{ppb} across the datasets.

\begin{table*}[t]
    \centering
    \caption{Muon distribution projections averaged over the main datasets.}
    \label{tb:wp:beamMoments}
\begin{tabular}{l *{5}{S[table-format=-1.2e-2]}}
        \toprule
& {\RunFournoRF} & {\RunFiveX} & {\RunFiveXY} & {\RunSixXY} & {\RunFourFiveSix} \\ 
         \midrule
        $k_1$    &  1.00e+00 &  1.00e+00 &  1.00e+00 &  1.00e+00 &  1.00e+00 \\
         $k_2$    &  8.39e-02 &  3.95e-02 &  4.79e-02 &  4.88e-02 &  6.06e-02 \\
         $k_3$    & -2.20e-02 & -1.77e-02 & -9.86e-03 & -2.38e-02 & -1.88e-02 \\
         $k_4$    &  4.76e-02 &  3.12e-02 &  3.56e-02 &  3.87e-02 &  3.99e-02 \\
         $k_5$    & -1.89e-03 &  3.65e-04 &  1.10e-03 & -3.54e-04 & -5.03e-04 \\
         $k_6$    & -6.15e-03 & -2.62e-03 & -2.90e-03 & -1.65e-03 & -3.93e-03 \\
         $k_7$    & -2.58e-03 & -1.37e-03 & -7.30e-04 & -2.46e-03 & -1.91e-03 \\
         $k_8$    &  1.84e-03 &  2.89e-03 &  3.25e-03 &  4.23e-03 &  2.77e-03 \\
         $k_9$    &  1.12e-03 &  5.98e-04 &  5.73e-04 &  6.26e-04 &  8.07e-04 \\
         $k_{10}$ & -1.05e-05 &  6.40e-04 &  7.23e-04 &  7.06e-04 &  4.08e-04 \\
         $k_{11}$ & -2.74e-04 & -2.61e-04 & -1.95e-04 & -5.16e-04 & -2.96e-04 \\
         $k_{12}$ &  6.73e-04 & -2.70e-04 &  6.13e-05 &  2.40e-04 &  2.64e-04 \\
        \bottomrule
    \end{tabular}
\end{table*}

\section{Magnetic field corrections}
The mapped, time-tracked, and muon-weighted magnetic field, $\langle \omega_p^\prime\times M\rangle$, is corrected for transient magnetic field perturbations that are not captured during the mapping process.  
These transient fields, which occur on a microsecond timescale and are synchronized with muon injection, arise from two main sources: eddy currents induced by the kicker magnets and the pulsed operation of the \ac{ESQ} system.
The kicker magnets generate eddy currents in the kicker plates and vacuum chambers that persist after the initial injection pulse.
Similarly, mechanical oscillations in the pulsed \ac{ESQ} system create time-varying field perturbations.
These transient fields affect the overall magnetic field experienced by the muons and are corrected for as indicated by $B_{K}$ and $B_{Q}$ in Eq.~\eqref{eq:R}.

\subsection{Transient fields from Kickers: $B_K$}
\label{sec:BK}

The strong $\sim$\SI{220}{G} pulsed magnetic field with a pulse length FWHM of $\sim$\SI{150}{\nano\second}, used to steer the muons, ideally only on their first passage, onto storage orbit, introduces eddy currents in the kicker plates and vacuum chambers. The kicker system operates by sending high-voltage pulses through a Blumlein transmission line, which generates $\sim$\SI{53}{kV} pulses that drive current flow (about \SI{4}{kA}) through aluminum kicker plates. The complete kicker pulse sequence consists of three components: a Blumlein charging pulse occurring approximately \SI{400}{\micro\second} before muon arrival, the fast steering kick itself, and the subsequent slow transient from eddy current decay. These eddy currents decay with time constants on the order of \SI{60}{\micro\second}. They produce transient magnetic fields of approximately \SI{15}{mG} that are still present during the times the muon spin precession is measured (starting around \SI{30}{\micro\second} after injection). This effect is not captured in field maps, which are obtained without the kicker active, nor by the asynchronous \ac{NMR}-based field-tracking system, and hence needs to be corrected for.

These magnetic transients are measured with Faraday-based magnetometers, exploiting the magnetic-field-dependent polarization rotation in \ac{TGG} crystals. Two different setups were used: a free-space laser system entering the vacuum chambers through a window and guided to the kicker center through mirrors, and a fiber-optic system using light guides to transport the laser to the TGG crystal. The free-space system employed a \SI{3}{mW}, \SI{630}{nm} linearly polarized laser, while the fiber-optic-based system used a \SI{405}{nm} diode laser with adjustable power output \SIrange{25}{35}{mW}. Additional measurements were taken in dedicated campaigns after the last muons were stored.

Both magnetometer systems were calibrated by exploiting the main magnet ramp-up and ramp-down cycles, using the large field change between \SI{0}{T} and the operational \SI{1.45}{T}. Drifts in magnetometer sensitivity over time were monitored and corrected by tracking the magnetic field produced by the Blumlein charging pulse, which provides a consistent, clear field as a stable reference signal. Measurements were performed both with and without the main \SI{1.45}{T} magnetic field. Large oscillations in the magnetometer signals were observed when the magnet was on that vanished when the magnet was off. The observed oscillations had two origins: currents induced in the kicker plates and cage producing real transient magnetic fields, and mechanical vibrations of the kicker structure driven by Lorentz forces causing artifacts in the optical measurements. In the free-space system, these were canceled using a quarter-wave plate technique that measured the transient at different circular polarization states, thereby largely suppressing vibrational artifacts through appropriate linear combinations while preserving the genuine magnetic transient signal. In the fiber-optic system, vibrations were identified in alternate measurements using the main magnetic field at \SI{1.3}{T}, at which the polarimetric response operates on the opposite slope, and were mitigated by a damping rig that suppressed the motion of the fiber-optic cables.

\begin{figure}
    \centering
    \includegraphics[width=0.98\linewidth]{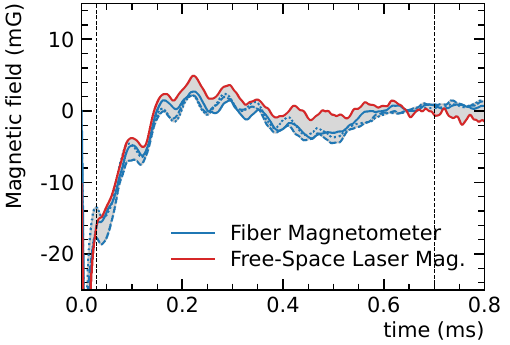}
    \caption{Magnetic perturbations due to kicker eddy currents as measured by the two magnetometers: Free laser magnetometer in red and fiber magnetometer in blue. The different blue lines correspond to measurements at different kicker plates and different measurement campaigns. The vertical dashed lines delimit the analysis window of the muon precession frequency measurement. The gray area indicates the spread between the measurements and is added to guide the eye.}
    \label{fig:kicker}
\end{figure}

Measurements at the kicker center show good agreement between different kicker segments and magnetometers, as shown in Fig.~\ref{fig:kicker}, as well as good agreement with earlier measurements. However, measurements at larger radial offsets, specifically at \SI{17.5}{mm}, showed a significant radial dependence, with transient amplitudes approximately $2.2$ times those at the central position.

Measurements using a mock-up of the kicker plates and vacuum chamber were made to develop a model of the transverse spatial dependence of the kicker transient field. The mock-up used the same materials and relative dimensions as the real kicker, though smaller by a factor of $1.21$. Magnetic fields in the mock-up were created by both running current pulses mimicking the kick shape through the mock kicker plates, and by running AC currents at \SIrange{2}{5}{MHz}. The resulting magnetic fields were measured over a grid of points in $x$ and $y$ using pickup coils. The contributions of eddy currents induced in the plates were isolated from the primary driving current by comparing the measured field maps with predictions from the Biot-Savart law for a uniformly distributed current through the kicker plates. The eddy currents produce strong current peaks at the edges of the kicker plates and lower currents in the midplane, resulting in significant changes in the field distribution relative to the uniform-current Biot-Savart law prediction. These eddy currents are expected to persist into the muon storage interval and decay over a time scale of \SIrange{50}{100}{\micro\second}.

\begin{figure*}
    \centering
    \includegraphics[width=0.98\linewidth]{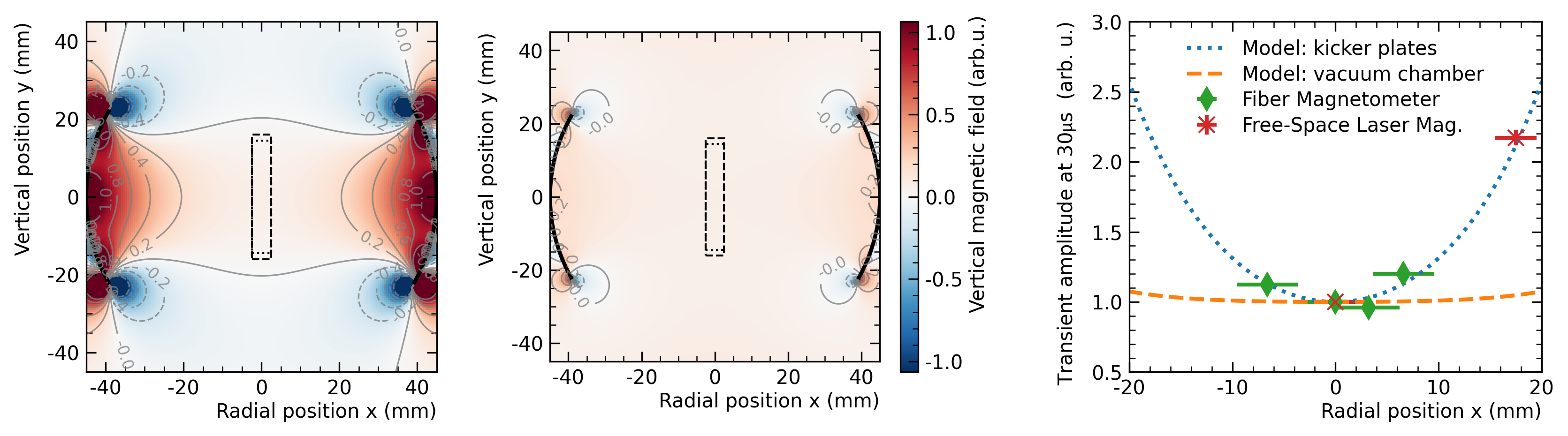}
    \caption{Spatial models for the transient magnetic field generated by eddy currents from the kicker system. The left panel shows the model based on eddy currents induced in the kicker plates alone, while the middle panel includes eddy currents in both the plates and the vacuum chamber. The color map indicates the relative vertical magnetic field with corresponding contour lines. The geometry of the kicker plates is outlined in solid black. The dashed and dotted rectangular regions represent the positions of the TGG crystals used for the measurements when centered. The right panel shows the effect integrated over the crystal volume in combination with measurements at crystals located at different radial positions.}
    \label{fig:kickerModel}
\end{figure*}

Using these measurements, two spatial models were developed: one based on eddy currents induced in the kicker plates alone, and another for eddy currents induced in both the kicker plates and the vacuum chamber, both shown in Fig.~\ref{fig:kickerModel}. The plates-only model agreed with the radial-offset measurements at FNAL of the kicker transient field, while the plates-and-chamber model disagreed (it predicted a stronger but flatter radial dependence). This was unexpected, as the real system is expected to have eddy currents in the plates and the chamber, and no explanation was found for why the chamber eddy currents might not contribute significantly. Rather than providing a definitive model of the spatial dependence, these results are used to set limits on the possible spatial dependence. The plates-only model, which is consistent with the observations, is used to calculate the central value of $B_{k}$. The change in $B_{k}$ predicted by the plates-and-chamber model is used as a spatial-modeling uncertainty.

The total correction $B_{k}$ is derived from these dedicated measurements and spatial models, weighted by the muon distribution. The correction is calculated by modeling how the time-dependent transient field affects the measured $\omega_a$ value through simulated positron count modulation. The muon storage region is divided into a spatial grid, with the positron count rate in each element given by 
\begin{eqnarray}
    N(t,x,y) &=& M(x,y)e^{-t/\tau_\mu} \times \\ 
    &&[1 + A \cos(\omega_a t + \omega_a \cdot K(t,x,y) + \phi_0)] \nonumber.
\end{eqnarray}
The term $K(t,x,y)$ quantifies the cumulative magnetic influence of the kicker transient. It is calculated by integrating the measured time-dependent field $B_k(t')$, as shown in Fig.~\ref{fig:kicker}, to determine the total magnetic impulse up to time $t$. This temporal profile is then mapped to the transverse coordinates $(x,y)$ via the spatial model and scaled to account for the azimuthal extent.
The total count rate is obtained by summing over all grid elements weighted by the measured muon distribution $M(x,y)$. This modified count rate is fitted to extract an altered $\omega_a$ value, and $B_k$ is computed to correct for this effect.

The systematic uncertainties in the $B_k$ correction are summarized in Table~\ref{tab:bk}. The dominant uncertainty of \SIrange{14}{17}{ppb} arises from differences between spatial models used to extrapolate the measured transients across the storage region. Variations in transient waveforms between kicker locations, different bunches, and measurement campaigns contribute \SIrange{9}{10}{ppb}, while residual and unsuppressed mechanical vibrations add \SIrange{10}{11}{ppb}. Potential misalignment of the \ac{TGG} crystals (\SI{2}{mm} radially, \SI{5}{mm} vertically) contributes \SI{2}{ppb}, and uncertainties in azimuthal coverage beyond the three kicker segments add \SIrange{6}{7}{ppb}. The absolute calibration precision using magnet ramps and Blumlein reference signals contributes less than \SI{1}{ppb}.

The better understanding of the magnetic transients from the kickers is applied to the \RunTwoThree $B_k$ values and uncertainties when combined with \RunFourFiveSix, leading to total shifts of \SIrange{-19}{-14}{ppb} and uncertainty increases of \SIrange{17}{24}{ppb}, respectively.

\begin{table}[htbp]
\centering
\caption{$B_k$ correction values and uncertainties for Run-4/5/6 datasets. All values are in ppb.}
\label{tab:bk}
\begin{tabular}{lrrrrr}
\toprule
 & noRF & xRF & xyRF5 & xyRF6 \\
 & (ppb) & (ppb) & (ppb) & (ppb) \\
\midrule
$B_k$ correction & -38.1 & -34.8 & -35.8 & -36.7 \\
\midrule
Uncertainties & & & & \\
Spatial model & 16.8 & 13.7 & 14.6 & 15.5 \\
Position & 2.0 & 1.8 & 1.9 & 1.9 \\
Trace shape & 9.9 & 9.1 & 9.3 & 9.6 \\
Vibration & 11.1 & 10.2 & 10.4 & 10.7 \\
Calibration & 0.5 & 0.4 & 0.4 & 0.4 \\
Azimuthal factor & 6.6 & 6.0 & 6.2 & 6.4 \\
\hline
Total & 23.5 & 20.3 & 21.2 & 22.1 \\
\bottomrule
\end{tabular}
\end{table}

\subsection{Transient fields from ESQ: $B_{Q}$}
\label{sec:BQ}

Transient fields induced by the pulsing of the \ac{ESQ} system were mapped out using vacuum-sealed \ac{NMR} probes mounted on a movable frame. 
This setup allowed for detailed measurements of the transient fields around the storage ring. 
The radial dependence of the effect was checked at a handful of locations around the ring.

The observed field perturbations exhibit characteristics of a driven oscillator with multiple overlapping intrinsic frequencies. 
The introduced bias is proportional to the amplitude of the oscillations at the time of muon injection and the local rate of field change over the muon fill (\SI{700}{\micro\second}), times the dilated muon lifetime, $\tau_{\mu}$. 

While the effect is clearly observable in the fixed installed \ac{NMR} probes due to skin-depth effects in the \SI{}{\centi\meter} thick aluminum vacuum chambers, it is not possible to achieve the required resolution to assess the effect without {\it in situ} mapping. 
The fixed probes confirm that the effect is constant over time between datasets. 
Since the effect is present in the fixed probe data, it is also present in the tracked field. 
However, the asynchronous operation of the \ac{NMR} probes relative to muon injection causes the effect to average out over time.

The resulting correction of
\begin{equation}
    B_Q = -21.0 \pm 19.5 \,{ppb}
\end{equation}
is the same as derived for the \RunTwoThree data and presented in Ref.~\cite{PhysRevD.110.032009}.  A detailed breakdown of the uncertainties is provided therein. 
Further analysis for this publication focused on two issues. The transverse dependence was scrutinized, but did not lead to any adjustments to the assigned uncertainties. Furthermore, the \ac{RF} modulation applied to the \ac{ESQ} system was found to have no effect on the observed magnetic field transient.

\section{Determination of \amu} \label{sec:amu}
The Muon \gm Experiment at \ac{FNAL} measures the ratio
\begin{eqnarray} \label{eq:RCopy}
    \Rmuprime(T_r) = \frac{\oa}{\opprimetilde(T_r)}
\end{eqnarray}
as defined in Eq.~\eqref{eq:Rmu} with
\begin{equation}
\oa = \oam\left(1+C^{}_{e}+C^{}_{p}+C^{}_{m l}+C^{}_{dd}+C^{}_{pa}\right)
\end{equation} and
\begin{equation}
\opprimetilde(T_r) = \langle \opprimeatTexp\times M\rangle(1+B^{}_K+B^{}_Q ) .
\end{equation}
The ratio \Rmuprime is evaluated for each dataset and then averaged over the full period under the assumption 
that systematic uncertainties of the same category are fully correlated among different datasets. 
The corresponding correlation matrix is reported in Table~\ref{tab:Rmu-ds-corr}, while the individual \oa, \opprimetilde and their ratio \Rmuprime are reported in Table~\ref{tab:Rmu-results}.
The ratios are statistically consistent and are fit to obtain the combined
\begin{equation}
\Rmuprime(T_r)_{\RunFourFiveSix} = 0.00370730090(42)_{\text{stat}}(28)_{\text{syst}},
\end{equation}
at the reference temperature $T_r=\SI{25}{\celsius}$; the fit yields $\chi^2/\text{dof}=0.97/3$ corresponding to $p(\chi^2)=\SI{81}{\percent}$.

\begin{table}
\caption{Correlation matrix of the \RunFourFiveSix datasets measurements of $\Rmuprime(T_r)$.}
\label{tab:Rmu-ds-corr}
\begin{center}
\begin{tabular}{lrrrr}
\toprule
\Rmuprime & 
\RunFournoRF & 
\RunFiveX & 
\RunFiveXY & 
\RunSixXY\\
\midrule
\RunFournoRF & 1.00 & 0.116 & 0.115 & 0.103\\
\RunFiveX & 0.116 & 1.00 & 0.086 &0.077\\
\RunFiveXY & 0.115&0.086 & 1.00 &0.076\\
\RunSixXY & 0.103&0.077&0.076 & 1.00\\
\bottomrule
\end{tabular}
\end{center}
\end{table}

\begin{table}
\begin{center}
\caption{\RunFourFiveSix datasets measurements of $\omega_a$, $\Tilde{\omega}'_p(T_r)$, and their ratios $\Rmuprime(T_r)$ multiplied by 1000.}
\label{tab:Rmu-results}
\begin{tabular}{lrrr}
\toprule
 Dataset & $\omega_a/2\pi$\,(Hz) & $\Tilde{\omega}'_p(T_r)/2\pi$\,(Hz) & \Rmuprime $\times 1000$\\
\midrule
\RunFournoRF & 229077.504(43) & 61790920.0(3.5) & 3.70730043(73)\\
\RunFiveX & 229077.626(56) & 61790938.9(3.5)  & 3.70730126(94)\\
\RunFiveXY & 229077.500(58) & 61790910.9(3.5)  & 3.70730090(96)\\
\RunSixXY & 229077.509(65) & 61790903.6(3.6) & 3.70730148(107)\\
\RunFourFiveSix  &  &  & 3.70730090(51)\\
\bottomrule
\end{tabular}
\end{center}
\end{table}
Table~\ref{tab:Run456Syst} summarizes the uncertainty and correction terms corresponding to Eq.~\eqref{eq:R}. The total shift applied to the raw measured frequency ratio is \SI{572}{ppb}, dominated by the electric field correction, $C_e$ (see Sec.~\ref{sec:efield}), followed by the pitch correction, $C_p$ (see Sec.~\ref{sec:pitch}). The total systematic uncertainty on the \RunFourFiveSix measurement is \SI{76}{ppb}, which surpasses the goal set in the 2018 Technical Design Report~\cite{grange2018muong2technicaldesign} of \SI{100}{ppb}.

\begin{table}
\caption{Values and uncertainties of the $\mathcal{R}_\mu'$ terms in Eq.~\eqref{eq:R} and uncertainties due to the external parameters in Eq.~\eqref{eq:amueqnewcodata} for $a_{\mu}$. Positive $C_i$ increase $a_{\mu}$; positive $B_i$ decrease $a_{\mu}$. The $\omega_{a}^{m}$ uncertainties are decomposed into statistical and systematic contributions.}
\label{tab:Run456Syst}
\begin{center}
\begin{ruledtabular}
\begin{tabular}{lrrr}
Quantity & Correction & Uncertainty & Section\\
          & (ppb) & (ppb) \\
\hline
$\omega_a^{m}$ statistical & - & 114 & \ref{sec:oam}\\
$\omega_a^{m}$ systematic & - & 30 &  \ref{sec:wa-systematic-uncertainties-subsection}\\
$C_{e}$ & 347 & 27 & \ref{sec:efield}\\
$C_{p}$ & 175 & 9 & \ref{sec:pitch} \\
$C_{ml}$ & 0 &  2 & 
\ref{sec:MuonLoss}\\
$C_{dd}$ & 26 & 27 & \ref{sec:DiffDec}\\
$C_{pa}$ & -33 & 15 & \ref{sec:Cpa} \\
\hline
$\langle \omega_p^\prime\times M\rangle$ & - & 48 & \ref{sec:op}\\
$B_{K}$ & -37 & 22 & \ref{sec:BK}\\
$B_{Q}$ & -21 & 20 & \ref{sec:BQ}\\
\hline
$\mu_{p}'(\SI{25}{\celsius})/\mu_{B}$ & - & 4 & \\
$m_{\mu}/m_{e}$ & - & 22 & \\
\hline
Total systematic & - & 76\\
Total external parameters & - & 22\\
Totals & 572 & 139 \\
\end{tabular}
\end{ruledtabular}
\end{center}
\end{table}

The enhanced statistical precision of this larger dataset, combined with dedicated measurements, led to the identification of three corrections with corresponding uncertainty adjustments to the previously published result: 
the sensitivity of \oa to small, slow gain shifts, improved understanding of spatial dependencies in the transient magnetic fields from kicker system eddy currents, and a sign error correction in one component of the differential decay correction. 
These corrections, determined independently, share the same sign
and combined to shift the \RunOne and \RunTwoThree results by
\SI{50}{ppb} and \SI{98}{ppb}, respectively, and result in a total systematic uncertainty of \SI{78}{ppb} for the adjusted \RunTwoThree
result. These corrections were finalized before unblinding the \RunFourFiveSix measurement. 
On top of that, 
the  $\Rmuprime$ values measured in \RunOne~\cite{Run1PRL} and \RunTwoThree~\cite{PhysRevLett.131.161802} need to be corrected to the new reference temperature yielding
$\Rmuprime(T_r=\SI{25}{\celsius})_{\text{Run-1}} = 0.0037073003(16)_{\text{stat}}(6)_{\text{syst}},$
$\Rmuprime(T_r=\SI{25}{\celsius})_{\text{Run-2/3}} = 0.00370730087(75)_{\text{stat}}(29)_{\text{syst}}.$
 Assuming that the systematic uncertainties are fully correlated, the combined Fermilab experimental measurement is 
 \begin{equation}
 \Rmuprime(T_r)_{\text{FNAL}} = 0.00370730088(36)_{\text{stat}}(29)_{\text{syst}}~.
 \end{equation} 

This value can be combined with the \ac{BNL} measurement of $R_\mu$ for free protons in vacuum~\cite{PhysRevD.73.072003} $R_\mu = 0.0037072063(20)$, after converting  to the shielded proton reference using the measured diamagnetic shielding correction $\sigma_{p'}(T_r)$~\cite{phillips_magnetic_1977} to obtain
\begin{align}
\Rmuprime(T_r)_{\text{BNL}} =
\frac{R_\mu}{1 - \sigma_{p'}(T_r)} = 0.0037073015(17)_{\text{stat}}(10)_{\text{syst}}.
\end{align}
Due to the significant changes in the beam characteristics and detectors between the experiments, the systematic uncertainties for the \ac{BNL} and \ac{FNAL} measurements are treated as uncorrelated.
The resulting experimental average is 
\begin{equation}
\Rmuprime(T_r)_{\text{Exp}} = 0.00370730091(36)_{\text{stat}}(28)_{\text{syst}}~~.
\end{equation}

The values of the muon magnetic anomaly,  obtained for each dataset from the observable \Rmuprime through  external factors  following Eq.~\eqref{eq:amueqnewcodata} are

\begin{widetext}
\begin{eqnarray*} 
         a_\mu^{\text{FNAL Run-1}} &=& 116\,592\,0506(539)_{\text{tot}}  (506)_{\text{stat}}  (185)_{\text{syst}}  (26)_{\text{ext}}  \times 10^{-12}~(\SI{460}{ppb}), \\
         a_\mu^{\text{FNAL Run-2/3}} &=& 116\,592\,0701(253)_{\text{tot}}  (235)_{\text{stat}}   (91)_{\text{syst}}  (26)_{\text{ext}} \times 10^{-12}~(\SI{210}{ppb}), \\
    a_\mu^{\text{FNAL Run-4/5/6}} &= &116\,592\,0710(162)_{\text{tot}} (133)_{\text{stat}}   (89)_{\text{syst}}  (26)_{\text{ext}} \times 10^{-12}~(\SI{139}{ppb}), \\
    a_\mu^{\text{FNAL}} &= &116\,592\,0705(148)_{\text{tot}} (114)_{\text{stat}}   (91)_{\text{syst}}  (26)_{\text{ext}} \times 10^{-12}~(\SI{127}{ppb}), \\
    a_\mu^{\text{Exp}} &= &  116\,592\,0715(145)_{\text{tot}} (112)_{\text{stat}}   (88)_{\text{syst}}  (26)_{\text{ext}} \times 10^{-12}~(\SI{124}{ppb}).
\end{eqnarray*}
\end{widetext}

\begin{figure}[h]
\centering
\includegraphics[width=\linewidth]{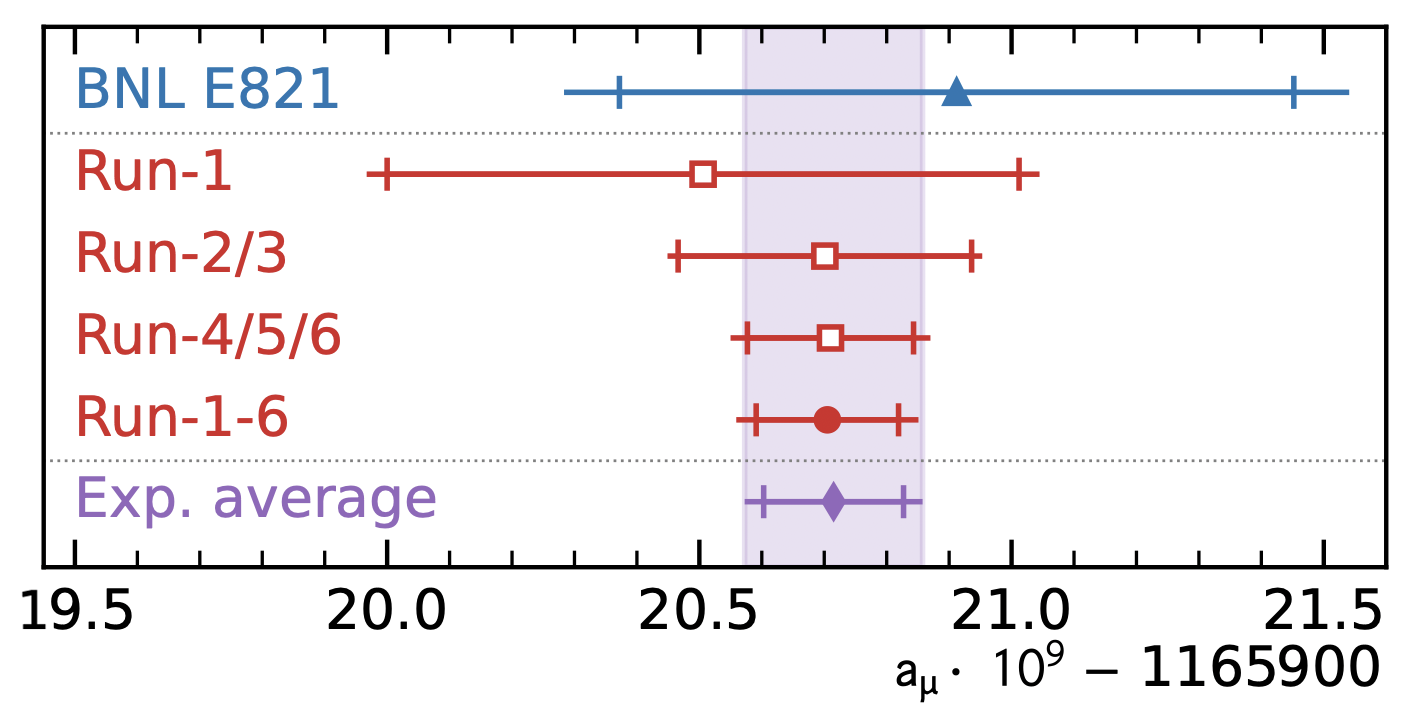}
\caption{From top to bottom:  experimental values of \amu from \ac{BNL} E821, \RunOne, \RunTwoThree, this measurement \RunFourFiveSix, the Fermilab Muon \gm combined measurement Run-1-6, and the combined experimental average (Exp. average). The inner tick marks indicate the statistical contribution to the total uncertainties.}\label{fig:results}
\end{figure}

The total uncertainty is the sum in quadrature of the statistical, systematic, and external-factor contributions. 
These results are displayed in Fig.~\ref{fig:results}. 
Values of $\Rmuprime(T_r)$ and $a_{\mu}$ with extra digits to facilitate further calculations without loss of precision due to rounding are provided in the supplemental material.

\section{Consistency Checks} \label{sec:consistencyOfAmu}

To verify the stability and consistency of the \amu
measurement, extensive cross-checks of \Rmu are performed across various experimental conditions and dataset subdivisions.

The consistency of the extracted \oa values across the \RunFournoRF, \RunFiveX, \RunFiveXY, \RunSixXY datasets provides a particularly stringent cross-check. The \ac{RF} system substantially modifies the beam dynamics, reducing the \ac{CBO} amplitude and muon loss rate each by approximately a factor of five. Despite these large changes in the beam phase-space evolution, the values of \oa remain statistically consistent across all \ac{RF} configurations, demonstrating that residual \ac{CBO} effects and loss-related distortions do not bias the result at the achieved precision.

Within each of the main datasets, the data are also divided into $5$ groups according to each of the following metrics: i) magnet current, ii) total weighted field, iii) horizontal and vertical field gradients, iv) time of day, and v) a random selection.
The value of \Rmu was computed for each of these new datasets individually, and no correlation was found between \Rmu and any of these variables.

\begin{figure}[htbp]
    \centering

    \includegraphics[width=0.98\linewidth]{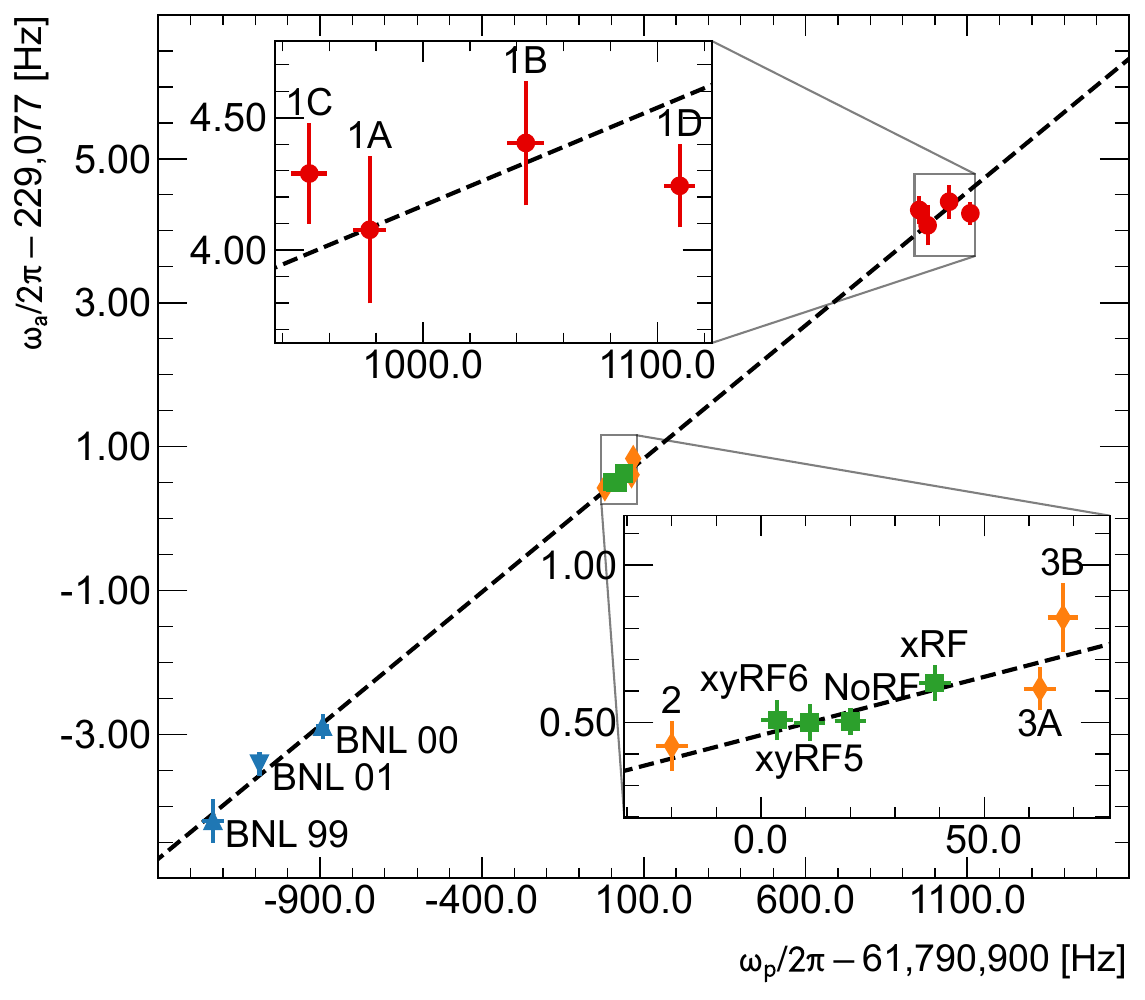}
    \caption{\oa vs. \op for all major Fermilab \gm datasets, together with the three published results from \ac{BNL}. The dashed line indicates the world average value of \Rmu and intercepts $y=0$. The $\chi^2$/NDF of this combination is $12.27/13$.}
    \label{fig:waVsWpAllRuns}
\end{figure}

\begin{figure}[htbp]
    \centering
    \includegraphics[width=0.98\linewidth]{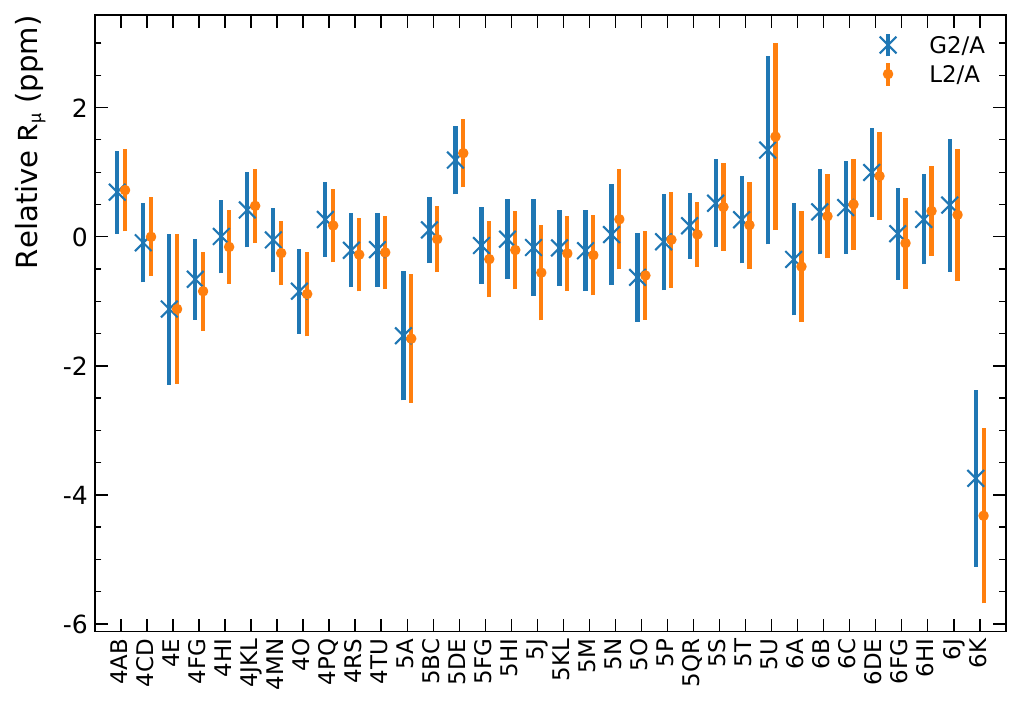}
    \caption{Comparison of relative, raw $\mathcal{R}_{\mu}=\frac{\omega_{a}^{m}}{\langle \op\times M\rangle}$ versus lettered dataset for two analyses (G2 and L2) for the same A-method without beam dynamic correction terms. The error bars represent statistical uncertainties from the \oam-fits only, which dominate by far. Note that these uncertainties are highly correlated between the two analyses.}
    \label{fig:amuVsLetteredDataset}
\end{figure}

In addition to the major dataset values of \amu detailed in Section \ref{sec:amu}, the values of \Rmu and \amu were calculated for each lettered dataset. This enables a series of consistency checks across a wide range of experimental conditions. 
Figure~\ref{fig:waVsWpAllRuns} shows $\omega_a$ vs. $\omega_p$ for all major Fermilab Muon \gm datasets along with the three published \ac{BNL} results, demonstrating consistency with a constant value of \Rmu across a significant range of magnetic field conditions.
The value of \Rmu vs. the lettered sub-datasets for two of the \oa analysis groups (\wagroupTyler and \wagroupSJTU) can be seen in Figure \ref{fig:amuVsLetteredDataset}. 
In this figure, excellent consistency is observed across all lettered sub-datasets, with the possible exception of lettered dataset \texttt{6K}. Once this outlier was identified, a thorough investigation of all experimental parameters was conducted to ensure that no uncontrolled changes in experimental conditions could have affected the result.
The next dataset, \texttt{6LMN} \footnote{This dataset consisted mostly of systematic runs, and had previously been removed from the final production pool due to its low production-quality statistics.}, showed a reversion to the mean, which indicated that this fluctuation was not due to an ongoing change in experimental conditions. The inclusion of lettered dataset \texttt{6K} results in a \SI{24}{ppb} shift in the value of \Rmu in \RunFourFiveSix.

\section{Comparison to Theory} \label{sec:theorycomparison}
\begin{figure}
    \centering
    \includegraphics[width=0.98\linewidth]{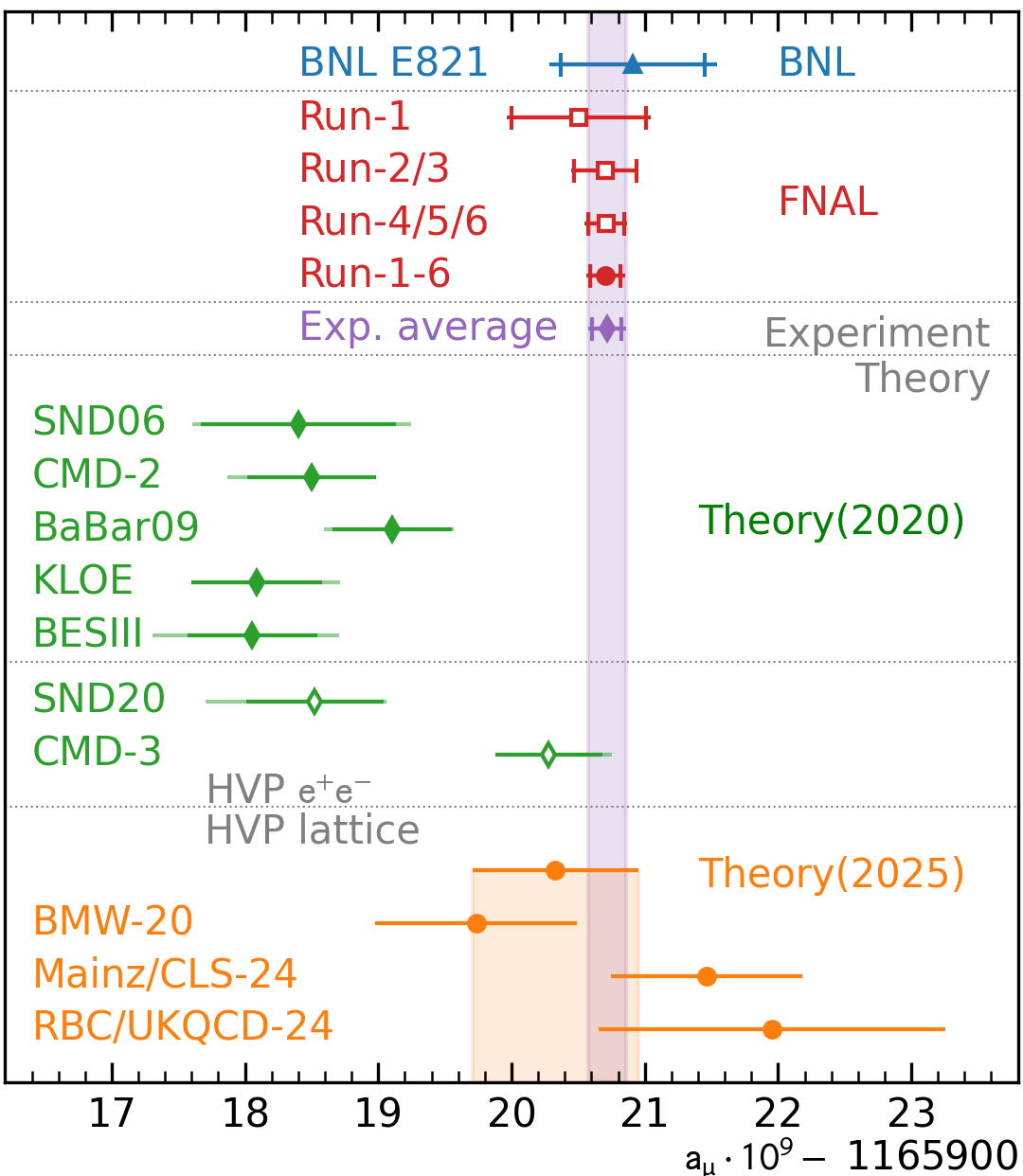}
\caption{Comparison of the experimental value of $a_\mu$ with the Standard Model predictions.}
    \label{fig:DataTheoryComp}
\end{figure}

In recent years, all aspects of the SM theory prediction \amuSM have been scrutinized and refined through continued theoretical and computational efforts. 
These were summarized by the \gm Theory Initiative  White Paper 2025 (WP25)~\cite{ALIBERTI20251}, based on results from~\cite{\SMref}, which supersedes a previous paper published in 2020 (WP20) \cite{Aoyama:2020ynm} by the same community.
While the QED and electroweak contributions are widely considered non-controversial, the
SM prediction of the muon \gm is limited by knowledge of the vacuum fluctuations involving strongly interacting particles, comprising effects called hadronic vacuum polarization (HVP) and hadronic light-by-light scattering. 
The current estimation~\cite{ALIBERTI20251} of the latter term is  $115.5(9.9)$, in units $10^{-11}$, which, compared to the leading HVP term of $7045(61)$, renders the uncertainty of this contribution sub-dominant.
In WP20, the HVP contribution was calculated by means of a dispersion relation technique involving experimental data on the cross-section of electron-positron
annihilation into hadrons. 
In the last 20 years, worldwide efforts using $e^+e^- \rightarrow$ hadrons data at energies below a few GeV have made it possible to achieve a remarkable uncertainty of $0.6\%$ on \amuHLO.
However, a recent result from the CMD-3 collaboration \cite{PhysRevD.109.112002} is in tension with the previous measurements used in \cite{Aoyama:2020ynm}. 
The HVP predictions using different experimental datasets are shown as green diamonds in Fig.~\ref{fig:DataTheoryComp}, with the open symbols representing the datasets published after the Theory Initiative WP20
by CMD-3 and by the SND collaboration \cite{SND20}
.
On the other hand, in recent years, significant progress has been made in first-principles calculations of \amuHLO using lattice QCD. 
In 2021, the BMW collaboration published the first lattice calculation of \amuHLO with sub-percent precision~\cite{Borsanyi:2020mff}. 
Following that result, other lattice groups have verified a partial calculation of \amuHLO in an intermediate window, less prone to systematic effects, and have found internal agreement and tension with the dispersion integral. 
Finally, several groups have produced a full calculation, which has been summarized in the \gm Theory Initiative's 2025 White Paper WP25~\cite{ALIBERTI20251}.
In that paper, only the lattice QCD calculations contribute to \amuHLO, while no attempt has been made to average over the $e^+e^-$ cross section due to significant scatter in the data, in particular in the critical region of the $\rho$ resonance.
The full \amu calculation, making use of lattice QCD for the HLO contribution, is fully compatible with the experimental result
 \begin{eqnarray*} 
    a_\mu^{\text{Exp}} &= &\amuworld ~(\SI{124}{ppb}), \\
    a_\mu^{\text{Theo}} &= &\amuWP ~(\SI{530}{ppb}).
\end{eqnarray*}
Figure \ref{fig:DataTheoryComp} shows the experimental result (purple band) and the WP25 theoretical prediction (orange band), with the hadronic contribution based on lattice calculations.
The three published full lattice calculations averaged to obtain the Standard Model reference value are shown in orange.
In the same figure, the \amu prediction, which uses $e^+e^-\rightarrow$ hadrons data from different experiments, is also shown.
A tension between all previous determinations and the last one from CMD-3 is evident.

In conclusion, the final measurement of the muon magnetic anomaly from the Fermilab Muon \gm experiment is fully compatible with the Standard Model prediction in WP25.
The experimental result is 4.3 times more precise than the theoretical evaluation.
However, the tension between the \amuHLO prediction based on $e^+e^- \rightarrow$ hadrons and the one based on lattice calculation has to be clarified for a firm comparison with the theory to be established.

\section{Conclusion}\label{sec:conclusion}
The Muon \gm experiment at \ac{FNAL} has completed its data taking, and the result for the muon magnetic anomaly, \amu, based on the analysis of all six datasets from \RunOne to \RunSix, is presented. 
The final precision of \amuFNALprecisionppb represents an improvement compared to the \ac{BNL} E821 experiment of a factor of $4.2$ and exceeds the original precision goal of \SI{140}{ppb}. 
The key ingredients for this achievement were a significant increase in statistics, corresponding to about 21 times as many stored muons, and a significant reduction in many systematic uncertainties compared to the \ac{BNL} E821 experiment.
The new world average of  
\begin{eqnarray*} 
    a_\mu^{\text{Exp}} &= &\amuworld ~(\SI{124}{ppb})
\end{eqnarray*}
is a stringent test for any \ac{BSM} scenarios. 
With this final result, the Muon \gm experiment at \ac{FNAL} has made a further step in the long history of the ``magic-momentum'' storage ring technique. Given the relatively evenly distributed final systematic uncertainties, no immediate path exists to achieve another significant improvement in precision with this technique.

The initial phase of the J-PARC Muon \gm/EDM experiment~\cite{10.1093/ptep/ptz030} anticipates a total precision of \SI{400}{ppb}; reaching $\mathcal{O}(\SI{100}{ppb})$ precision, comparable to the \ac{FNAL} result, would require significant improvements in muon yield and additional upgrades~\cite{Venanzoni:2025rsd}.

On the other hand, over the next few years, more progress is expected in determining \amuSM.
Additional results are expected from the lattice evaluation of the hadronic contribution to \amu, which will reduce the current uncertainty to a level similar to the experimental one.
In parallel, several experiments are either reviewing or analysing additional data on the $e^+ e^- \rightarrow hadrons$ cross section to pin down the current tension among the different results, evident in Fig.~\ref{fig:DataTheoryComp}. 
This experimental effort is complemented by theoretical work to clarify the existing data through refined Monte Carlo event generators and the evaluation of missing higher-order radiative corrections~\cite{Aliberti:2024fpq}.
Of interest is the new measurement expected from the Belle II experiment at KEK (Japan). Belle II already published the $e^+ e^- \rightarrow \pi^+ \pi^- \pi^0$ channel~\cite{PhysRevD.110.112005} in the region $0.62$--$3.50$~\unit{GeV} using only part of their collected data (\SI{191}{\per\femto\barn}) observing a cross section that is $2.5 \, \sigma$ larger than the one observed by BaBar.
However, the two-pion cross section, which is by far the dominant contribution, is currently being analysed.

In parallel to the continuing efforts to resolve the current discrepancy in the s-channel $e^+ e^- \rightarrow hadrons$ cross section, a new experiment, MUonE~\cite{CarloniCalame:2015obs,MUonE:2016hru} is being designed to take data at \ac{CERN}. 
The  experiment aims to precisely measure muon-electron scattering to provide an independent determination of the leading-order hadronic contribution to the muon anomaly.

While the main objective of the Muon \gm experiment at \ac{FNAL}, to measure the muon anomaly more than four times more precisely than the \ac{BNL} E821 experiment, was achieved, the recorded data can be used to perform searches for various \ac{BSM} physics. 
The collaboration is working to perform such analyses in three different areas. 
First, a muon EDM, $d^{}_\mu$, would tilt the spin precession plane, producing an up-down oscillation out of phase with the \gm precession.
The analysis of the \RunTwoThree dataset is expected to yield more than an order of magnitude improvement over the existing \ac{BNL} E821 limit of $|d^{}_\mu| < 1.8 \times 10^{-19}\,e\cdot\mathrm{cm}$~\cite{Bennett:2009cx}. 
Second and third, searches for CPT and Lorentz-violating physics and for ultra-light muonphilic scalar and pseudo-scalar Dark Matter candidates are underway.
Both effects would manifest as modulations of \amu over time.

\begin{acknowledgments}

We thank the Fermilab management and staff for their strong support of this experiment,
as well as our university and national laboratory engineers, technicians, and workshops
for their tremendous support. Greg Bock, Lee Lykken, and Rick Ford set the blinding
clock and diligently monitored its stability. We also thank members of the J-PARC Muon
\gmtwo/EDM experiment for the cross-collaboration efforts.

The Muon \gmtwo\ Experiment was performed at the Fermi National Accelerator Laboratory,
a U.S. Department of Energy, Office of Science, HEP User Facility. Fermilab is managed
by Fermi Forward Discovery Group, LLC, acting under Contract No. 89243023CSC000002.
Additional support for the experiment was provided by the U.S. DOE Office of Science
through the offices of HEP, NP, ASCR, by the U.S.-Japan Science and Technology
Cooperation Program in HEP, by the National Science Foundation (USA), by the Istituto
Nazionale di Fisica Nucleare (Italy), by the Science and Technology Facilities Council
(UK) and the Royal Society (UK) under Grant No.~URF$\backslash$R1$\backslash$231503,
by the National Natural Science Foundation of China (Grant No. 12475108, 12305217, 12075151),
by the Ministry of Science, ICT and Future Planning (MSIP), the National Research
Foundation of Korea (NRF), and the Institute for Basic Science under Project Code
No.~IBS-R017-D1 (Republic of Korea), by the German Research Foundation (DFG)
through the Cluster of Excellence PRISMA+ (EXC 2118/1, Project ID 39083149) and the
Cluster of Excellence PRISMA++ (EXC 2118/2, Project ID 39083149), by the European Union
Horizon 2020 research and innovation programme under the Marie Sk\l{}odowska-Curie
Grant Agreements No. 101006726 and No. 734303, by the European Union STRONG 2020 project
under Grant Agreement No. 824093, and by the Leverhulme Trust, LIP-2021-014.

 \end{acknowledgments}

\appendix

\section{Calibration Constants of Trolley Probes}
\begin{table}[!htbp]
    \centering
    \caption{Overview of $\delta^{\text{calib}}$ values (Val.), correlated (Corr. Unc.) and uncorrelated (Uncorr. Unc.) uncertainties for each trolley probe.}
    \label{tb:PPCorrections}
    \begin{tabular}{lrrr}
        \hline
        \multirow{2}{*}{Probe} & \multicolumn{3}{c}{$\delta^{\text{calib}}$ (ppb)} \\
& Val. & Uncorr. Unc. & Corr. Unc. \\
        \hline
        1 & 1555.8 & 4.2 & 32.5 \\
        2 & 1429.5 & 4.7 & 32.9 \\
        3 & 1633.4 & 4.7 & 32.1 \\
        4 & 1439.6 & 6.0 & 32.2 \\
        5 & 1603.5 & 7.0 & 32.6 \\
        6 & 1800.9 & 7.1 & 36.9 \\
        7 & 1997.2 & 9.3 & 36.4 \\
        8 & 1311.7 & 8.1 & 33.6 \\
        9 & 1496.7 & 6.2 & 33.8 \\
        10 & 502.3 & 7.4 & 32.9 \\
        11 & 2915.2 & 8.7 & 37.9 \\
        12 & 1822.6 & 7.2 & 35.9 \\
        13 & 2075.9 & 7.2 & 32.4 \\
        14 & 1359.0 & 4.9 & 33.8 \\
        15 & 1292.1 & 4.5 & 32.1 \\
        16 & 432.0 & 3.3 & 32.7 \\
        17 & 2807.8 & 9.0 & 41.3 \\
        \hline
        total noRF & & 1.7 & 34.2 \\
        total xRF & & 1.6 & 34.2 \\
        total xyRF5 & & 1.6 & 34.2 \\
        total xyRF6 & & 1.6 & 34.2 \\
        \hline
    \end{tabular}
\end{table} \section{Examples of Representative Fits to \oam} \label{app:representative_fit}
This section shows the full fit results for two different fitting approaches: an A-method fit (Table~\ref{tab:fullFitA}), and a ratio A-method (RA-method) fit (Table~\ref{tab:fullFitRA}).  Both fits are to the noRF data, and  both express \oam in terms of the parameter $R$ as in Eq.~\ref{eq:Rdef}.

As noted earlier, the linearized fit model takes the form of ~\ref{eq:waform2} with, in this case, the addition of a term to potentially handle any additional slow drift term (which is consistent with zero)
\begin{multline}
N(t) =  N_{0}e^{-t/\gamma\tau_{\mu}}[1-\kappa\Lambda(t)][1+A\cos(\oam t - \phi_0)] \\
     \times \sum_{\omega_i} \xi_{\omega_i}(t)  + N_A T_A(t).
     \label{eq:fullLinearA}
\end{multline}
The sum ranges over the beam dynamical frequencies and sidebands
\begin{multline}
\omega_i \in \{\omega_{\rm CBO}, 
\omega_{{\rm CBO}\pm  a}, 
\omega_{\rm VW}, 
\omega_{\rm{VW}\pm a},
\omega_y, 
\omega_{y\pm a}, \\
\omega_{{\rm VW}-{\rm CBO}}, 
2\omega_{\rm CBO},
\omega_{y-{\rm CBO}}  \}.
\end{multline}
The damped oscillatory terms driven by the beam motion take the form
\begin{multline}
    \xi_{\omega_i}(t) = e^{-t/\tau_{\alpha_i}}\cos(\omega_i t)\sum_{k \le 3}p_k^{\alpha_i}\left(\frac{t}{\tau_{\alpha i}}\right)^k \\
    + e^{-t/\tau_{\beta_i}}\sin(\omega_i t)\sum_{k \le 3}p_k^{\beta_i}\left(\frac{t}{\tau_{\beta_i}}\right)^k.
\end{multline}
The parameter $\kappa$ normalizes the muon loss term $\Lambda(t)$, accounting for the unknown efficiency of the lost muon detection algorithm.  Finally, the shape of the residual slow gain drift term takes the form
\begin{equation}
    T_A \propto e^{-t/\tau_a}[1+A_a\cos(\omega_a t - \phi_a)], 
\end{equation}
weighted appropriately over the energy spectrum as a function of time.  It sufficed to calculate $T_A(t)$ as a correction to the observed time dependence of the positron rate.

\begin{table*}[tb]
    \centering
    \begin{tabular}{r@{ = }l@{\hspace{8mm}}r@{ = }l@{\hspace{8mm}}r@{ = }l}
    \hline\hline
    parameter                          & value                       & parameter                              & value                       & parameter                                & value \\ \hline
$R$ [ppm]                              & $-98.729    \pm 0.180     $ & $f_{\rm VW}$                           & $2.27707    \pm 0.00044   $ & $p_1^{\beta_{\rm y}}$                    & $-0.00044   \pm 0.00017   $ \\ 
$A_a$                                  & $0.3613200  \pm 0.0000077 $ & $\tau_{\alpha_{\rm VW}}$ [$\mu$s]      & $47         \pm 10        $ & $p_0^{\alpha_({\rm y-a})}$               & $-0.000057  \pm 0.000011  $ \\ 
$\phi_a$                               & $4.1193669  \pm 0.0000297 $ & $p_0^{\alpha_{\rm VW}}$                & $-0.00023   \pm 0.00011   $ & $p_0^{\beta_{({\rm y-a})}}$              & $0.000010   \pm 0.000013  $ \\ 
$N$                                    & $69993830   \pm 1940      $ & $p_2^{\alpha_{\rm VW}}$                & $0.000134   \pm 0.000039  $ & $p_0^{\alpha_{({\rm y+a})}}$             & $0.000012   \pm 0.000010  $ \\ 
$\gamma\tau$                           & $64.416     \pm 0.012     $ & $\tau_{\beta_{\rm VW}}$                & $19         \pm 6         $ & $p_0^{\beta_{({\rm y+a})}}$              & $-0.000001  \pm 0.000012  $ \\ 
$\kappa$                               & $0.00028    \pm 0.00023   $ & $p_0^{\beta_{\rm VW}}$                 & $-0.0023    \pm 0.0011    $ & $f_{({\rm VW-CBO})}$                     & $1.91560    \pm 0.00019   $ \\ 
$N_A$                                  & $0.000026   \pm 0.000045  $ & $p_1^{\beta_{\rm VW}}$                 & $-0.0007    \pm 0.0013    $ & $\tau_{\alpha_{({\rm VW-CBO})}}$         & $28         \pm 4         $ \\ 
$f_{\rm CBO}$                          & $0.370388   \pm 0.000023  $ & $p_0^{\alpha_{({\rm VW-a})}}$          & $0.000034   \pm 0.000023  $ & $p_0^{\alpha_{({\rm VW-CBO})}}$          & $0.00065    \pm 0.00014   $ \\ 
$\tau_{\alpha_{\rm CBO}}$              & $175        \pm 12        $ & $p_0^{\beta_{({\rm VW-a})}}$           & $0.0000057  \pm 0.0000059 $ & $p_3^{\alpha_{({\rm VW-CBO})}}$          & $-0.000067  \pm 0.000026  $ \\ 
${p_0}^{\alpha_{\rm CBO}}$             & $0.001790   \pm 0.000036  $ & $p_0^{\alpha_{({\rm VW+a})}}$          & $-0.000074  \pm 0.000029  $ & $\tau_{\beta_{({\rm VW-CBO})}}$          & $117        \pm 23        $ \\ 
${p_2}^{\alpha_{\rm CBO}}$             & $0.00045    \pm 0.00010   $ & $p_0^{\beta_{({\rm VW+a})}}$           & $-0.00019   \pm 0.00014   $ & $p_0^{\beta_{({\rm VW-CBO})}}$           & $0.000361   \pm 0.000043  $ \\ 
$\tau_{\beta_{\rm CBO}}$               & $18         \pm 5         $ & $f_{\rm y}$                            & $2.21087    \pm 0.00014   $ & $p_1^{\beta_{({\rm VW-CBO})}}$           & $-0.00109   \pm 0.00013   $ \\ 
$p_0^{\beta_{\rm CBO}}$                & $-0.00094   \pm 0.00046   $ & $\tau_{\alpha_{\rm y}}$                & $142        \pm 28        $ & $p_2^{\beta_{({\rm VW-CBO})}}$           & $0.00061    \pm 0.00016   $ \\ 
$p_3^{\beta_{\rm CBO}}$                & $-0.000161  \pm 0.000074  $ & $p_0^{\alpha_{\rm y}}$                 & $-0.000479  \pm 0.000036  $ & $p_0^{\alpha_{\rm 2CBO}}$                & $0.0000065  \pm 0.0000063 $ \\ 
$p_0^{\alpha_{({\rm CBO-a})}}$         & $0.0000294  \pm 0.0000059 $ & $p_1^{\alpha_{\rm y}}$                 & $-0.00079   \pm 0.00026   $ & $p_0^{\beta_{\rm 2CBO}}$                 & $0.0000271  \pm 0.0000062 $ \\ 
$p_0^{\beta_{({\rm CBO-a})}}$          & $-0.00012   \pm 0.00011   $ & $p_2^{\alpha_{\rm y}}$                 & $0.00179    \pm 0.00041   $ & $p_0^{\alpha_{({\rm y-CBO})}}$           & $-0.0000109 \pm 0.0000068 $ \\ 
$\tau_{\alpha_{({\rm CBO+a})}}$        & $14       \pm 15        $   & $p_3^{\alpha_{\rm y}}$                 & $-0.00065   \pm 0.00025   $ & $\tau_{\beta_{({\rm y-CBO})}}$           & $40         \pm 10        $ \\ 
$p_0^{\alpha_{({\rm CBO+a})}}$         & $0.0005     \pm 0.0014    $ & $\tau_{\beta{\rm y}}$                  & $101        \pm 22        $ & $p_0^{\beta_{({\rm y-CBO})}}$            & $0.000093   \pm 0.000035  $ \\ 
$p_0^{\beta_{({\rm CBO+a})}}$          & $-0.000025  \pm 0.000090  $ & $p_0^{\beta{\rm y}}$                   & $0.001021   \pm 0.000074  $ & $p_3^{\beta_{({\rm y-CBO})}}$            & $0.000078   \pm 0.000023  $ \\ 
$\chi^2$                               & $4002 / 4098              $      \\
 \hline\hline
    \end{tabular}
    \caption{Full fit results for a linearized A-method fit to the noRF dataset using the model of Eq.~\eqref{eq:fullLinearA}  All frequencies $f_i = \omega_i/2\pi$ are shown in MHz and all lifetimes $\tau_i$ in \SI{}{\micro\second}.}
    \label{tab:fullFitA}
\end{table*}

The full fit model for the RA-method fit simplifies considerably because of the time-randomization of the data to suppress the vertical beam motion frequencies, for which the constructed ratio has reduced sensitivity. 
The RA method constructs the same ratio $R(t)$ that is used to combine data based on a similar model of the observed rate, with
\begin{equation}
R(t) = \frac{2N(t) - N(t+T_a/2) - N(t-T_a/2)}{2N(t) + N(t+T_a/2) + N(t-T_a/2)},
\label{eq:fullRAModel}
\end{equation}
where $T_a$ is the $\oa$ precession period.  The rate model $N(t)$ uses the nonlinear form
\begin{equation}
\begin{aligned}
N(t) &= [1-                       \kappa\Lambda(t)]e^{-t/\gamma\tau_{\mu}} \\
     & \times [1+A\cdot A_{\rm CBO}^{\oa}(t)\cos(\oa t + \phi_{\rm CBO}^{\oa}(t) - \phi_0)] \\
     &\times [1+A^N_{\rm CBO}\cos(\omega_{\rm CBO}(t)t-\phi^N_{\rm CBO})D_{\rm CBO}(t)] \\
     &\times [1+A^N_{2\rm CBO}\cos(2\omega_{\rm CBO}(t)t-\phi^N_{2\rm CBO})D^2_{\rm CBO}(t)].
\end{aligned}
\end{equation}
The variations in the $\oa$ amplitude and phase from the coupling of the CBO and acceptance are modeled as
\begin{equation}
\begin{aligned}
    A_{\rm CBO}(t)    &= 1 + A_A\cos(\omega_{\rm CBO}(t) t - \phi_A) D_{\rm CBO}(t) \\
    \phi_{\rm CBO}(t) &= A_\phi \cos(\omega_{\rm CBO}(t) t - \phi_\phi) D_{\rm CBO}(t) \\
    D_{\rm CBO}(t)    &= e^{-t/\tau_{\rm CBO}} + C_{\rm CBO}
\end{aligned}
\end{equation}
The time evolution of the frequency $\omega_{\rm CBO}(t)$ due to the varying ensemble average over the muon beam is modeled as
\begin{equation}
\omega_{\rm CBO}(t)\cdot t = \omega_{\rm CBO}^0\cdot t + A_{\omega_{\rm CBO}} e^{-t/\tau_A}.
\end{equation}
Table~\ref{tab:fullFitRA} presents the parameter values for this fit.

\begin{table*}[tb]
    \centering
    \begin{tabular}{r@{ = }l@{\hspace{1cm}}r@{ = }l@{\hspace{1cm}}r@{ = }l}
    \hline\hline
    parameter         & value                 & parameter                         & value                 & parameter                & value \\ \hline
$R$ [ppm]             & $-98.746 \pm 0.187$   & $\omega^0_{\rm CBO}$ [rad/$\mu$s] & $2.327 \pm 0.0002$    & $\phi_{\rm CBOP}$ [rad]  & $4.21 \pm 0.29$        \\
$A$                   & $0.35662 \pm 0.00001$ & $\phi_{\rm CBO}^{N}$ [rad]        & $6.281 \pm 0.016$     & $A^N_{\rm 2CBO}$           & $0.00003 \pm 0.00001$  \\
$\phi_0$ [rad]        & $4.11950 \pm 0.00003$ & $\tau_{\rm CBO}$ [$\mu$s]         & $136 \pm 88$          & $\phi^N_{\rm 2CBO}$ [rad]  & $0.99 \pm 0.31$        \\
$\gamma\tau$ [$\mu$s] & $64.403$ (fixed)      & $A_{\rm A}$                       & $0.00013 \pm 0.00004$ & $C_{\rm CBO}$            & $0.262$ (fixed)        \\
$\kappa$              & $0.003 \pm 0.009$     & $\phi_{\rm A}$ [rad]              & $0.05 \pm 0.31$       & $A_{\omega_{\rm CBO}}$          & $3.3 \pm 4.2$          \\
$A_{\rm CBO}^{N}$   & $0.00147 \pm 0.00002$ & $A_{\rm \phi}$                      & $0.00015 \pm 0.00004$ & $\chi^2$                 & 4125 / 4140 \\
\hline\hline
    \end{tabular}
    \caption{Fit results for a blinded representative RA-method fit to the noRF dataset using the model of Eq.~\eqref{eq:fullRAModel}. 
}
    \label{tab:fullFitRA}
\end{table*}

\providecommand{\noopsort}[1]{}\providecommand{\singleletter}[1]{#1}

\begin{acronym}
  \acro{2D}{two-dimensional}
  \acro{3D}{three-dimensional}

\acro{ADC}{analog-to-digital converter}  
  \acro{ANL}{Argonne National Laboratory}

\acro{BNL}{Brookhaven National Laboratory}
  \acro{BSM}{Beyond the Standard Model}
  
\acro{CBO}{coherent betatron oscillation}
  \acro{COD}{closed orbit distortion}
  \acro{CPU}{central processing unit}
  \acro{CPT/LV}{CPT and Lorentz violation}
  \acro{ctag}{calorimeter tag}
  \acro{CERN}{the European Organization for Nuclear Research}
  \acro{CCW}{counterclockwise}
  \acro{CW}{clockwise}

\acro{DAC}{digital-to-analog converter}  
  \acro{DAQ}{data acquisition}  
  \acro{DQC}{data quality cuts}  
  \acro{DQM}{data quality monitoring}  
  \acro{DFT}{discrete Fourier transform}
  \acro{DM}{dark matter}

\acro{EDM}{electric dipole moment}
  \acro{ESQ}{electrostatic quadrupole}

\acro{FFT}{fast Fourier transform}
  \acro{FID}{free induction decay}
  \acro{FNAL}{Fermi National Accelerator Laboratory}
  \acro{FPGA}{field programmable gate arrays}
  \acro{FR}{Fast Rotation}

\acro{GPS}{global positioning system}
  \acro{GPU}{graphic processing unit}
  \acro{GUI}{graphic user interface}

\acro{HV}{high voltage}
  \acro{HVP}{hadronic vacuum polarization}
  
\acro{IBMS}{Inflector Beam Monitoring System}
  \acro{IRIG-B}{inter-range instrumentation group code B}
  \acro{IC}{integrated circuit}
  \acro{J-PARC}{Japan Proton Accelerator Research Complex}

\acro{LVDS}{low-voltage differential signaling}
  \acro{LED}{light emitting diode}

\acro{MIDAS}{maximum integrated data acquisition system}
 \acro{MRI}{magnetic resonance imaging}
 \acro{MiniSciFi}{Minimally Intrusive Scintillating Fiber}

\acro{NMR}{nuclear magnetic resonance}
  \acro{NNLS}{Non-Negative Least Squares}

\acro{ODB}{online database}

\acro{ppb}{parts per billion}
  \acro{ppm}{parts per million}
  \acro{ppt}{parts per trillion}
  \acro{PEEK}{polyether ether ketone}
  \acro{PLL}{phase-locked loop}
  \acro{POT}{potentiometer}
  \acro{QCD}{quantum chromodynamics}
  \acro{QED}{quantum electrodynamics}
  \acro{PID}{proportional–integral–derivative}
  \acro{PBSC}{polarizing beam splitter cube}
  \acro{PLA}{polylactic acid}

\acro{RF}{radio frequency}
  \acro{RMS}{root mean square}

\acro{SCC}{surface correction coils}
  \acro{SPI}{serial peripheral interface}
  \acro{SM}{Standard Model}
  \acro{SiPM}{Silicon Photo Multiplier}
  \acro{SR}{storage ring}

\acro{TDR}{technical design report}
  \acro{TI}{Theory Initiative}
  \acro{TTL}{transistor–transistor logic}
  \acro{TGG}{terbium gallium garnet}

\acro{UTC}{universal time coordinated}

\acro{VME}{Versa Module European}
  \acro{VTM}{virtual trolley measurement}

\acro{WP2020}{2020 White Paper}
  \acro{WP2025}{2025 White Paper}

\end{acronym}
 \end{document}